\documentclass[]{mn2e}
\usepackage{epsfig}

\title[DIB maps across the Magellanic Clouds]{Mapping atomic and diffuse
interstellar band absorption across the Magellanic Clouds and the Milky Way}
\author[Bailey et al.]{Mandy Bailey$^{1,2}$,
                       Jacco Th.\ van Loon$^{1}$,
                       Peter J.\ Sarre$^{3}$
                       and
                       John E.\ Beckman$^{4,5,6}$\\
$^{1}$Lennard-Jones Laboratories, Keele University, ST5 5BG, UK\\
$^{2}$Astrophysics Research Institute, Liverpool John Moores University, IC2,
     Liverpool Science Park, Liverpool L3 5RF, UK\\
$^{3}$School of Chemistry, The University of Nottingham, University Park,
     Nottingham, NG7 2RD, UK\\
$^{4}$Instituto Astrof\'{\i}sica de Canarias, E-38205 La Laguna, Tenerife,
     Spain\\
$^{5}$Department of Astrophysics, University of La Laguna, E-38205 La Laguna,
     Tenerife, Spain\\
$^{6}$C.S.I.C., E-28040 Madrid, Spain}
\date{Submitted: 28 August 2015; accepted: 17 September 2015}
\pagerange{\pageref{firstpage}--\pageref{lastpage}}
\pubyear{2015}
\begin{document}
\maketitle
\label{firstpage}
\begin{abstract}
Diffuse interstellar bands (DIBs) trace warm neutral and weakly-ionized
diffuse interstellar medium (ISM). Here we present a dedicated, high
signal-to-noise spectroscopic survey of two of the strongest DIBs, at 5780 and
5797 \AA, in optical spectra of 666 early-type stars in the Small and Large
Magellanic Clouds, along with measurements of the atomic Na\,{\sc i}\,D and
Ca\,{\sc ii}\,K lines. The resulting maps show for the first time the
distribution of DIB carriers across large swathes of galaxies, as well as the
foreground Milky Way ISM. We confirm the association of the 5797 \AA\ DIB with
neutral gas, and the 5780 \AA\ DIB with more translucent gas, generally
tracing the star-forming regions within the Magellanic Clouds. Likewise, the
Na\,{\sc i}\,D line traces the denser ISM whereas the Ca\,{\sc ii}\,K line
traces the more diffuse, warmer gas. The Ca\,{\sc ii}\,K line has an
additional component at $\sim200$--220 km s$^{-1}$ seen towards both
Magellanic Clouds; this may be associated with a pan-Magellanic halo. Both the
atomic lines and DIBs show sub-pc-scale structure in the Galactic foreground
absorption; the 5780 and 5797 \AA\ DIBs show very little correlation on these
small scales, as do the Ca\,{\sc ii}\,K and Na\,{\sc i}\,D lines. This
suggests that good correlations between the 5780 and 5797 \AA\ DIBs, or
between Ca\,{\sc ii}\,K and Na\,{\sc i}\,D, arise from the superposition of
multiple interstellar structures. Similarity in behaviour between DIBs and
Na\,{\sc i} in the SMC, LMC and Milky Way suggests the abundance of DIB
carriers scales in proportion to metallicity.
\end{abstract}

\section{Introduction}

The interstellar medium (ISM) plays a critical role in the star formation and
chemical evolution of galaxies. Yet a full understanding remains elusive, with
it being seen only in part or under special conditions. This is true in
particular for the molecular component, as molecular hydrogen is virtually
invisible, challenging our ability to investigate how and when atomic gas
becomes molecular. Even though the exact nature of their presumed molecular
carriers is unknown\footnote{Campbell et al.\ (2015) have recently reported
confirmation of two near-IR diffuse bands as due to gas-phase C$_{60}^+$.},
the diffuse interstellar bands (DIBs) they generally cause in the spectra of
stars are a powerful probe of the molecular ISM and the atomic--molecular
interfaces.

The Magellanic Clouds are two nearby ($\approx60$ and 50 kpc for the SMC and
LMC, respectively) interacting, gas-rich dwarf galaxies with sub-solar ISM
metallicities ($Z\approx\frac{1}{5}$ and $\frac{1}{2}$ Z$_\odot$ for the SMC
and LMC, respectively). The recession velocities ($\sim150$ km s$^{-1}$ for
the SMC and 250--300 km s$^{-1}$ for the LMC) result in Doppler shifts with
respect to absorption in the Milky Way of $\sim3.5$ \AA\ in the SMC and
$\sim5.5$ \AA\ in the LMC at visible wavelengths. While some DIBs are broader
than this, many are narrow enough to separate the Magellanic from the
foreground Galactic absorption (Jenniskens \& D\'esert 1994). This is the case
particularly for the 5780 and 5797 \AA\ DIBs, which are intrinsically strong
and their ratio is a clear diagnostic of exposure to UV irradiation and/or
high electron density.

Hutchings (1966) was the first to report a DIB in the Magellanic Clouds, viz.\
the broad 4430 \AA\ band. Subsequent detections and upper limits were obtained
by Blades \& Madore (1979) and Houziaux, Nandy \& Morgan (1980, 1985).
Ehrenfreund et al.\ (2002) first detected the 5780 and 5797 \AA\ DIBs in the
SMC. While the UV radiation field seems to be the dominant factor affecting
the relative strengths of DIBs in the Magellanic Clouds (Cox et al.\ 2006),
other local conditions matter too (Welty et al.\ 2006), with Cox et al.\
(2007) highlighting metallicity effects both through reduced shielding against
UV light and possibly directly through elemental depletions. Following
pioneering efforts by van Loon et al.\ (2009) to map DIBs across the sky, van
Loon et al.\ (2013) presented such maps towards the Tarantula Nebula in the
LMC, showing a clear relation of DIBs with the diffuse ISM and with the
radiation field.

Here we present high signal-to-noise spectra from a dedicated campaign to map
the 5780 and 5797 \AA\ DIBs and the Ca\,{\sc ii}\,K and Na\,{\sc i}\,D lines
across most of the SMC and a large part of the LMC. These are the first maps
of this kind, revealing how the DIB carriers relate to the neutral and weakly
ionized gas, the diffuse and star-forming ISM, and UV-shielded and irradiated
conditions in these metal-poor environments. At the same time, we also map
the Galactic foreground absorption, revealing small-scale structure in all
these tracers, as well as Ca\,{\sc ii}\,K absorption arising in intermediate-
and high-velocity clouds which are probably located in the Galactic Halo.

\section{Data}
\subsection{Observations}

%
\begin{figure*}
\centerline{\epsfig{figure=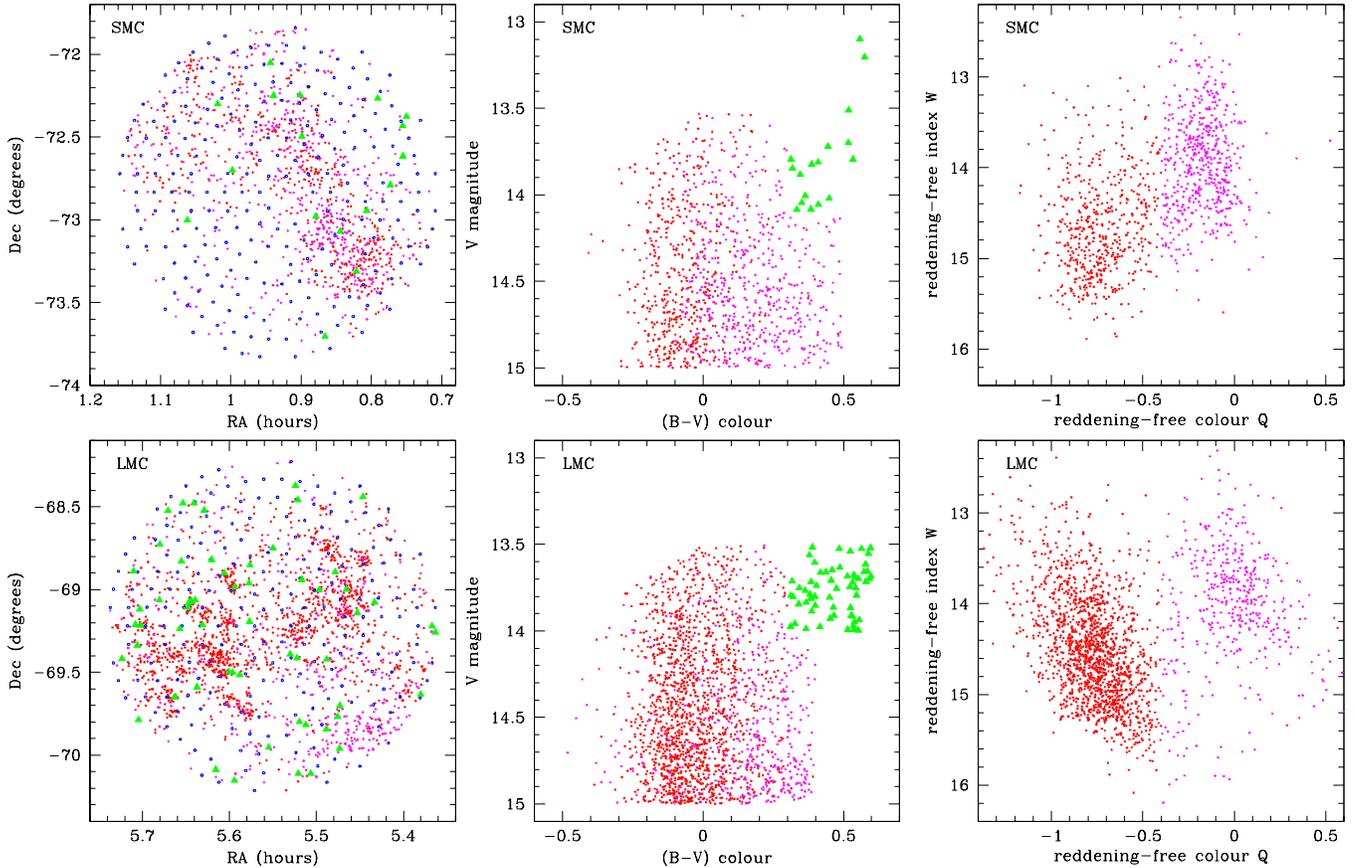,width=178mm}}
\caption[]{Distribution of SMC ({\it top row}) and LMC ({\it bottom row})
targets (high priority: red; low priority: magenta), fiducials (green
triangles) and sky positions (blue dots) on the sky ({\it left}), in optical
colour and brightness ({\it middle}) and in reddening free indices ({\it
right}).}
\end{figure*}

The observations were carried out with the 3.9-m Anglo-Australian Telescope
(AAT) at Siding Spring, Australia, during 3--6 November 2011 (programmes
ATAC/10B/51 and AAT/11B/17; PI: J.Th.\ van Loon). The two-degree field (2dF)
400-fibre positioning system was used in conjunction with the AA$\Omega$
spectrograph mounted at the Coud\'e focus, benefitting from a wide-field
corrector and an atmospheric dispersion compensator. The spectrograph has a
dichroic beam-splitter allowing simultaneous observation of a ``blue'' and a
``red'' spectrum created with volume phase holographic (VPH) reflection
gratings. Each spectrum was sampled with an e2v CCD with 2k pixels along the
spectral and 4k pixels along the spatial direction.

We used the 3200B and 2000R gratings with the 570-nm dichroic, with the
central wavelength set to 3935 \AA\ for the blue arm and 5840 \AA\ for the red
arm. This yielded a spectral resolving power of
$R=\lambda/\Delta\lambda\approx8,000$ (38 km s$^{-1}$), covering wavelength
ranges of 3779--4084 \AA\ in the central fibre and 3751--4054 \AA\ in the end
fibres in the blue, and 5569--6101 \AA\ in the central fibre and 5528--6056
\AA\ in the end fibres in the red.

The SMC field was observed in the first half of the night and the LMC field in
the second. The weather conditions deteriorated throughout the observing run,
with seeing from $1\rlap{.}^{\prime \prime}3$ to $\sim3^{\prime \prime}$ and
increasing levels of cloud cover. Eventually, the SMC and LMC fields received
total integration times of 12.5 hr and 13 hr, respectively, all split into
individual 30-minute integrations.

Before and after each night the following calibration measurements were
performed: bias (electronic offset), dark current level, arc lamp exposures
for wavelength calibration, and flat field observations on the sky and dome --
the latter employing 20-W lamps for the red and 75-W lamps for the blue.

\subsection{Target selection}

The targets, fiducial stars (for acquisition) and sky positions are shown in
figure 1; they were selected from the Magellanic Clouds Photometric Survey
(Zaritsky et al.\ 2002). While the best targets for the study of interstellar
absorption are early-type stars, which are bright and have relatively few and
weak spectral features, their colours may have been reddened by interstellar
dust and we must avoid too heavy a bias against such sight-lines. Therefore,
initially targets were selected to have $13.5<V<15$ mag, and $-1<(B-V)<0.5$
mag for the SMC but $-1<(B-V)<0.4$ mag for the LMC. A higher priority was
given to stars with a reddening-free colour (Johnson 1963; Massey 2007)
$Q=(U-B)-0.72\times(B-V)<-0.4$ mag (red symbols in Fig.\ 1). When plotted
using also the reddening-free Wesenheit index $W=V-3.1\times(B-V)$ (Madore
1982, using the canonical value $R_V=3.1$) the early- and later-type stars
separate much more clearly than in the unreddened colour--magnitude diagram.
Indeed, the early-type stars trace the star forming regions whereas the
later-type stars are concentrated towards the LMC bar in the South--West
corner of the $2^\circ$ field (Fig.\ 1). The later-type stars, while not ideal
targets for our study, were not excluded from selection entirely as they could
still fill some fibres that could not otherwise be allocated. In Appendix A we
describe the distribution over known spectral types, confirming the success of
our selection strategy. Fiducials (green triangles in Fig.\ 1) were selected
from 13th-mag (V) stars with $0.3<(B-V)<0.6$ mag. Targets or fiducial stars
were rejected if other stars were known to affect the signal within the fibre
significantly. Sky positions (blue dots in Fig.\ 1) were chosen to be clear of
known stars within or near to the fibre.

%
\begin{table*}
\caption{List of observed targets. Only the first three objects in each of the
SMC and LMC are shown here. The ID numbers are from our own target lists. Both
equatorial and Galactic coordinates are given as the latter are useful in
interpreting the foreground absorption. Spectral types and luminosity classes
are based on a literature search and only available for 174 SMC and 61 LMC
targets (see Appendix A). B- and V-band photometry are from Zaritsky et al.\
(2002). The full version of this table is
available on CDS.}
\begin{tabular}{lccccccl}
\hline\hline
ID & RA ($^{\rm h}\ ^{\rm m}\ ^{\rm s}$) & Dec ($^\circ\ ^\prime\ ^{\prime\prime}$) &
$l$ ($^\circ$) & $b$ ($^\circ$) & $B$ (mag) & $V$ (mag) & spectral type \\
\hline
\multicolumn{5}{l}{\it SMC:} \\
 983 & 1 05 49.02 & $-$72 48 18.0 & 301.44836 & $-$44.28177 & 14.58 & 14.79&-\\
 942 & 1 04 05.97 & $-$72 43 50.9 & 301.61804 & $-$44.36510 & 14.02 & 14.16&-\\
 762 & 0 58 11.26 & $-$72 36 10.0 & 302.22449 & $-$44.51616 & 14.36 & 14.21&-\\
\multicolumn{5}{l}{\it LMC:} \\
1856 & 5 41 16.56 & $-$69 14 02.2 & 279.59109 & $-$31.43029 & 14.31 & 14.20&-\\
1913 & 5 41 59.34 & $-$69 14 56.5 & 279.60104 & $-$31.36586 & 15.11 & 14.97&-\\
1281 & 5 36 09.39 & $-$69 13 58.0 & 279.65109 & $-$31.88117 & 14.58 & 14.50&-\\
\hline
\end{tabular}
\end{table*}

The {\sc configure} software was used to allocate fibres to targets, up to 8
fiducial stars, and 25 empty positions for sky measurements to be subtracted
from the target stars' spectra. An annealing scheme was employed to optimise
the target allocation, yielding $\approx350$ targets per 2dF field. The final
selection of observed targets -- 338 in the SMC and 328 in the LMC -- is shown
in figure 2 and table 1 (made available in full at CDS).

%
\begin{figure}
\centerline{\vbox{
\epsfig{figure=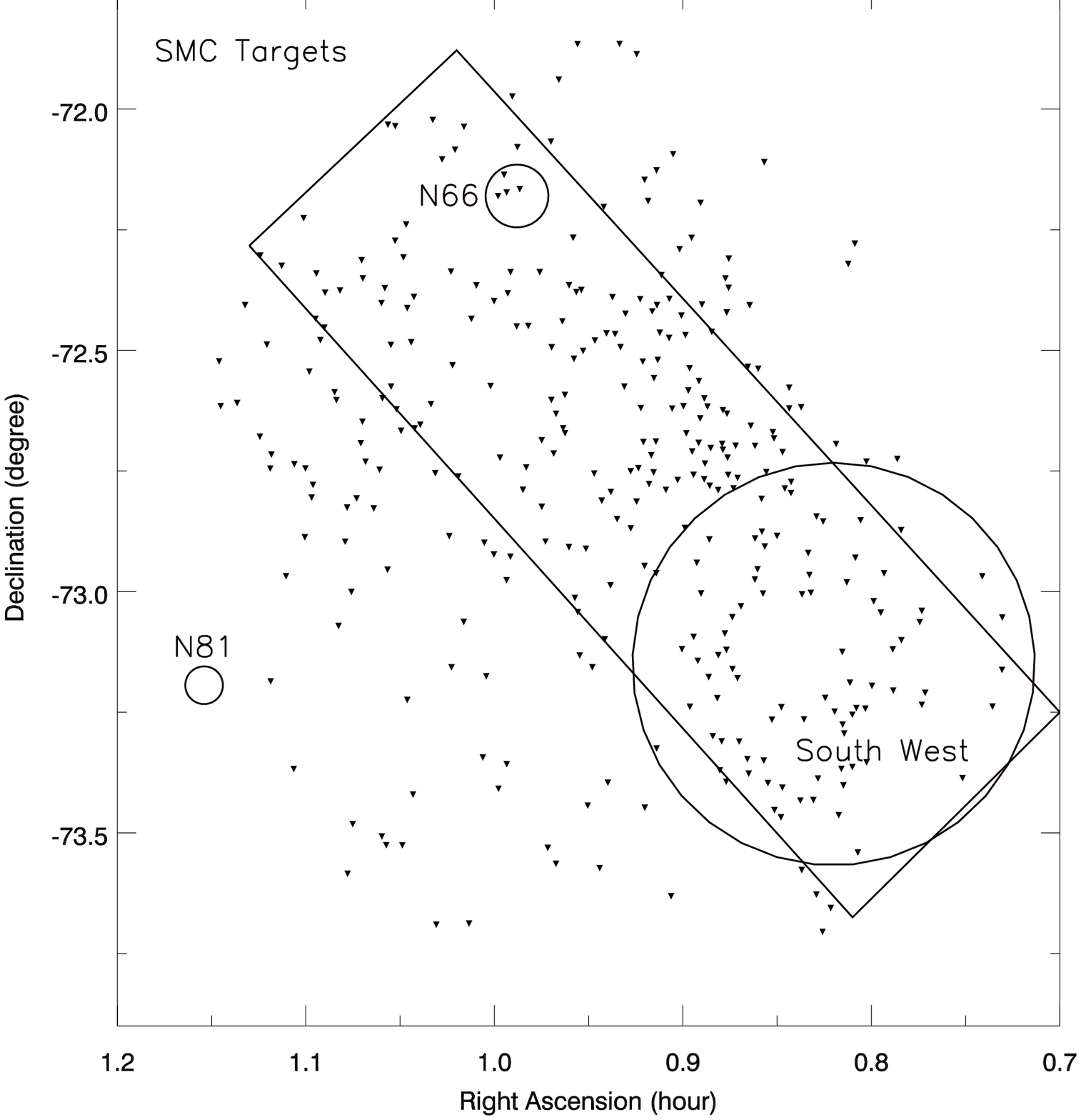,angle=90,width=85mm}
\epsfig{figure=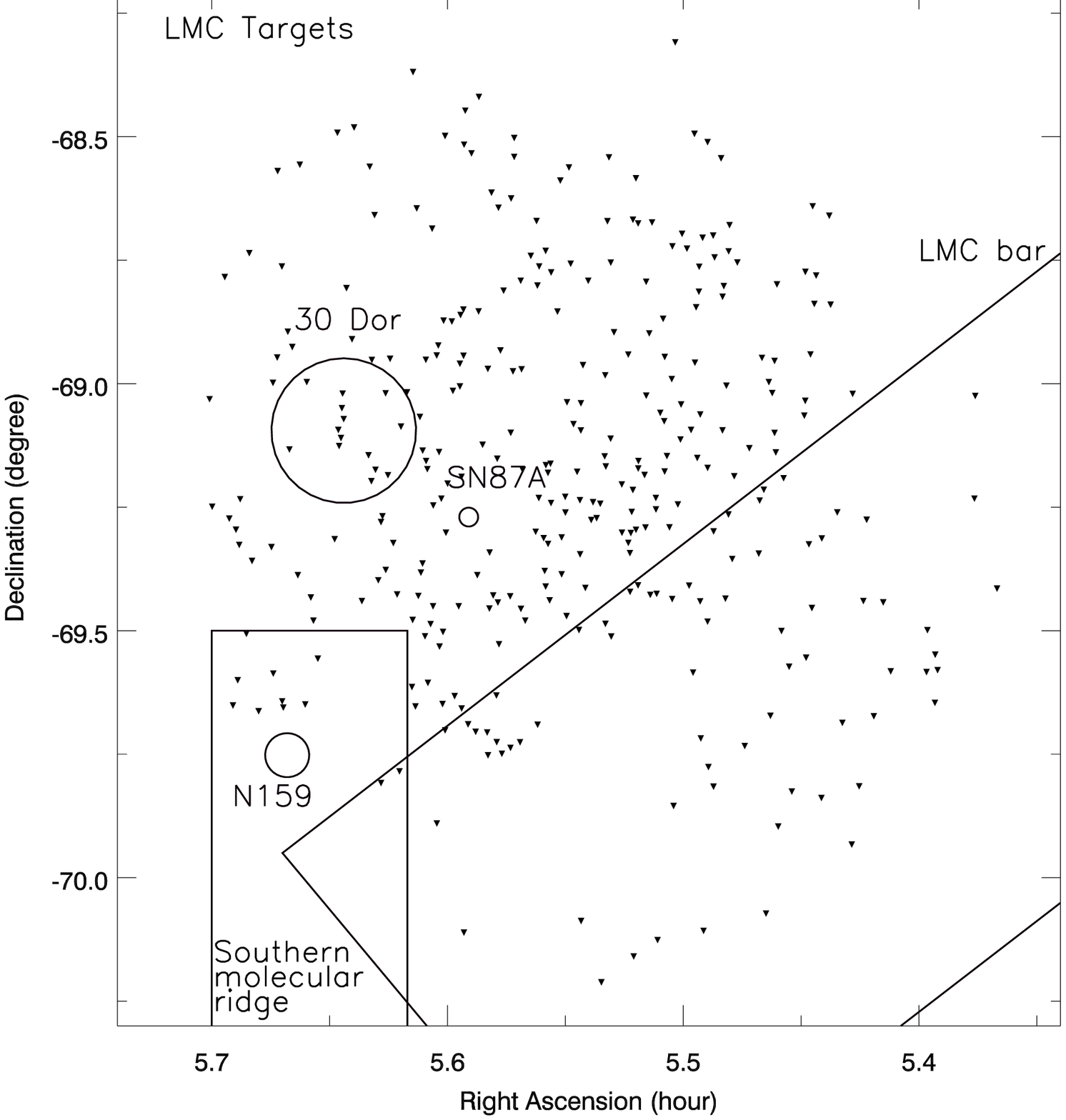,angle=90,width=85mm}
}}
\caption[]{Distribution of observed targets in the SMC ({\it top}) and LMC
({\it bottom}). Some features of interest are indicated, including
30\,Doradus, SN\,1987A, etc.}
\end{figure}

\subsection{Data processing}

The data were processed using the {\sc 2dFdr} tool (Lewis et al.\ 2002). The
reduction steps include: bias subtraction, dark subtraction, flat-fielding,
tram-line mapping of the (curved) spectra on the CCD, spectra extraction, arc
lamp spectral line identification, wavelength calibration, fibre throughput
calibration and sky subtraction. The blue and red spectra are processed
separately.

Accurate subtraction of sky emission lines can be a challenge for fibre-fed
spectrographs. While the standard sky subtraction in {\sc 2dFdr} produced good
results, the very strong telluric Na\,{\sc i}\,D lines were not always removed
entirely (see also the high resolution fibre-fed spectra in van Loon et al.\
2013). The residual telluric emission was accounted for at the analysis stage
(see next section).

The spectra from all of the nights were processed individually but then
combined using custom-written routines in {\sc idl}: first they were rectified
and normalised, then the root-mean-square noise level (standard deviation in a
line-free region) was determined, the reciprocal value of which acted as the
weight in averaging the spectra.

\section{Analysis}

In this work we concentrate on the ISM tracers in the spectra, viz.\ the
Ca\,{\sc ii}\,K (the H component 34.8 \AA\ further to the red is obliterated by
H$\epsilon$), Na\,{\sc i}\,D$_1$ and D$_2$, and the 5780 \AA\ and 5797 \AA\
DIBs, both for the SMC and LMC as well as the Galactic foreground.

%
\begin{table}
\caption{Overview of the Gaussian fitting constraints, in \AA.}
\begin{tabular}{llclc}
\hline\hline
Feature                &
\multicolumn{2}{c}{\llap{---}--------- $\lambda_0$ ---------\rlap{---}} &
\multicolumn{2}{c}{\llap{---}--------- $\sigma$ ---------\rlap{---}}    \\
                       &
guess                  &
range                  &
guess                  &
range                  \\
\hline
\multicolumn{5}{l}{\it SMC, internal:} \\
Ca\,{\sc ii}\,K (blue) &
 3935.6                           & $\pm2.00$ & 0.2 & 0.10\,\ --0.40 \\
Ca\,{\sc ii}\,K (red)  &
 3936.6                           & $\pm1.00$ & 0.2 & 0.10\,\ --0.40 \\
5780 DIB               &
 5783.6                           & $\pm1.00$ & 0.6 &    0.036--0.90 \\
5797 DIB               &
 $\lambda_{5780}$+16.5             & $\pm0.75$ & 0.5 & 0.25\,\ --1.00 \\
Na\,{\sc i}\,D$_2$     &
 5893.0                           & $\pm2.00$ & 0.5 & 0.25\,\ --1.00 \\
Na\,{\sc i}\,D$_1$     &
 $\lambda_{\rm D2}$\ \ +\,\ 5.97   &           & $\sigma_{\rm D2}$   & \\
\multicolumn{5}{l}{\it Galactic, in SMC direction:} \\
Ca\,{\sc ii}\,K        &
 3933.6                           & $\pm2.00$ & 0.3 & 0.15\,\ --0.60 \\
5780 DIB               &
 5780.1                           & $\pm1.00$ & 0.6 &    0.036--0.90 \\
5797 DIB               &
 $\lambda_{5780}$+16.5             & $\pm0.75$ & 0.5 & 0.25\,\ --1.00 \\
Na\,{\sc i}\,D$_2$     &
 5890.0                           & $\pm2.00$ & 0.5 & 0.25\,\ --1.00 \\
Na\,{\sc i}\,D$_1$     &
 $\lambda_{\rm D2}$\ \ +\,\ 5.97   &           & $\sigma_{\rm D2}$   & \\
\multicolumn{5}{l}{\it LMC, internal:} \\
Ca\,{\sc ii}\,K (blue) &
 3937.0                           & $\pm1.00$ & 0.2 & 0.12\,\ --0.30 \\
Ca\,{\sc ii}\,K (red)  &
 3937.5                           & $\pm0.50$ & 0.2 & 0.12\,\ --0.30 \\
5780 DIB               &
 5786.0                           & $\pm1.00$ & 0.6 &    0.036--0.90 \\
5797 DIB               &
 $\lambda_{5780}$+16.5             & $\pm0.75$ & 0.5 & 0.25\,\ --1.00 \\
Na\,{\sc i}\,D$_2$     &
 $\lambda_{\rm D1}$\ \ $-$\,\ 5.97 &           & $\sigma_{\rm D1}$   & \\
Na\,{\sc i}\,D$_1$     &
 5901.0                           & $\pm2.00$ & 0.5 & 0.25\,\ --1.00 \\
\multicolumn{5}{l}{\it Galactic, in LMC direction:} \\
Ca\,{\sc ii}\,K (blue) &
 3933.6                           & $\pm1.00$ & 0.2 & 0.12\,\ --0.30 \\
Ca\,{\sc ii}\,K (red)  &
 3935.0                           & $\pm0.50$ & 0.2 & 0.12\,\ --0.30 \\
5780 DIB               &
 5780.8                           & $\pm1.00$ & 0.6 &    0.036--0.90 \\
5797 DIB               &
 $\lambda_{5780}$+16.5             & $\pm0.75$ & 0.5 & 0.25\,\ --1.00 \\
Na\,{\sc i}\,D$_2$     &
 5890.0                           & $\pm2.00$ & 0.5 & 0.25\,\ --1.00 \\
Na\,{\sc i}\,D$_1$     &
 $\lambda_{\rm D2}$\ \ +\,\ 5.97   &           & $\sigma_{\rm D2}$   & \\
\hline
\end{tabular}
\end{table}

%
\begin{table}
\caption{Spectral line measurements for SMC sight-lines. $EW$ stands for
equivalent width; $e[EW]$ its uncertainty. Flags $f$ are 1 or 0 for a
detection or non-detection, respectively; The full, transposed version of this
table is available on CDS.}
\begin{tabular}{lccc}
\hline\hline
ID                             & 983         & 942         & 762         \\
\hline
\multicolumn{4}{l}{\it SMC, internal:}                                    \\
$EW_{\rm CaII\,K\,blue}$ (\AA)    & 0.224       & 0.185       & 0.310       \\
$e[EW]_{\rm CaII\,K\,blue}$ (\AA) & 0.019       & 0.005       & 0.006       \\
$(EW/e[EW])_{\rm CaII\,K\,blue}$  & \llap{1}1.6 & \llap{3}6.1 & \llap{4}8.5 \\
$EW_{\rm CaII\,K\,red}$ (\AA)     & 0.040       & 0.041       & 0.154       \\
$e[EW]_{\rm CaII\,K\,red}$ (\AA)  & 0.012       & 0.004       & 0.005       \\
$(EW/e[EW])_{\rm CaII\,K\,red}$   & 3.4         & 9.1         & \llap{3}0.5 \\
$EW_{5780}$ (\AA)               & -           & 0.006       & 0.052       \\
$e[EW]_{5780}$ (\AA)            & -           & 0.001       & 0.002       \\
$(EW/e[EW])_{5780}$             & -           & 6.5         & \llap{2}3.3 \\
$EW_{5797}$ (\AA)               & 0.016       & -           & 0.009       \\
$e[EW]_{5797}$ (\AA)            & 0.007       & -           & 0.001       \\
$(EW/e[EW])_{5797}$             & 2.3         & -           & \llap{1}0.7 \\
$EW_{\rm NaI\,D2}$ (\AA)         & 0.096       & 0.096       & 0.246       \\
$EW_{\rm NaI\,D1}$ (\AA)         & 0.129       & 0.064       & 0.164       \\
$e[EW]_{\rm NaI\,D2}$ (\AA)      & 0.008       & 0.004       & 0.005       \\
$e[EW]_{\rm NaI\,D1}$ (\AA)      & 0.008       & 0.004       & 0.004       \\
$(EW/e[EW])_{\rm NaI\,D2}$       & \llap{1}2.2 & \llap{2}1.7 & \llap{4}6.2 \\
$(EW/e[EW])_{\rm NaI\,D1}$       & \llap{1}6.0 & \llap{1}8.0 & \llap{3}8.2 \\
$f_{\rm CaII\,K\,blue}$           & 1           & 1           & 1           \\
$f_{5780}$                     & 0           & 0           & 1           \\
$f_{5797}$                     & 0           & 0           & 1           \\
$f_{\rm NaI\,D}$                 & 0           & 1           & 1           \\
                              &             &             &             \\
\multicolumn{4}{l}{\it Galactic, in SMC direction:}    \\
$EW_{\rm CaII\,K}$ (\AA)         & 0.190       & 0.126       & 0.088       \\
$e[EW]_{\rm CaII\,K}$ (\AA)      & 0.019       & 0.004       & 0.005       \\
$(EW/e[EW])_{\rm CaII\,K}$       & \llap{1}0.0 & \llap{3}1.4 & \llap{1}9.0 \\
$EW_{5780}$ (\AA)               & 0.033       & 0.024       & 0.022       \\
$e[EW]_{5780}$ (\AA)            & 0.005       & 0.002       & 0.003       \\
$(EW/e[EW])_{5780}$             & 6.7         & \llap{1}3.7 & 7.6         \\
$EW_{5797}$ (\AA)               & 0.016       & 0.013       & 0.025       \\
$e[EW]_{5797}$ (\AA)            & 0.002       & 0.002       & 0.002       \\
$(EW/e[EW])_{5797}$             & 6.8         & 5.6         & \llap{1}0.9 \\
$EW_{\rm NaI\,D2}$ (\AA)         & 0.136       & 0.086       & 0.209       \\
$EW_{\rm NaI\,D1}$ (\AA)         & 0.048       & 0.056       & 0.148       \\
$e[EW]_{\rm NaI\,D2}$ (\AA)      & 0.008       & 0.006       & 0.012       \\
$e[EW]_{\rm NaI\,D1}$ (\AA)      & 0.005       & 0.005       & 0.011       \\
$(EW/e[EW])_{\rm NaI\,D2}$       & \llap{1}6.1 & \llap{1}5.0 & \llap{1}7.8 \\
$(EW/e[EW])_{\rm NaI\,D1}$       & 9.2         & \llap{1}1.5 & \llap{1}3.9 \\
$f_{\rm CaII\,K}$                & 1           & 1           & 1           \\
$f_{5780}$                     & 0           & 0           & 0           \\
$f_{5797}$                     & 0           & 0           & 1           \\
$f_{\rm NaI\,D}$                 & 1           & 1           & 1           \\
\hline
\end{tabular}
\end{table}

%
\begin{table}
\caption{As table 3 but for LMC sight-lines.}
\begin{tabular}{lccc}
\hline\hline
ID                             & 1856        & 1913        & 1281        \\
\hline
\multicolumn{4}{l}{\it LMC, internal:}                                    \\
$EW_{\rm CaII\,K\,blue}$ (\AA)    & 0.243       & 0.255       & 0.035       \\
$e[EW]_{\rm CaII\,K\,blue}$ (\AA) & 0.005       & 0.007       & 0.003       \\
$(EW/e[EW])_{\rm CaII\,K\,blue}$  & \llap{4}5.1 & \llap{3}7.3 & \llap{1}1.3 \\
$EW_{\rm CaII\,K\,red}$ (\AA)     & 0.169       & 0.190       & 0.168       \\
$e[EW]_{\rm CaII\,K\,red}$ (\AA)  & 0.005       & 0.009       & 0.005       \\
$(EW/e[EW])_{\rm CaII\,K\,red}$   & \llap{3}1.4 & \llap{2}1.2 & \llap{3}2.2 \\
$EW_{5780}$ (\AA)               & 0.141       & 0.015       & 0.051       \\
$e[EW]_{5780}$ (\AA)            & 0.003       & 0.003       & 0.003       \\
$(EW/e[EW])_{5780}$             & \llap{5}5.5 & 4.8         & \llap{2}0.2 \\
$EW_{5797}$ (\AA)               & 0.030       & 0.017       & 0.012       \\
$e[EW]_{5797}$ (\AA)            & 0.003       & 0.002       & 0.004       \\
$(EW/e[EW])_{5797}$             & \llap{1}1.8 & \llap{1}0.2 & 3.5         \\
$EW_{\rm NaI\,D2}$ (\AA)         & 0.544       & 0.524       & 0.308       \\
$EW_{\rm NaI\,D1}$ (\AA)         & 0.424       & 0.216       & 0.232       \\
$e[EW]_{\rm NaI\,D2}$ (\AA)      & 0.013       & 0.049       & 0.017       \\
$e[EW]_{\rm NaI\,D1}$ (\AA)      & 0.006       & 0.005       & 0.005       \\
$(EW/e[EW])_{\rm NaI\,D2}$       & \llap{4}1.6 & \llap{1}0.8 & \llap{1}7.6 \\
$(EW/e[EW])_{\rm NaI\,D1}$       & \llap{6}6.5 & \llap{3}9.7 & \llap{4}6.1 \\
$f_{\rm CaII\,K\,blue}$           & 1           & 1           & 1           \\
$f_{5780}$                     & 1           & 0           & 1           \\
$f_{5797}$                     & 1           & 0           & 0           \\
$f_{\rm NaI\,D}$                 & 1           & 1           & 1           \\
                              &             &             &             \\
\multicolumn{4}{l}{\it Galactic, in LMC direction:}    \\
$EW_{\rm CaII\,K\,blue}$ (\AA)    & 0.128       & 0.118       & 0.127       \\
$e[EW]_{\rm CaII\,K\,blue}$ (\AA) & 0.005       & 0.004       & 0.006       \\
$(EW/e[EW])_{\rm CaII\,K\,blue}$  & \llap{2}8.5 & \llap{3}0.6 & \llap{2}2.6 \\
$EW_{\rm CaII\,K\,red}$ (\AA)     & 0.034       & 0.026       & 0.121       \\
$e[EW]_{\rm CaII\,K\,red}$ (\AA)  & 0.005       & 0.006       & 0.004       \\
$(EW/e[EW])_{\rm CaII\,K\,red}$   & 7.5         & 4.6         & \llap{3}3.7 \\
$EW_{5780}$ (\AA)               & 0.017       & 0.027       & 0.010       \\
$e[EW]_{5780}$ (\AA)            & 0.001       & 0.004       & 0.001       \\
$(EW/e[EW])_{5780}$             & \llap{1}1.6 & 6.6         & 9.5         \\
$EW_{5797}$ (\AA)               & 0.009       & 0.016       & 0.005       \\
$e[EW]_{5797}$ (\AA)            & 0.001       & 0.004       & 0.002       \\
$(EW/e[EW])_{5797}$             & 7.2         & 4.3         & 2.9         \\
$EW_{\rm NaI\,D2}$ (\AA)         & 0.195       & 0.171       & 0.312       \\
$EW_{\rm NaI\,D1}$ (\AA)         & 0.254       & 0.099       & 0.257       \\
$e[EW]_{\rm NaI\,D2}$ (\AA)      & 0.041       & 0.019       & 0.186       \\
$e[EW]_{\rm NaI\,D1}$ (\AA)      & 0.061       & 0.015       & 0.154       \\
$(EW/e[EW])_{\rm NaI\,D2}$       & 4.7         & 9.0         & 1.7         \\
$(EW/e[EW])_{\rm NaI\,D1}$       & 4.2         & 6.4         & 1.7         \\
$f_{\rm CaII\,K\,blue}$           & 1           & 1           & 1           \\
$f_{5780}$                     & 1           & 1           & 1           \\
$f_{5797}$                     & 0           & 0           & 0           \\
$f_{\rm NaI\,D}$                 & 1           & 1           & 1           \\
\hline
\end{tabular}
\end{table}

The spectral regions around the above absorption features were fitted using a
combination of a first-order polynomial (baseline) and Gaussian functions. The
actual fitting was coded in {\sc idl} using the {\sc mpfit} routines.

The constraints on the fitting are summarised in table 2. In particular, the
wavelength of the 5797 \AA\ DIB was guided by that of the stronger 5780 \AA\
DIB and allowed to vary less to avoid spurious results. Likewise, the
wavelength and width of the D$_1$ (red) component of the Na\,{\sc i} doublet
were tied to that of the stronger D$_2$ (blue) component -- except in the LMC
where the opposite strategy was used because its D$_2$ line coincides with the
Galactic foreground D$_1$ line. Overlapping components were fitted
simultaneously. The peaks of the Gaussian functions were forced to be
negative, but an additional positive-valued double-Gaussian function (with a
fixed separation of 5.97 \AA) was included around the rest wavelengths of the
Na\,{\sc i} D lines to account for telluric emission. It was noted that the
SMC and LMC components of the Ca {\sc ii} K line often showed a second
component. Towards the LMC, a second component was also seen in the Galactic
foreground.

The equivalent width $EW$ was determined by integrating the Gaussian function:
\begin{equation}
EW = \int_{-\infty}^\infty
  A \exp\left(-\ \frac{(\lambda-\lambda_0)^2}{2\sigma^2}\right)
= A \sigma \sqrt{2\pi},
\end{equation}
where $A$ is the peak intensity, $\lambda_0$ the central wavelength and
$\sigma$ the width (with the Full Width at Half Maximum, $FWHM=2.355\sigma$).
The errors $e_0$ resulting from {\sc mpfitexpr} were scaled according to the
$\chi^2$ deviations from the fit and the degrees of freedom $f$ (the number of
spectral elements minus the number of fitting parameters) as in:
\begin{equation}
e = e_0 \sqrt{\chi^2/f}
\end{equation}
Hence the error in the equivalent width is obtained from:
\begin{equation}
e[EW] = \sqrt{2\pi \left((e[A] \sigma)^2+(A e[\sigma])^2\right)}
\end{equation}

All spectra and fits were vetted by eye, guided by the value for $EW/e[EW]$,
and only fits deemed reliable were retained. Rejected data include a few
spectra with broadened Ca\,{\sc ii}\,K absorption arising in the stellar
photosphere (Appendix A presents an overview of the spectral types and their
mostly negligible effect on the measured equivalent widths). The full tables
of measurements are available at CDS, with the first three entries shown in
tables 3 and 4.

Three of our targets (all in the LMC) have also been subject of measurement of
the 5780 \AA\ DIB in previous studies. Our target 1491 (BI\,253) was measured
in a high resolution spectrum by Welty et al.\ (2006); their
$EW_{5780}=0.075\pm0.016$ \AA\ agrees with our $EW_{5780}=0.056\pm0.003$ \AA.
Our targets 1611 (VFTS\,532) and 1625 (VFTS\,604) had been measured in similar
high resolution spectra by van Loon et al.\ (2013); they got
$EW_{5780}=0.175\pm0.007$ \AA\ for star 1611 compared to our
$EW_{5780}=0.194\pm0.006$ \AA, and $EW_{5780}=0.123\pm0.009$ \AA\ for star 1625
compared to our $EW_{5780}=0.137\pm0.003$ \AA. All three comparisons are
consistent within 2--3$\sigma$. Reassuringly, the measurements of the sodium
lines also agreed very well: for star 1611, $EW_{\rm NaI\,D1}=0.463$ and 0.466
\AA\ in van Loon et al.\ (2013) and this work, respectively, while for star
1625 these values were $EW_{\rm NaI\,D1}=0.467$ and 0.451 \AA, respectively.

\section{Results}

%
\begin{figure*}
\centerline{\vbox{\hbox{
\epsfig{figure=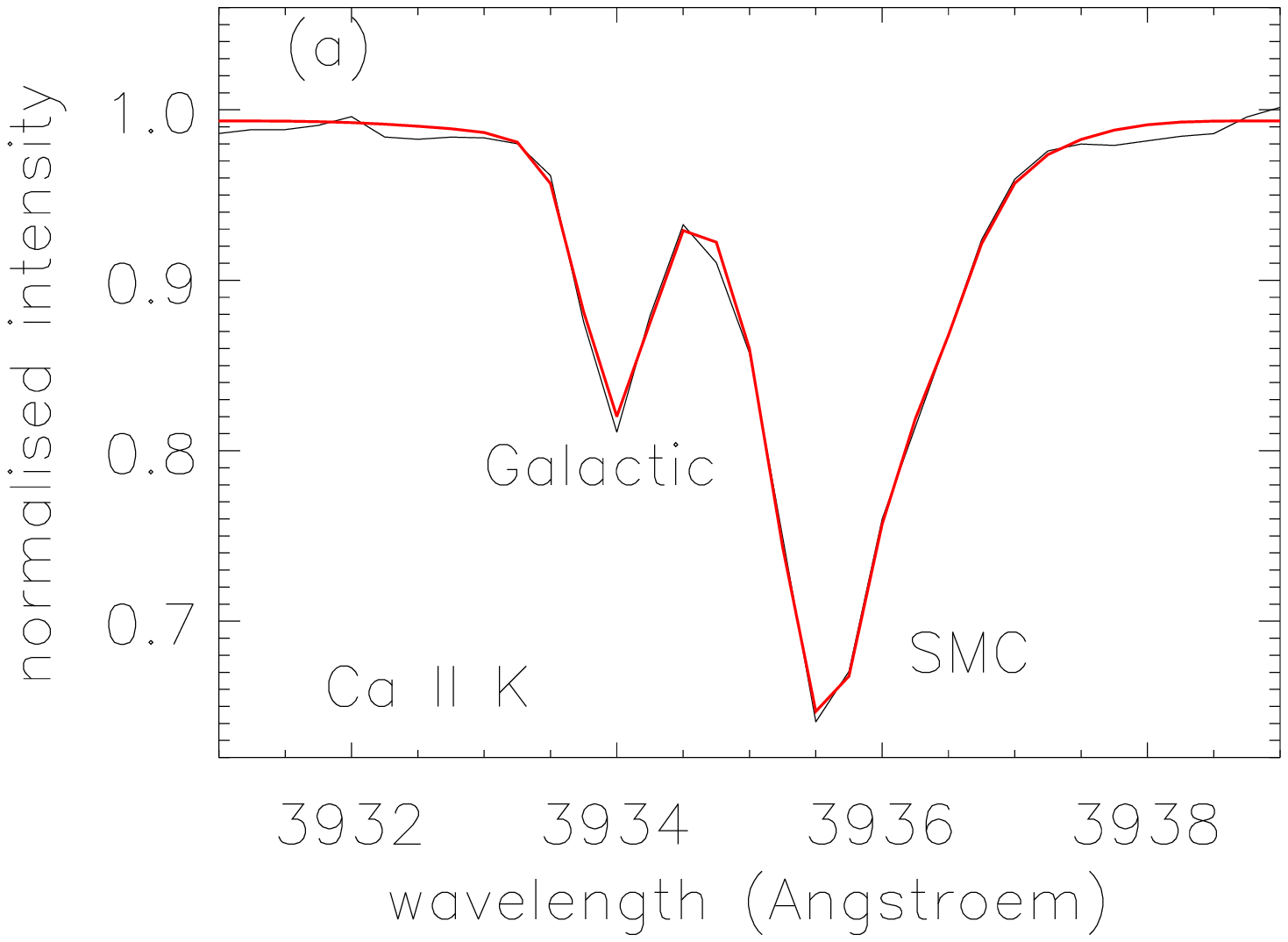,width=44mm}
\epsfig{figure=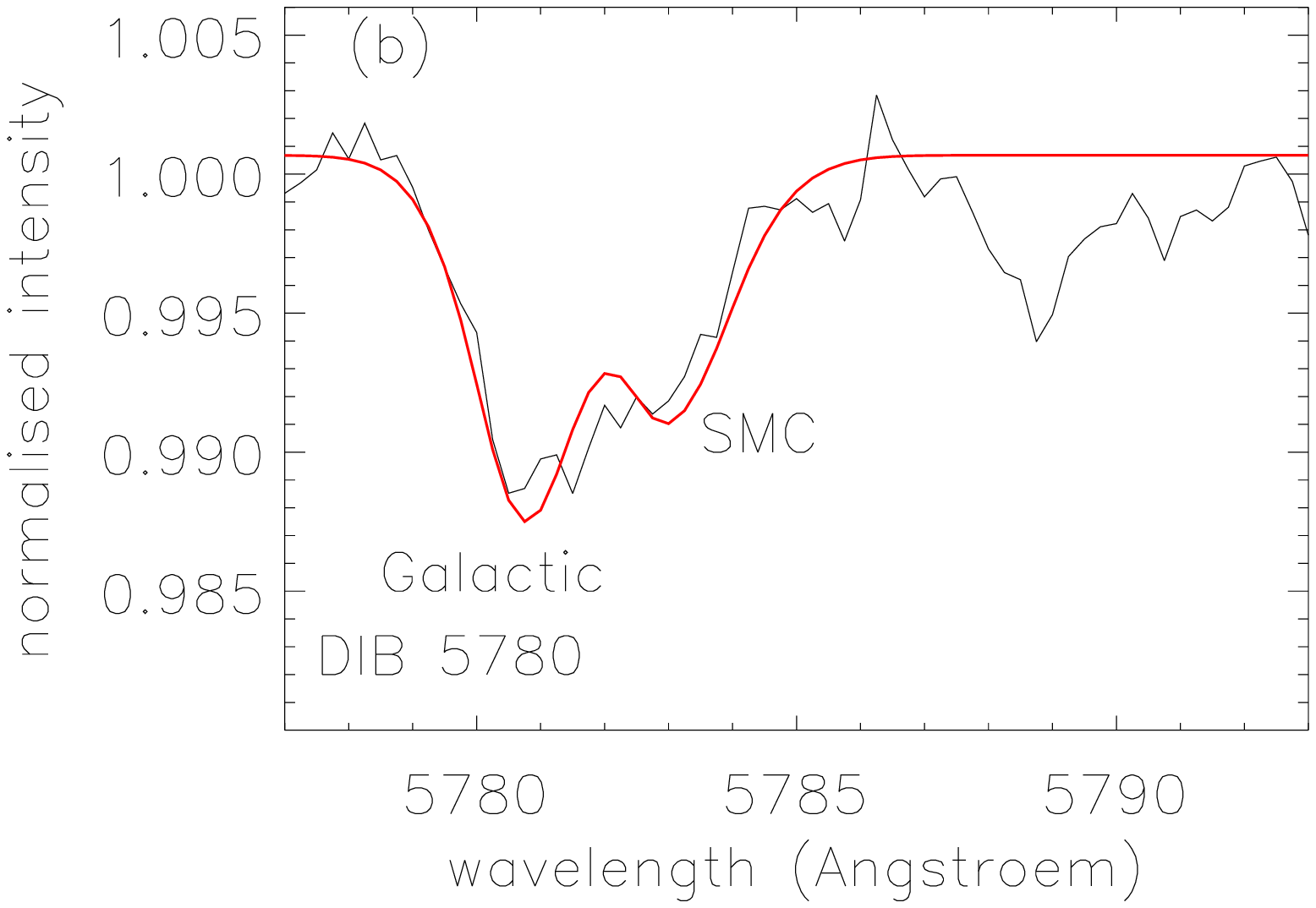,width=44mm}
\epsfig{figure=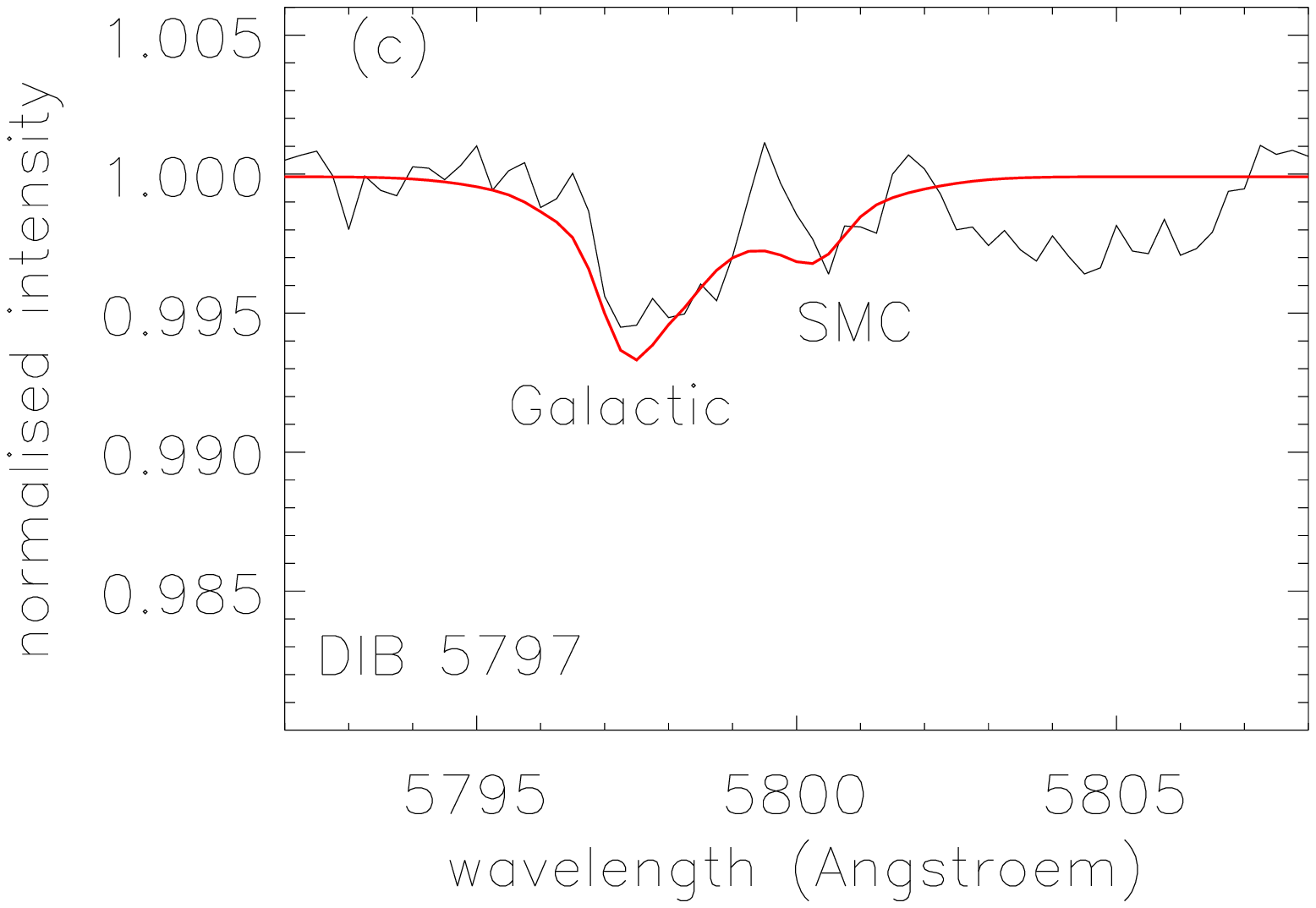,width=44mm}
\epsfig{figure=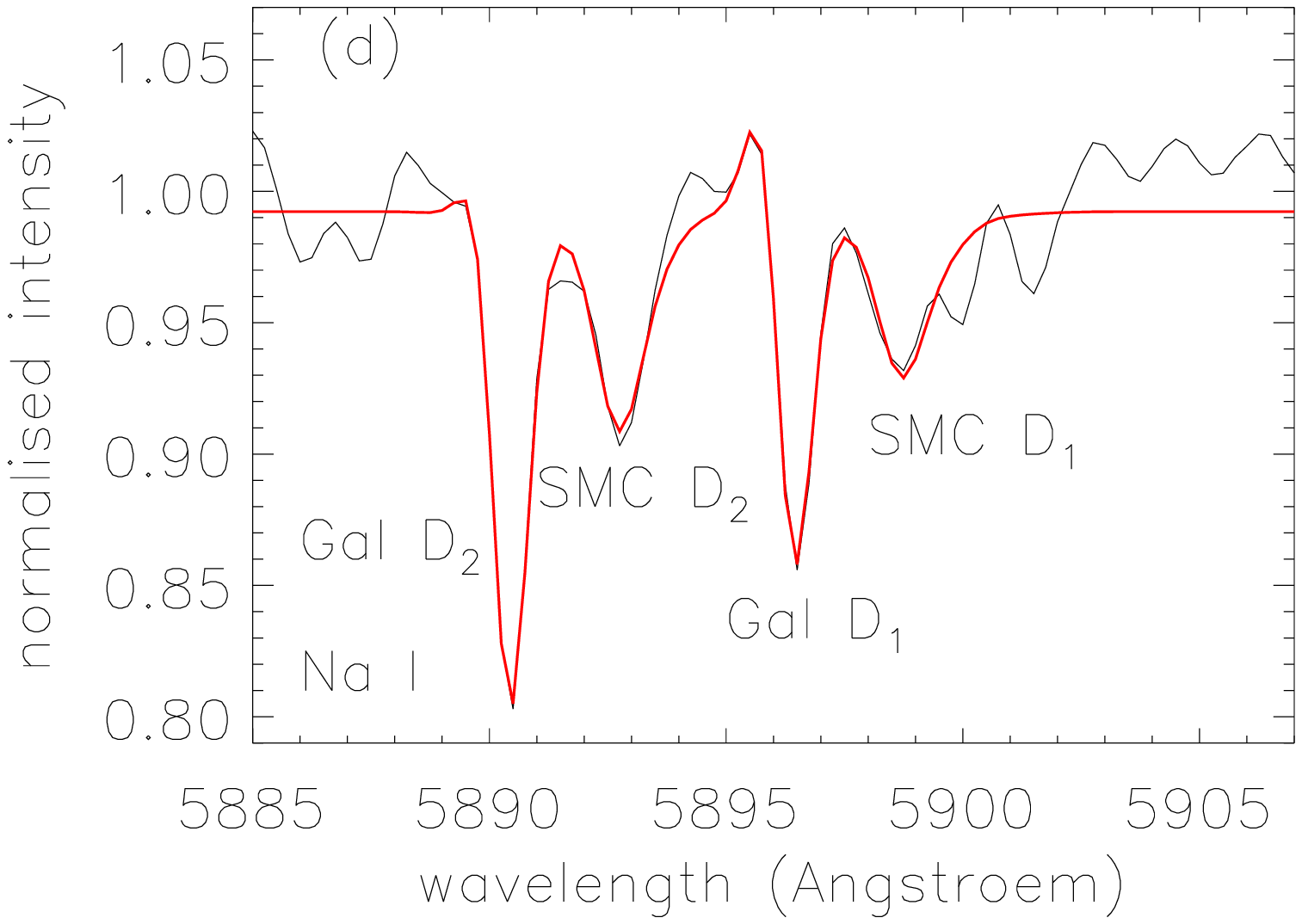,width=44mm}
}\hbox{
\epsfig{figure=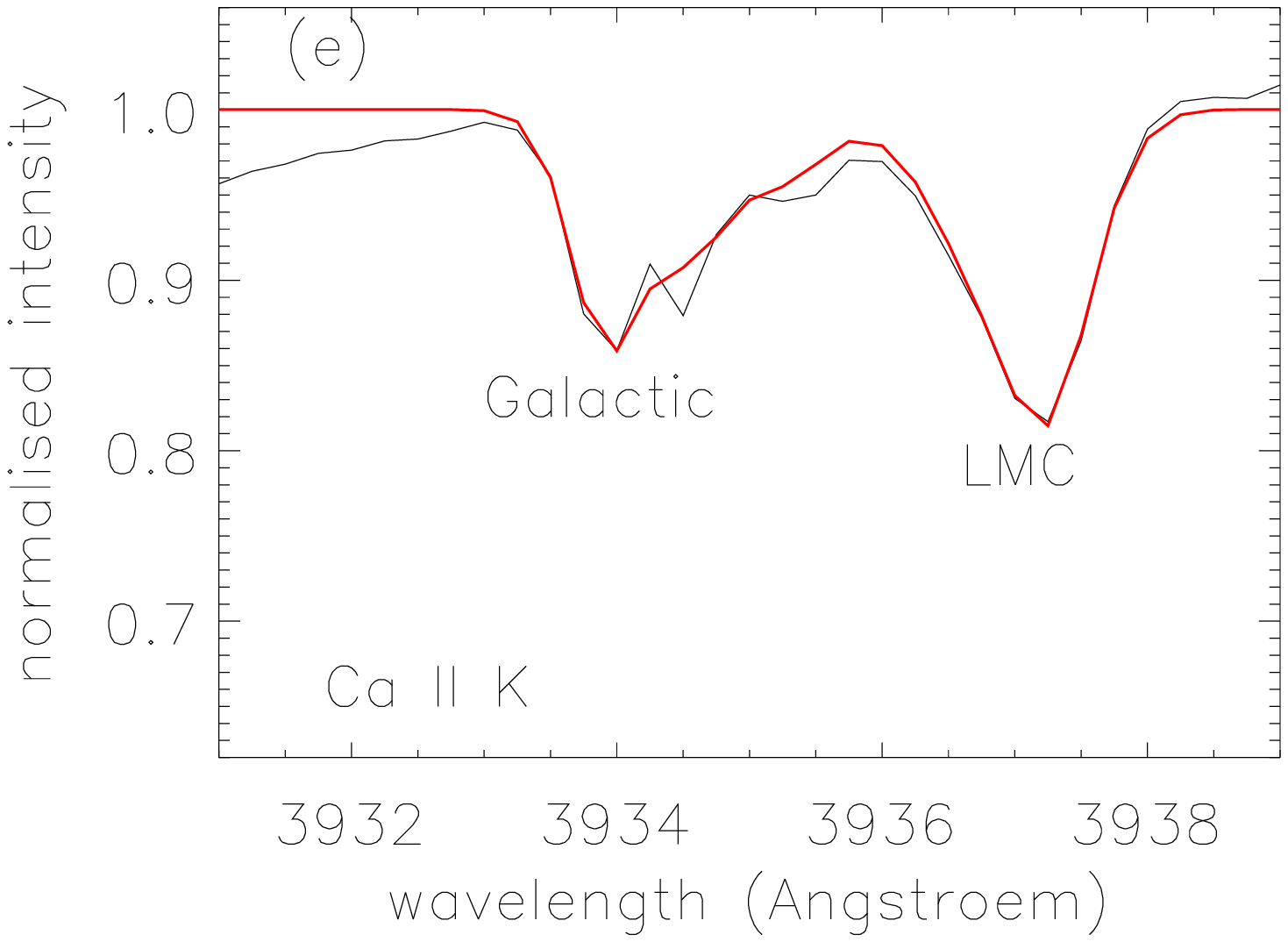,width=44mm}
\epsfig{figure=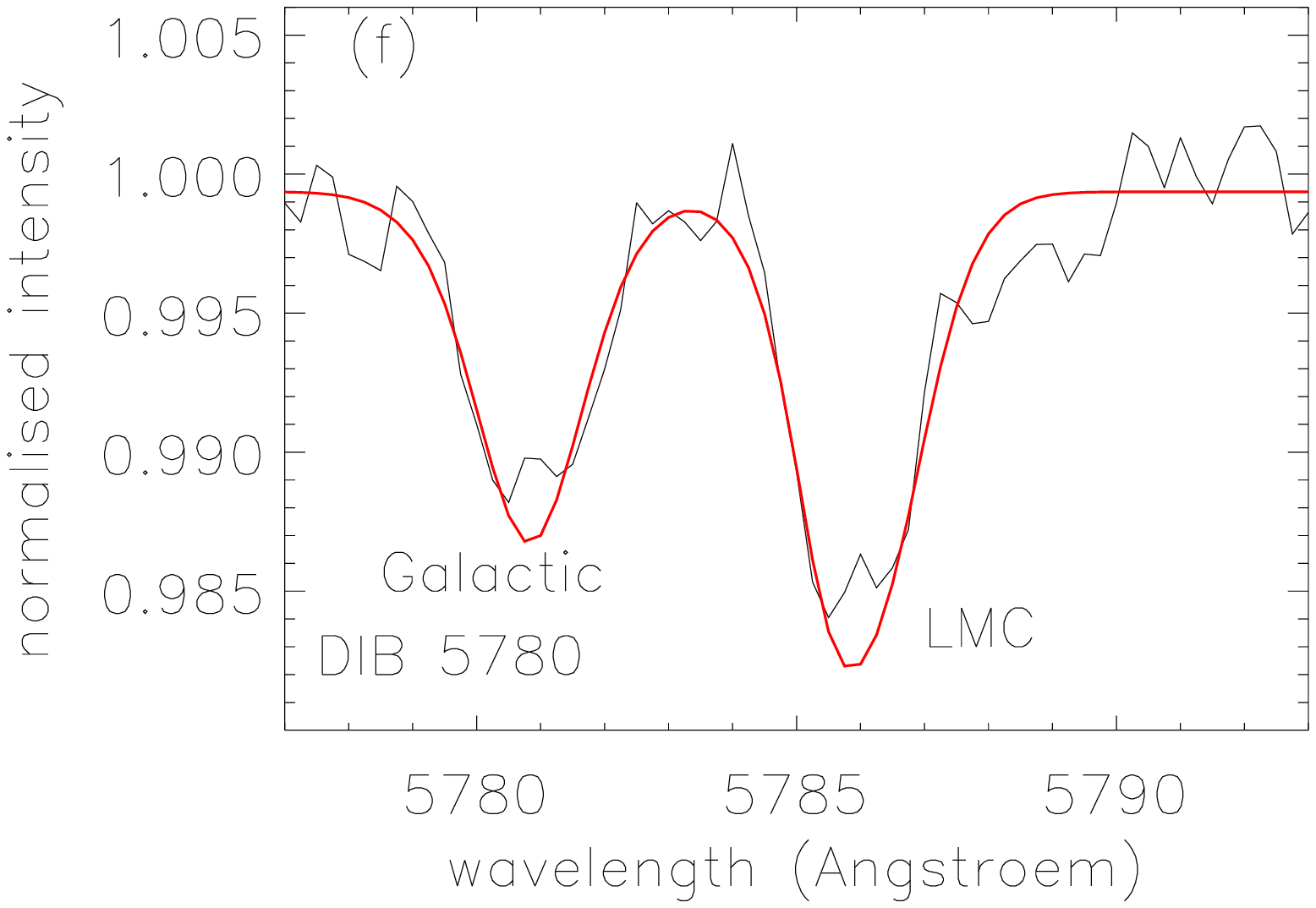,width=44mm}
\epsfig{figure=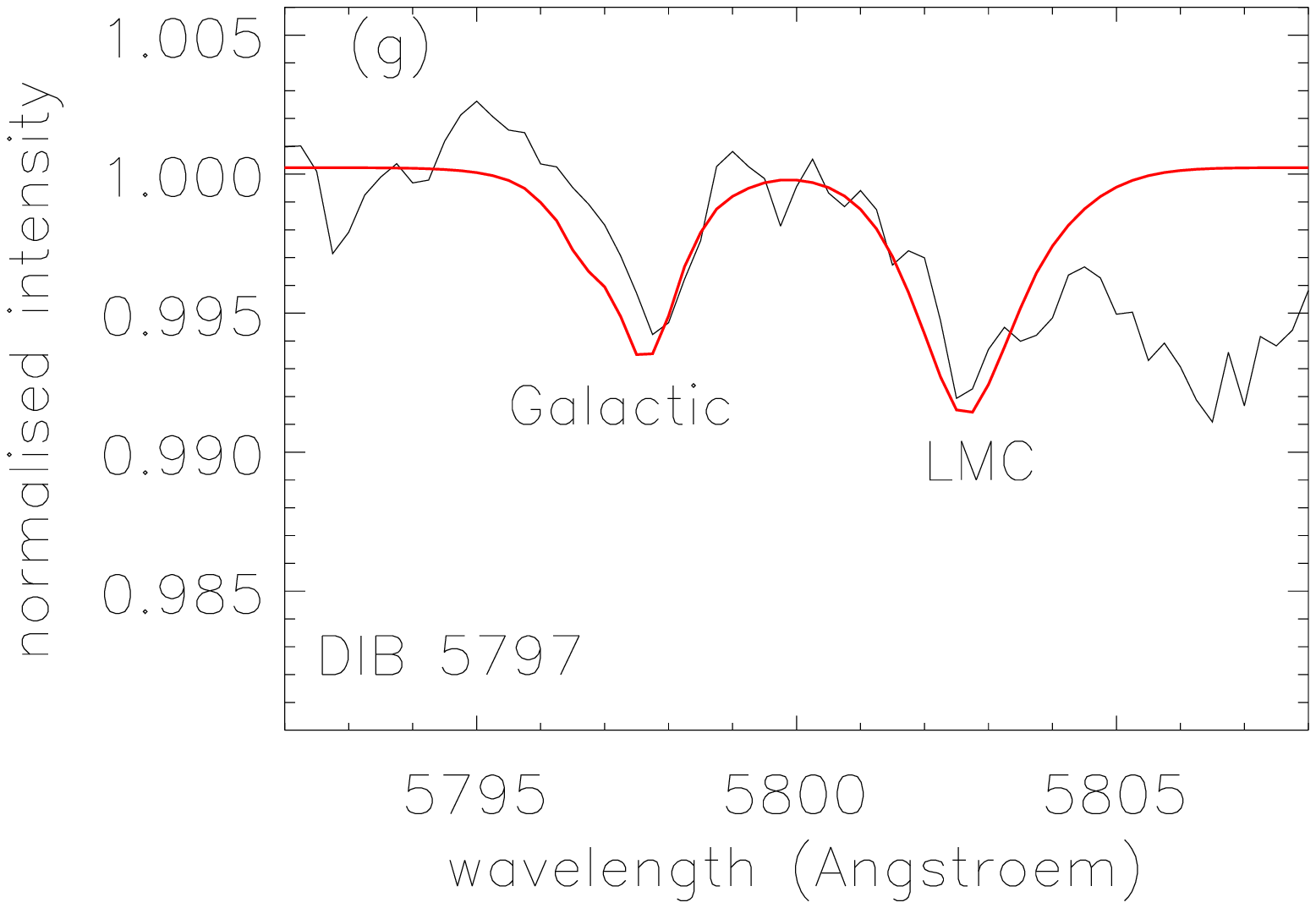,width=44mm}
\epsfig{figure=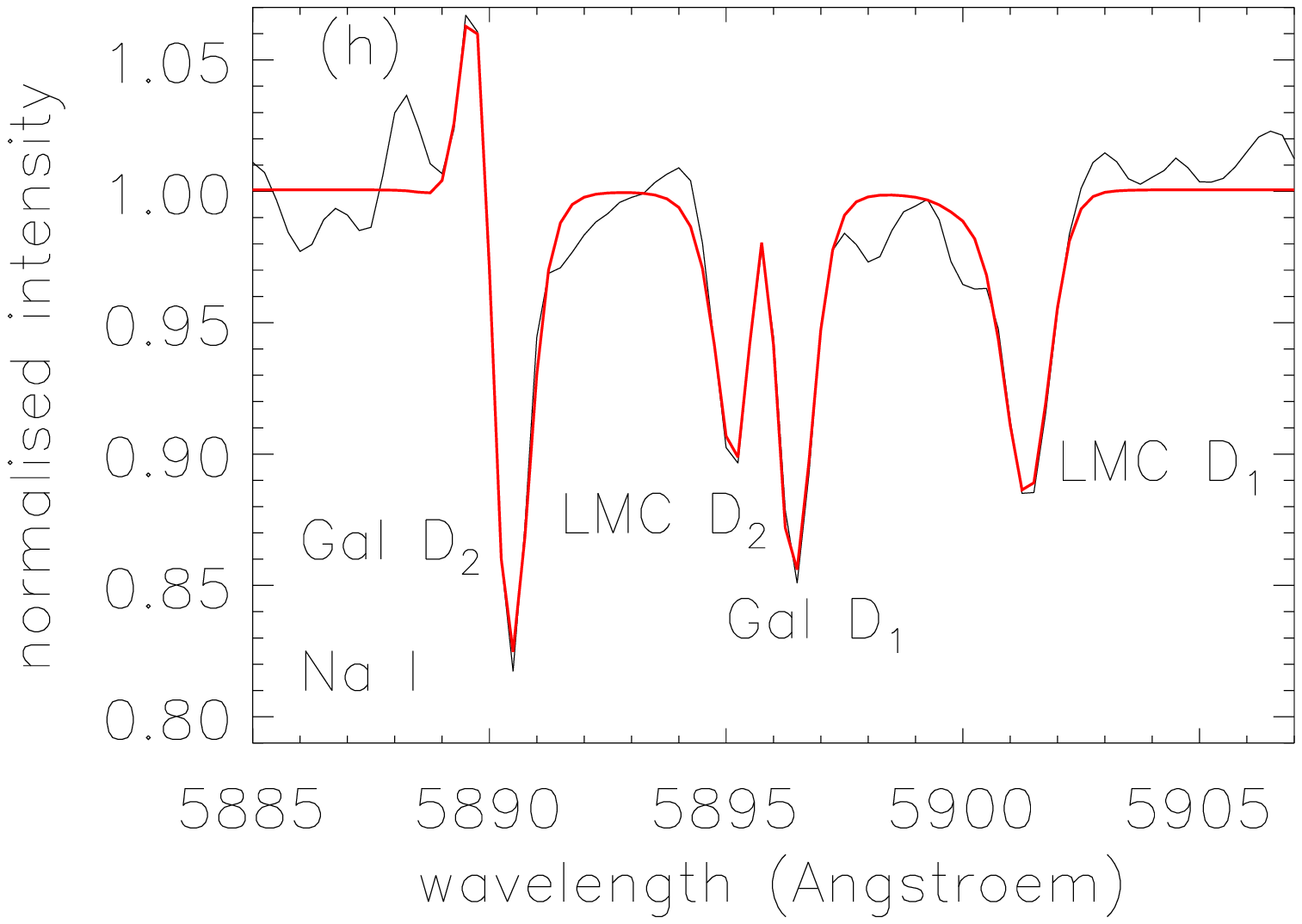,width=44mm}
}}}
\caption[]{Average spectra for the SMC ({\it top row}) and LMC ({\it bottom
row}) around the ({\it from left to right}:) Ca\,{\sc ii}\,K, 5780 \AA\ DIB,
5797 \AA\ DIB and Na\,{\sc i}\,D lines. The average fits are overplotted in
red.}
\end{figure*}

The average line profiles are shown in figure 3 (with the fits overplotted in
red). From the fitted line shape, two velocity components of the
Ca\,{\sc ii}\,K line are clearly visible in the Galactic foreground towards
the LMC (Fig.\ 3e). Many spectra recorded towards individual stars also showed
a second component internal to the SMC and LMC. The redshift of the LMC is
large enough for its 5780 \AA\ DIB to be separated from that of the Galactic
foreground (Fig.\ 3f). Towards the SMC the two velocity components are blended
(Fig.\ 3b), but simultaneous fitting of two Gaussian functions worked well in
accounting for their respective contributions. The narrower 5797 \AA\ DIB is
resolved both towards the LMC (Fig.\ 3g) and towards the SMC (Fig.\ 3c). The
Na\,{\sc i}\,D lines are well separated for the SMC (Fig.\ 3d) but towards the
LMC the LMC D$_2$ component is close to the D$_1$ component of the Galactic
foreground absorption (Fig.\ 3h). But again, simultaneous fitting worked well
in resolving the two components. This is confirmed in the maps we show later,
which -- as expected -- are very different for the different components.

While the spectral resolving power -- and intrinsic width of the DIBs -- did
not allow us to identify multiple kinematic components other than (sometimes)
a second one in the Ca\,{\sc ii}\,K, it is very likely that multiple kinematic
components are present in all absorption features and sight-lines (cf.\ van
Loon et al.\ 2013). Likewise, the Ca\,{\sc ii}\,K and Na\,{\sc i}\,D
absorption may be saturated in some of these components, and an increase of
their equivalent width can be caused by stronger line wings as well as
additional kinematic components. The D$_1$ line has half the intrinsic
strength of that of the D$_2$ line and should therefore be less affected by
saturation effects. While we caution against the use of the equivalent widths
of the  Ca\,{\sc ii}\,K and Na\,{\sc i}\,D lines in an absolute sense, a
comparative analysis can still be meaningful.

Remarkably, the Ca\,{\sc ii}\,K absorption towards the SMC is relatively
strong compared to the Galactic foreground and LMC sight-lines; the DIBs are
considerably weaker, though it must be said that the bulk of the absorption
within the SMC is concentrated in areas much smaller than the $2^\circ$ field
and individual sight-lines can show quite strong DIBs.

%
\begin{figure*}
\centerline{\vbox{\hbox{
\epsfig{figure=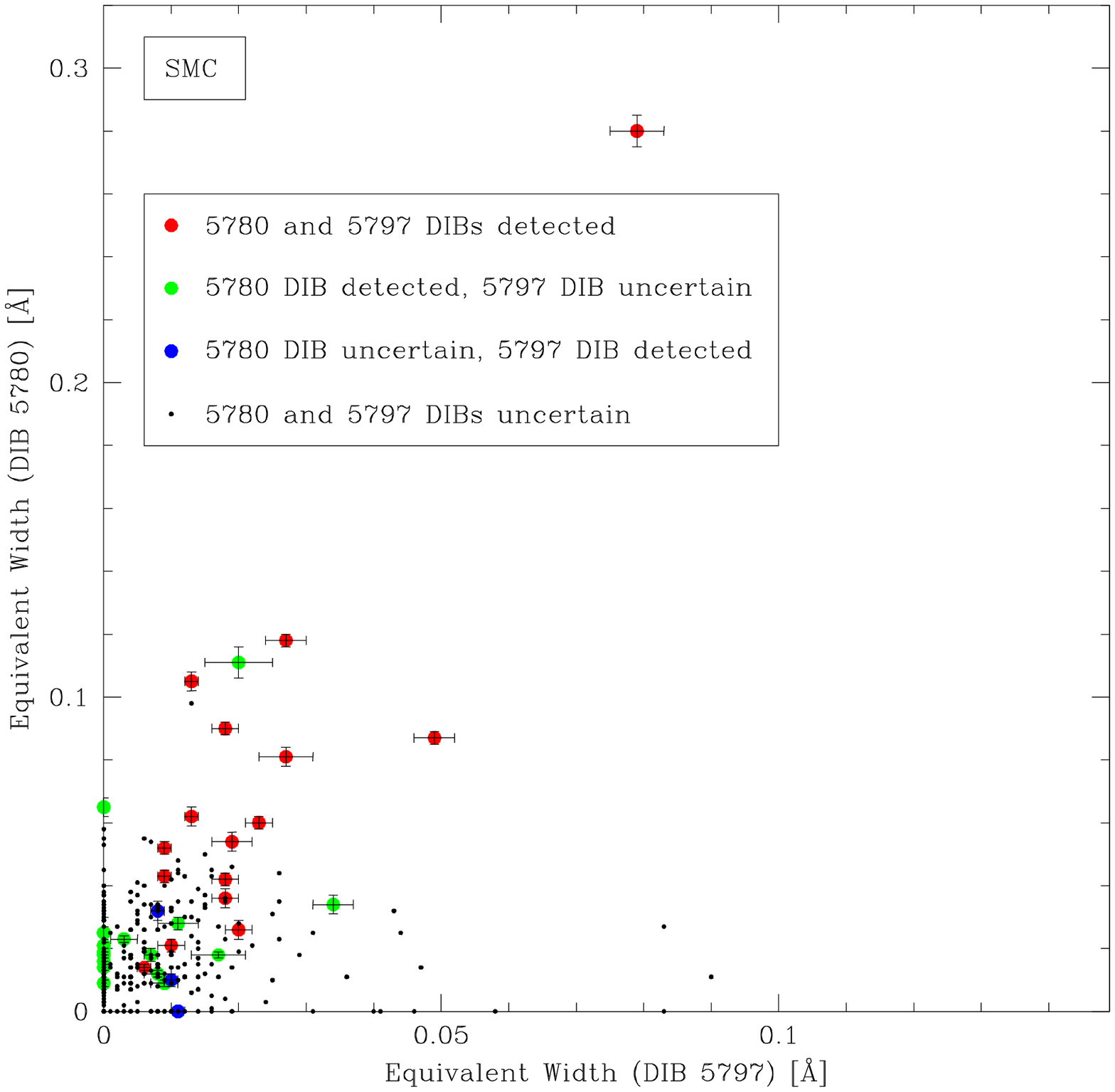,width=88mm}
\epsfig{figure=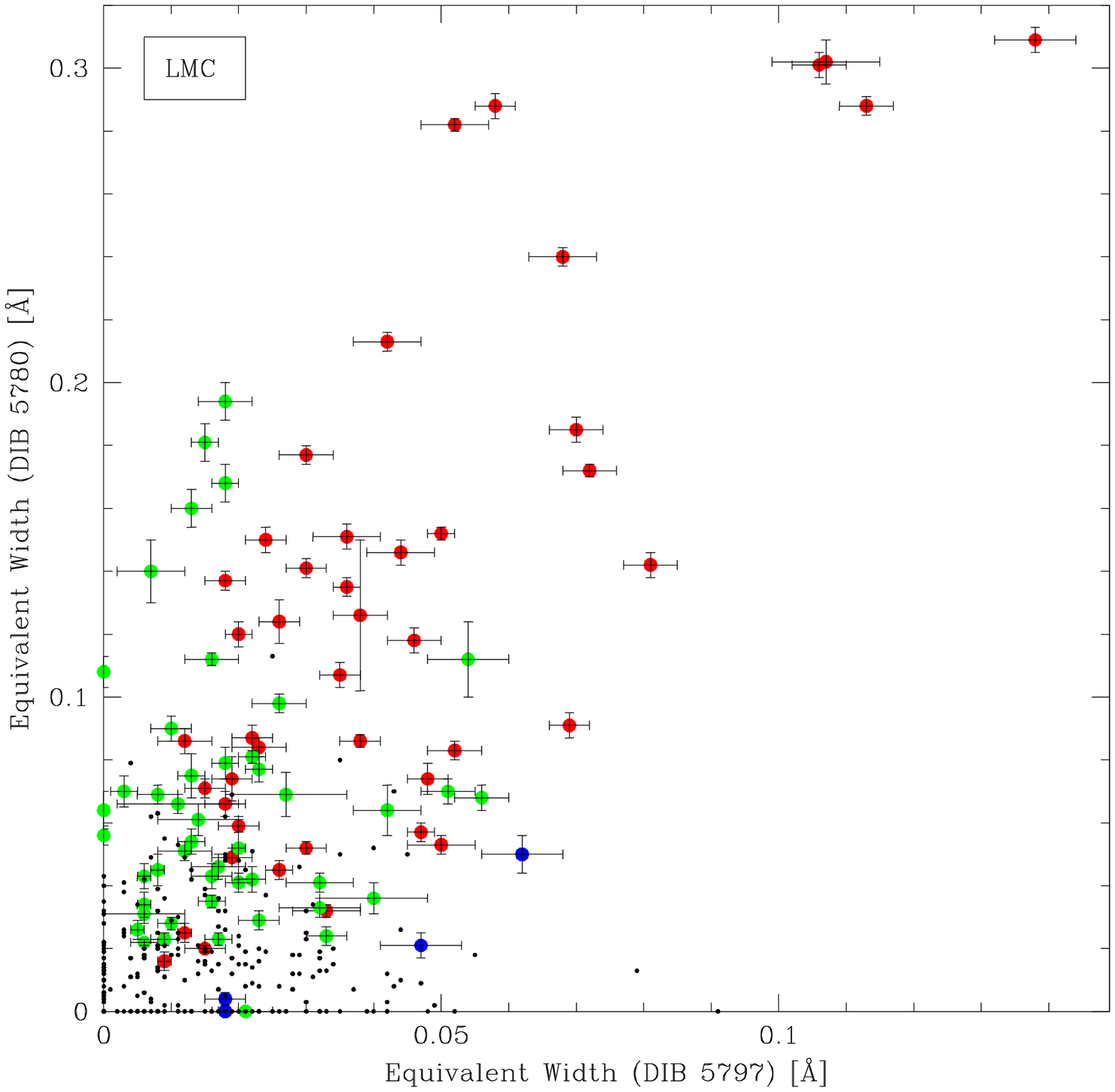,width=88mm}
}\hbox{
\epsfig{figure=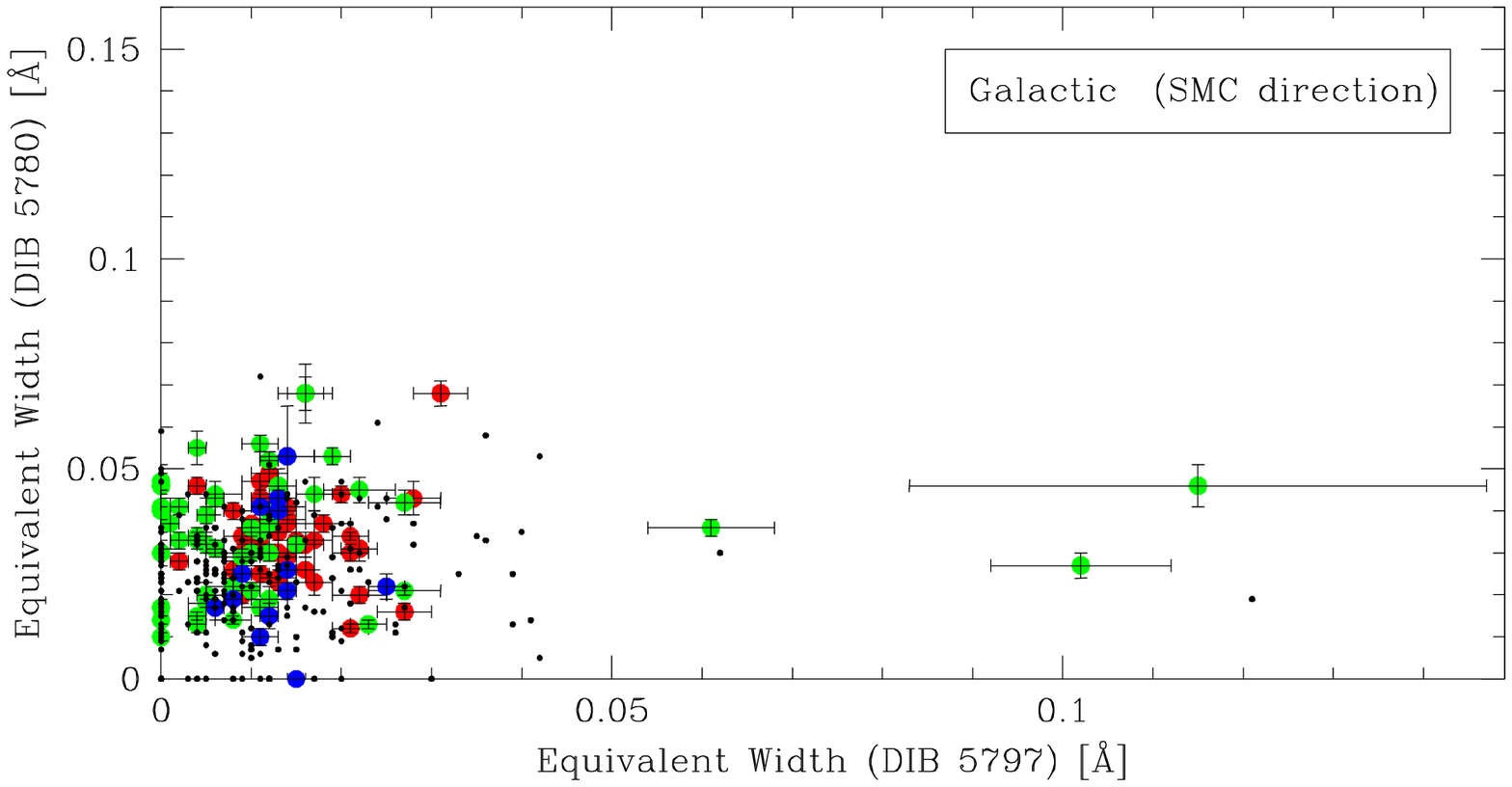,width=88mm}
\epsfig{figure=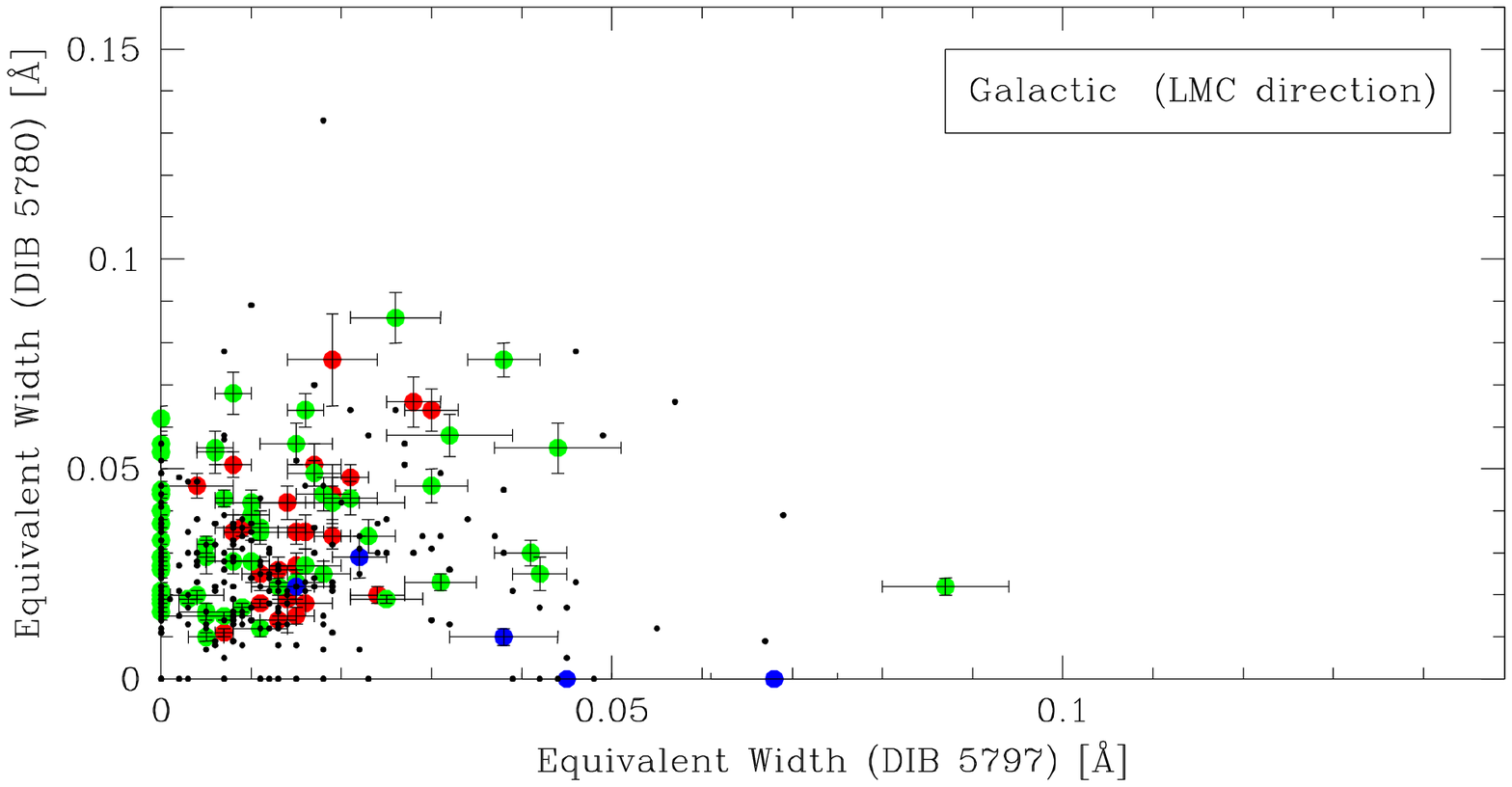,width=88mm}
}}}
\caption[]{The equivalent width of the 5780 and 5797 \AA\ DIBs for the SMC
({\it top left}), LMC ({\it top right}) and the Galactic foreground in the
direction of the SMC ({\it bottom left}) and LMC ({\it bottom right}).}
\end{figure*}

\subsection{Correlations}

%
\begin{figure*}
\centerline{\vbox{\hbox{
\epsfig{figure=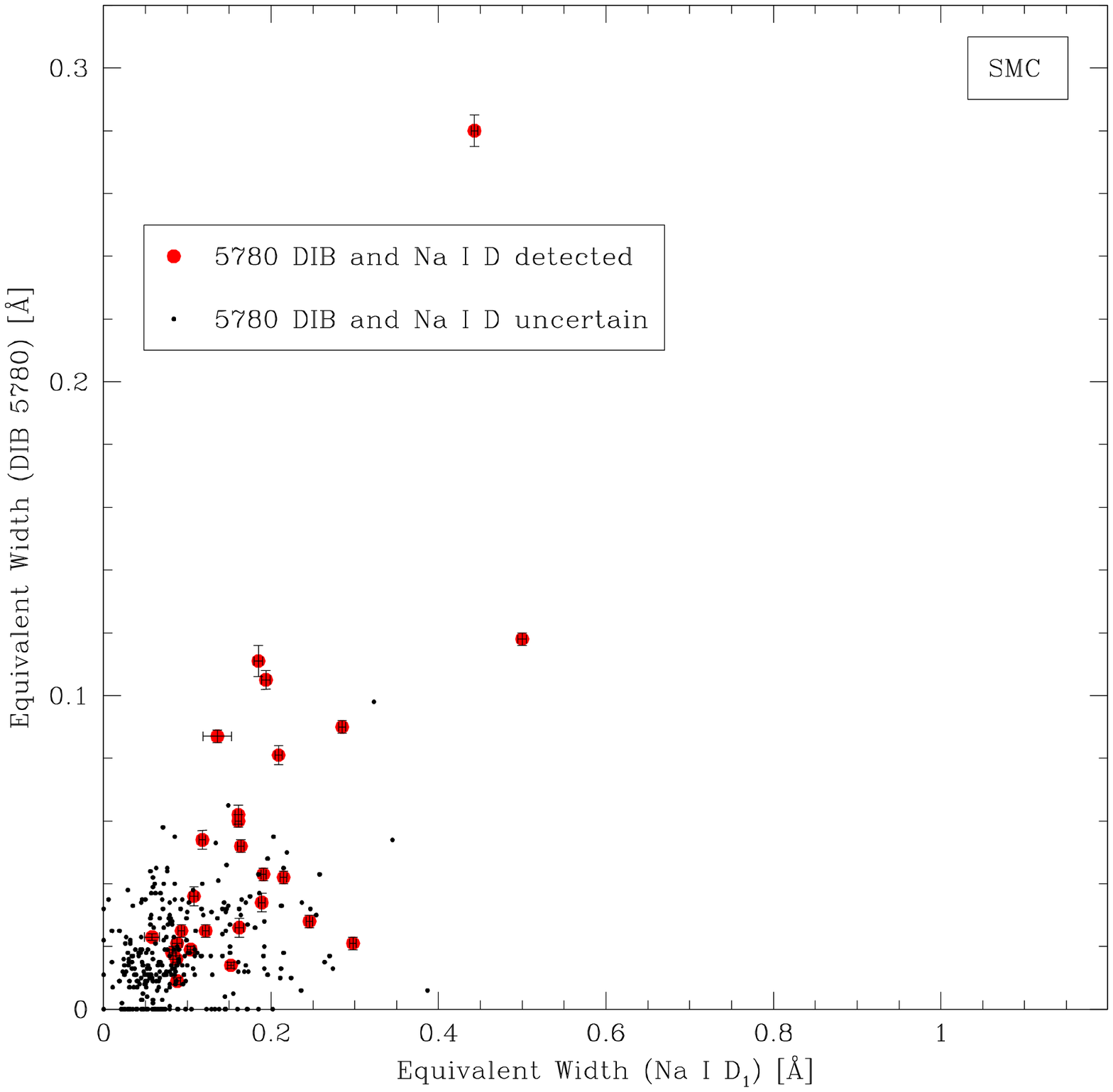,width=88mm}
\epsfig{figure=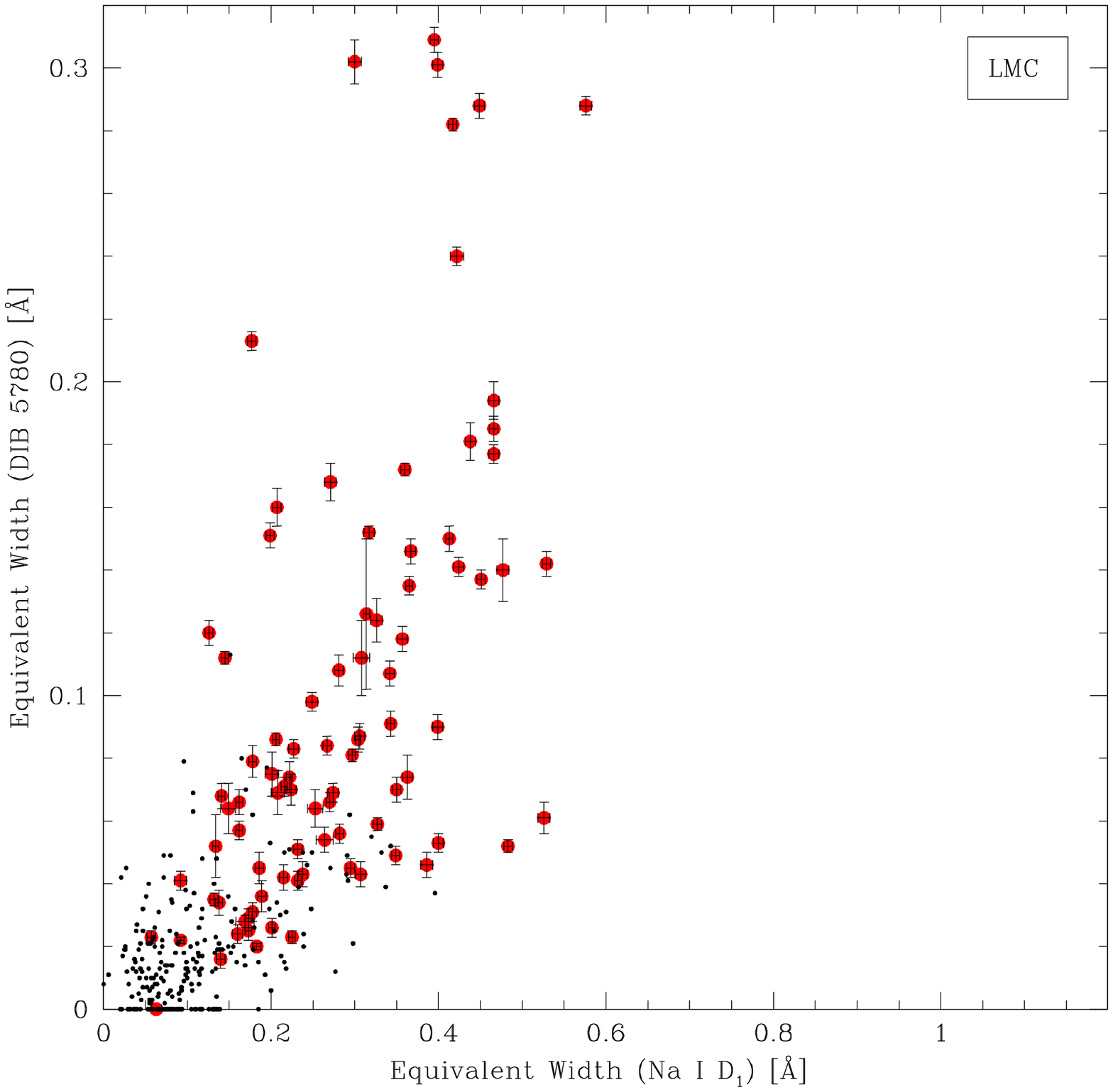,width=88mm}
}\hbox{
\epsfig{figure=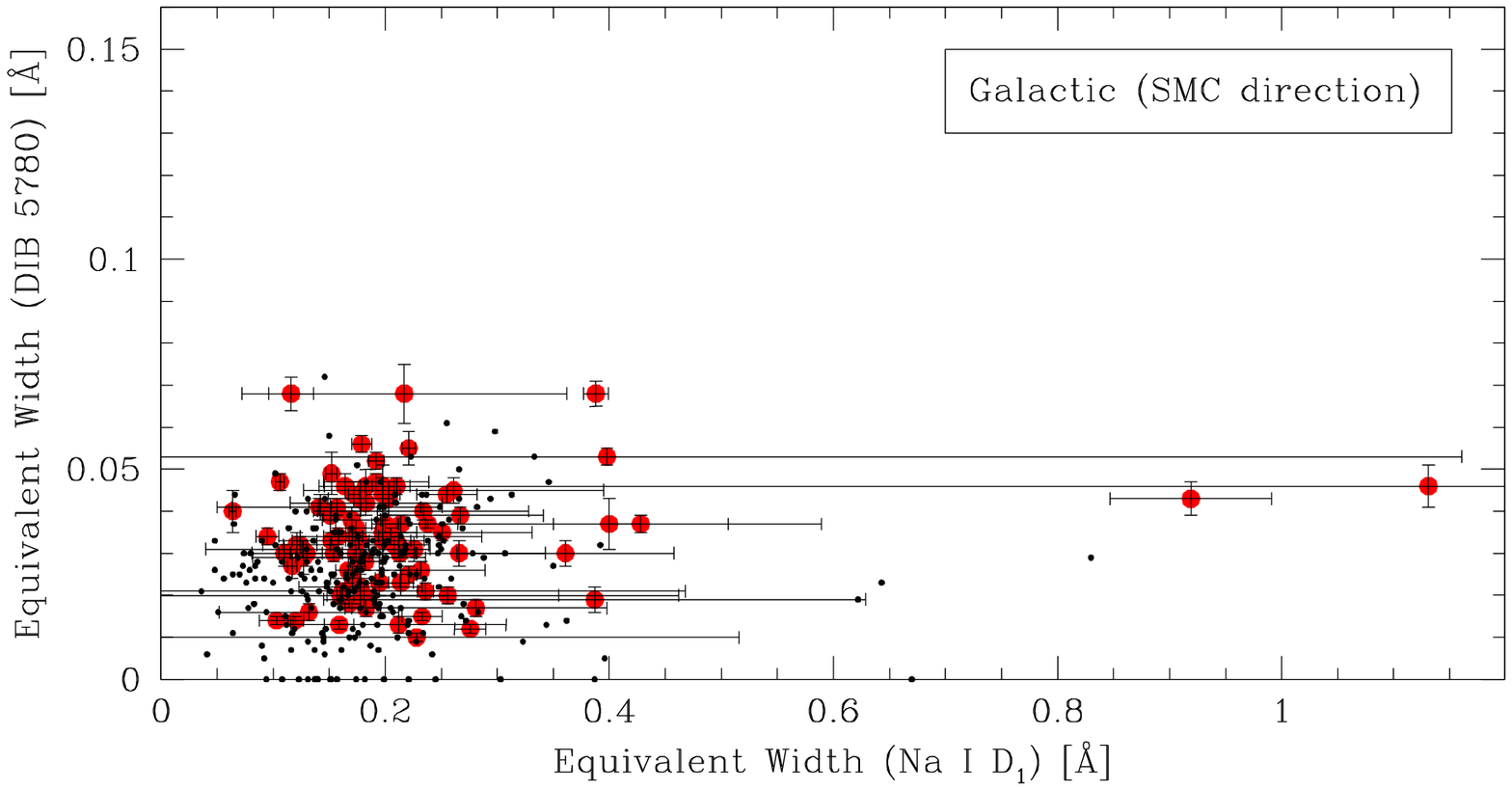,width=88mm}
\epsfig{figure=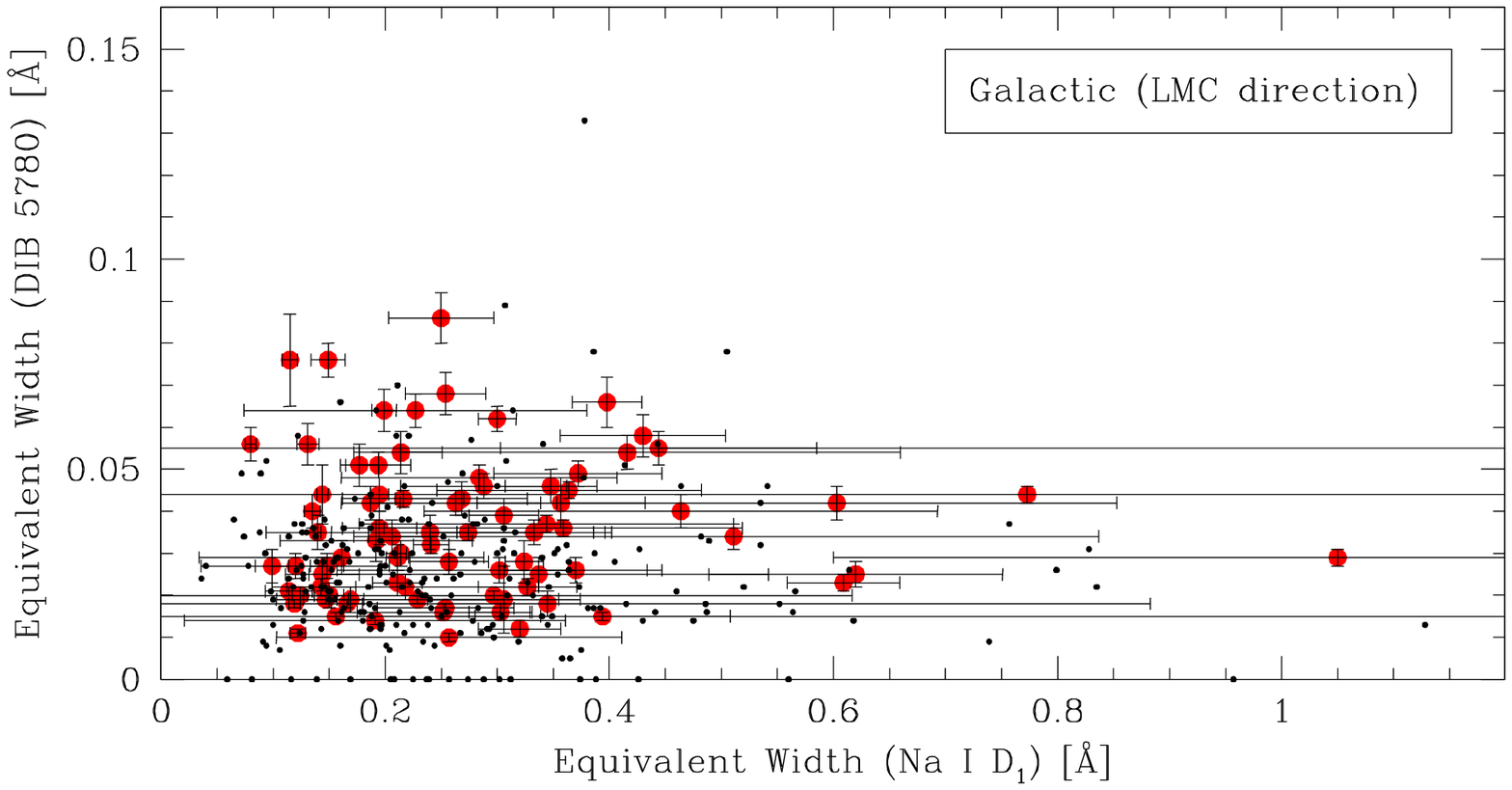,width=88mm}
}}}
\caption[]{The equivalent width of the 5780 \AA\ DIB versus that of the
Na\,{\sc i}\,D$_1$, for the SMC ({\it top left}), LMC ({\it top right}) and
the Galactic foreground in the direction of the SMC ({\it bottom left}) and
LMC ({\it bottom right}).}
\end{figure*}

Figures 4--6 show plots of the EW of the 5780 vs.\ 5797 \AA\ DIBs, 5780 \AA\
DIB vs.\ Na\,{\sc i}\,D$_1$ and 5797 \AA\ DIB vs.\ Na\,{\sc i}\,D$_1$,
respectively. (Almost) no correlation was discernible between the DIBs and
Ca\,{\sc ii}\,K. Plotted are not only the secure detections in both relevant
features, but also (with different colours/symbols) less certain detections
and non-detections. While some outliers and some of the scatter are due to
uncertainties, there are clear positive correlations between the 5780 and 5797
\AA\ DIBs and with Na\,{\sc i}\,D but with real (intrinsic) scatter also. The
equivalent width of the Na\,{\sc i}\,D$_1$ line grows slower than in
proportion to column density because of saturation effects, giving the false
impression of accelerated growth of the DIBs.

The Galactic foreground shows little difference between the SMC and LMC
directions, but the SMC itself clearly has far fewer sight-lines with strong
DIB absorption than the LMC (Fig.\ 4). Strong DIB absorption is seen internal
to the SMC towards star 135 (SMC\,10196 in Massey 2002) in the centre of the
South--Western star-forming complex: $EW_{5797}=0.079$ \AA\ and
$EW_{5780}=0.280$ \AA. The Na\,{\sc i}\,D absorption is also strong in that
direction. This sight-line passes in between the LHA-115\,N25 OB assocation
and the SMC-B2\,3 molecular cloud; possibly it crosses the irradiated skin of
a dense cloud, which may explain the strong DIB absorption. Similar strength
of DIB absorption is seen at several locations within the LMC; the correlation
appears to be saturating at $EW_{5780}\approx0.30$ \AA\ while the 5797 \AA\
DIB continues to strengthen beyond $EW_{5797}>0.1$ \AA. The two stars that
show the strongest 5797 \AA\ DIB in figure 4 are both located within the
Tarantula Nebula (cf.\ van Loon et al.\ 2013).

%
\begin{table*}
\caption{Linear regression analysis of DIB--DIB correlations of the form
$EW_{5780}=a+b\times EW_{5797}$ with correlation coefficient $r$. Where the
Student $t$ statistic is given, the critical value of $t$ is for a 1\% chance
that the null hypothesis is true.}
\begin{tabular}{lrcccccl}
\hline\hline
location        &
$N$             &
$a$ (\AA)       &
$b$             &
$r$             &
$t$             &
$t_{\rm critical}$ &
verdict         \\
\hline
SMC, internal   &
16              &
$0.007\pm0.005$ &
$2.93\pm0.47$   &
0.86            &
6.3             & 
2.6             & 
significant     \\
LMC, internal   &
44              &
$0.036\pm0.013$ &
$2.17\pm0.28$   &
0.77            &
7.8             & 
2.4             & 
significant     \\
Galactic, in SMC direction &
38              &
$0.032\pm0.005$ &
$0.16\pm0.31$   &
0.09            &
0.5             & 
2.4             & 
insignificant   \\
Galactic, in LMC direction &
25              &
$0.016\pm0.003$ &
$1.24\pm0.53$   &
0.44            &
2.3             & 
2.5             & 
very marginal   \\
\hline
\end{tabular}
\end{table*}

%
\begin{figure*}
\centerline{\vbox{\hbox{
\epsfig{figure=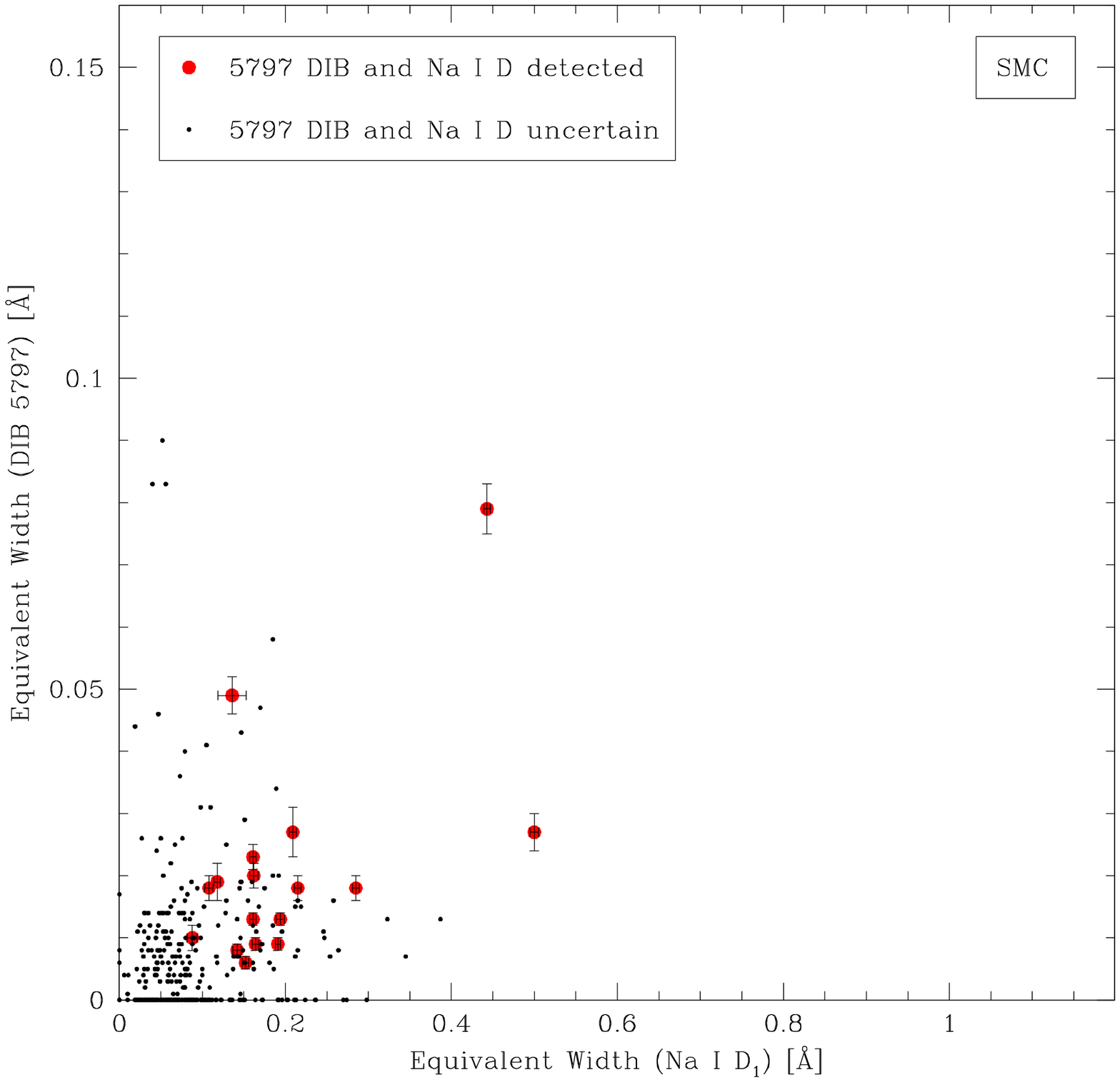,width=88mm}
\epsfig{figure=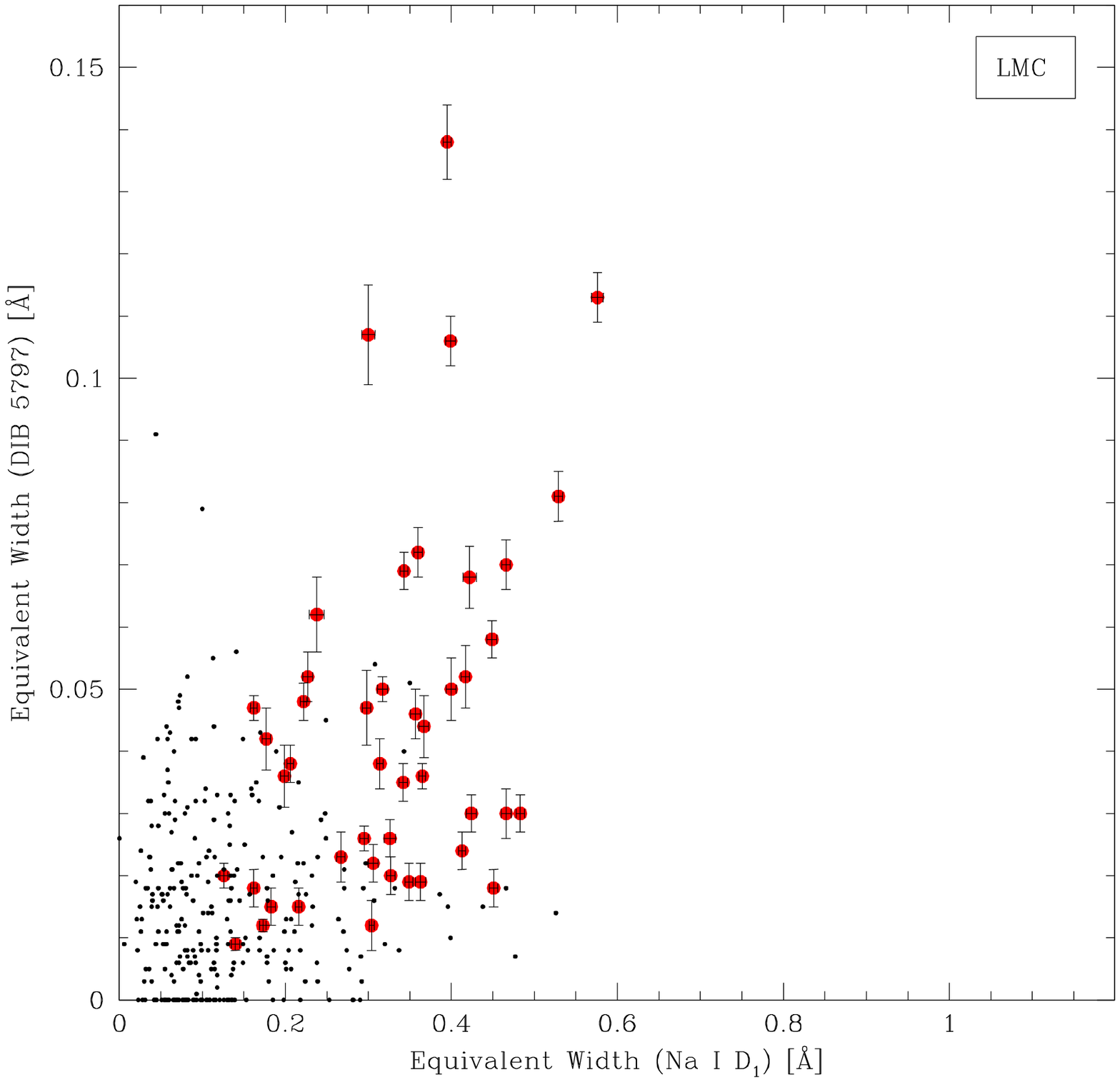,width=88mm}
}\hbox{
\epsfig{figure=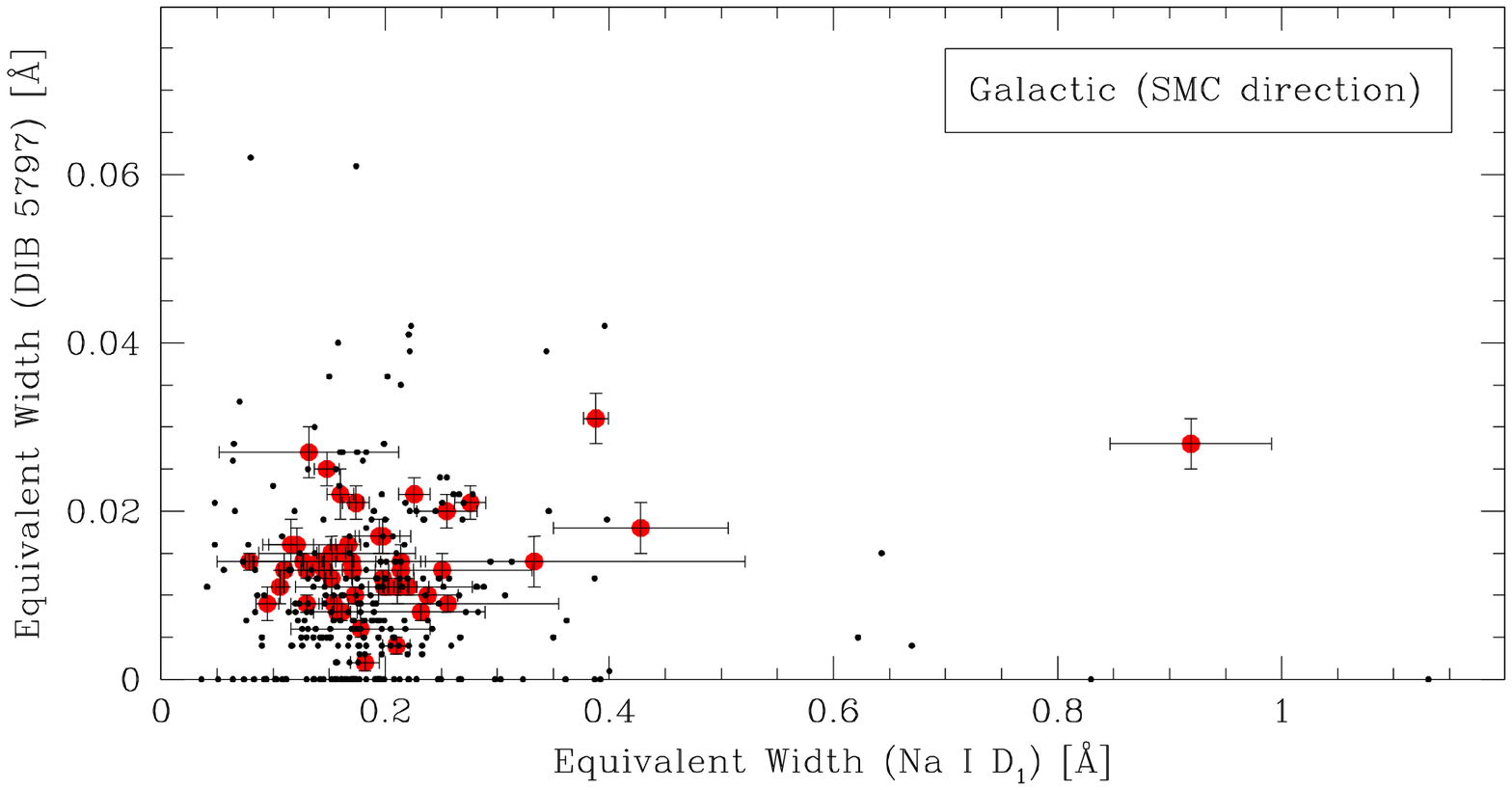,width=88mm}
\epsfig{figure=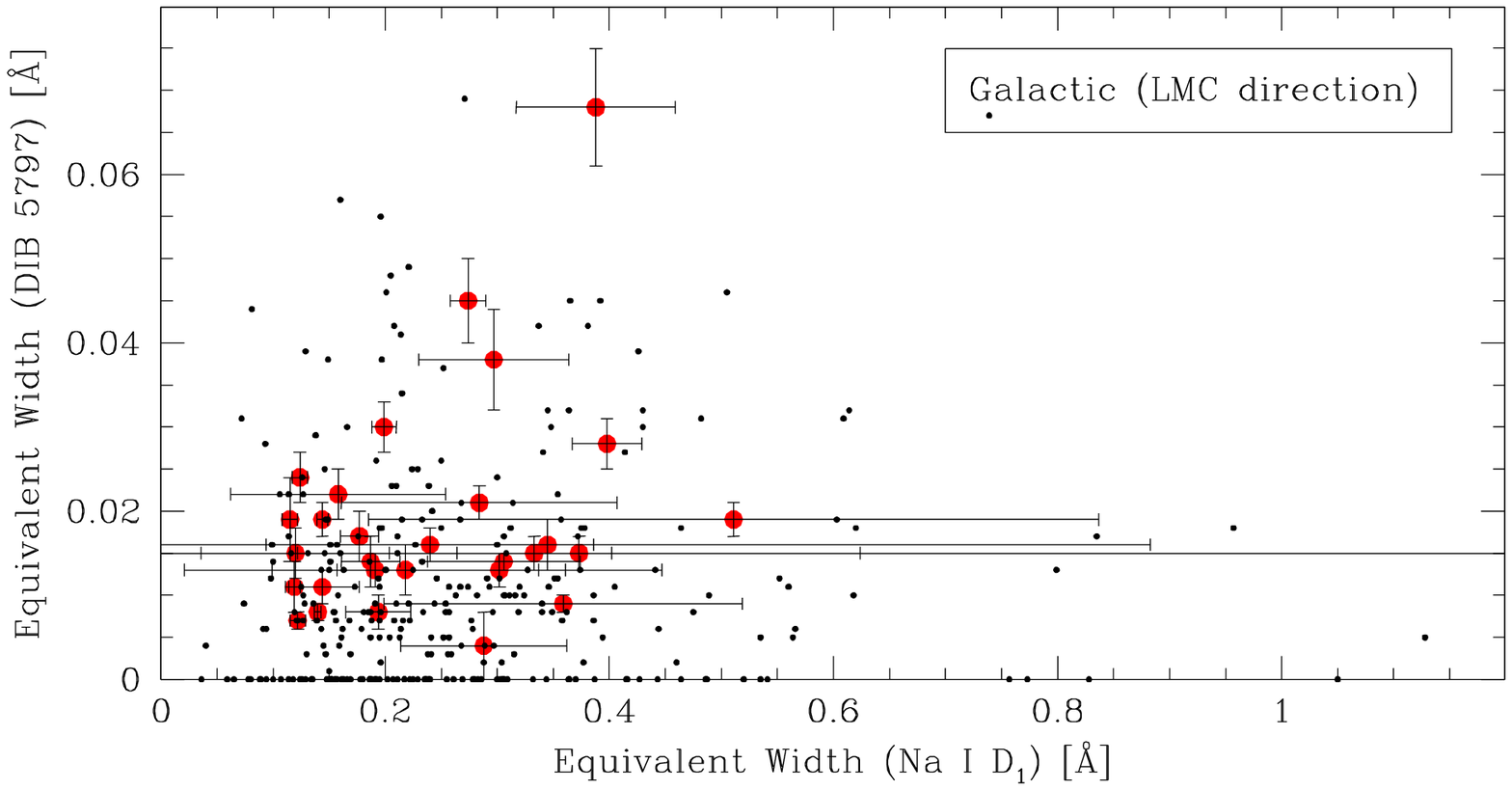,width=88mm}
}}}
\caption[]{The equivalent width of the 5797 \AA\ DIB versus that of the
Na\,{\sc i}\,D$_1$, for the SMC ({\it top left}), LMC ({\it top right}) and
the Galactic foreground in the direction of the SMC ({\it bottom left}) and
LMC ({\it bottom right}).}
\end{figure*}

In Table 5 the results are summarised from a linear regression analysis and a
Student $t$ test for the 5780 and 5797 \AA\ DIBs: the correlation is
significant if $t>t_{\rm critical}$, where
\begin{equation}
t=r\sqrt{\frac{N-2}{1-r^2}}
\end{equation}
with correlation coefficient $r$ and $N$ data points ($N-2$ degrees of freedom
in this case). Hence we conclude that a good DIB--DIB correlation is seen
within the SMC and LMC but that the Galactic foreground is probably too
transparent to show a significant correlation. The 5780/5797 DIB ratio is very
similar between the SMC, LMC and the Local Bubble -- where
$EW_{5780}=0.023(\pm0.005)+2.48(\pm0.20)\times EW_{5797}$ (Bailey et al.\ 2015).
There is a hint that the relation is steeper (relatively stronger 5780 \AA\
DIB) in the SMC, which could be explained as due to a harsher radiation field
from metal-poor stars (reduced line-blanketing) in a metal-poor ISM (reduced
attenuation by dust), but this is only marginally significant.

Both the DIBs and Na\,{\sc i}\,D are weaker in the SMC than in the LMC (Figs.\
5 \& 6), but they behave similarly. Both show what appears to be a threshold,
or a more slowly picking up of DIB strength as Na\,{\sc i}\,D increases,
around $EW_{\rm NaI\,D1}\sim0.1$ \AA. A similar behaviour was noticed in the
Local Bubble (Bailey et al.\ 2015); it probably is due at least in part to the
Na\,{\sc i}\,D saturating. The Galactic ISM is clearly more transparent in
front of the SMC than in front of the LMC, which is probably related to the
larger distance to the Galactic plane of the former. Quite strong absorption
is seen in sight-lines close to the LMC in the Local Bubble survey (Bailey et
al.\ 2015) though no sight-lines were observed by these authors within
$10^\circ$ of the SMC.

\subsection{Maps}

%
\begin{figure*}
\centerline{\vbox{\hbox{
\epsfig{figure=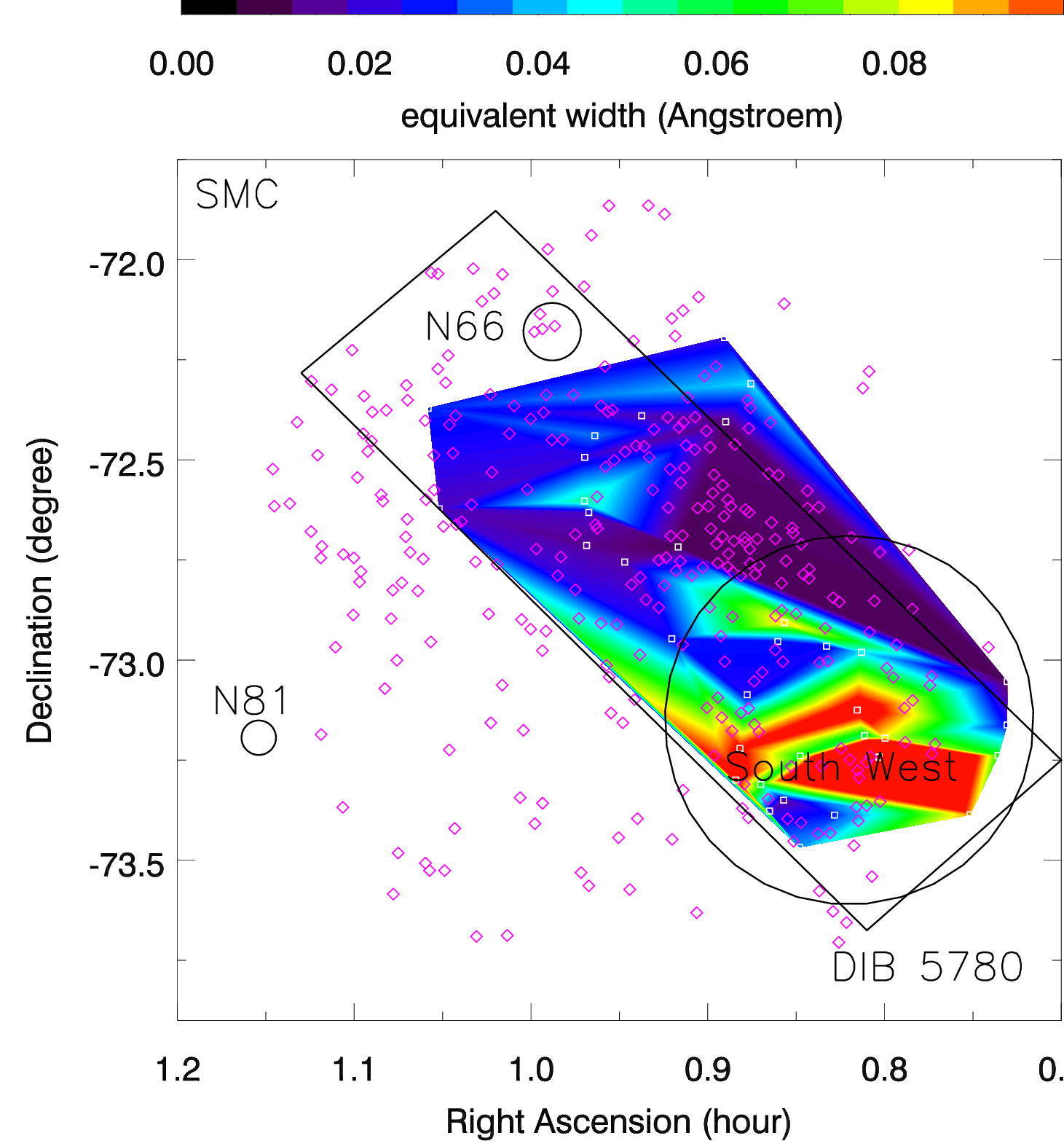,width=58mm}
\epsfig{figure=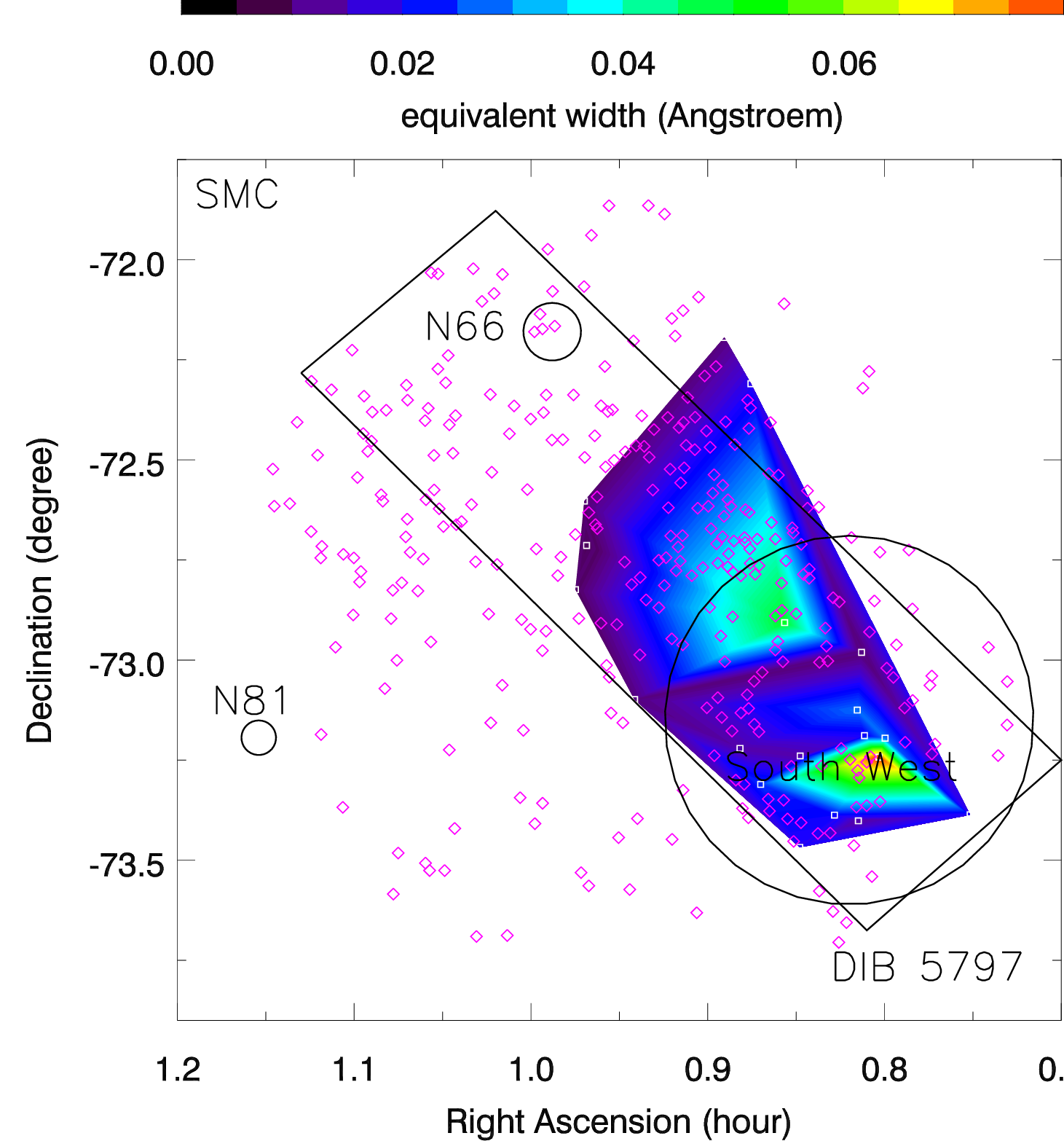,width=58mm}
\epsfig{figure=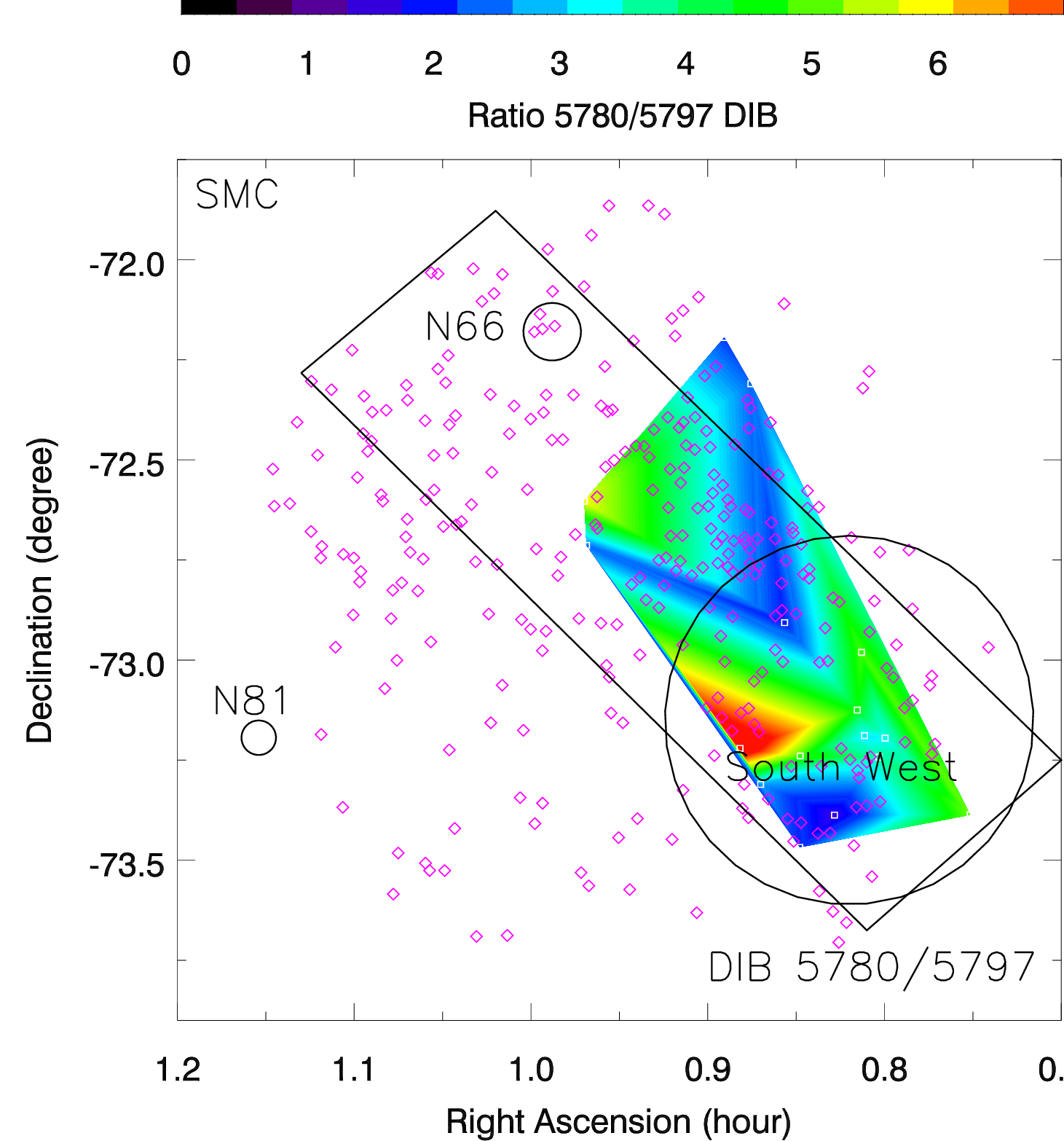,width=58mm}
}\vspace{2mm}\hbox{
\epsfig{figure=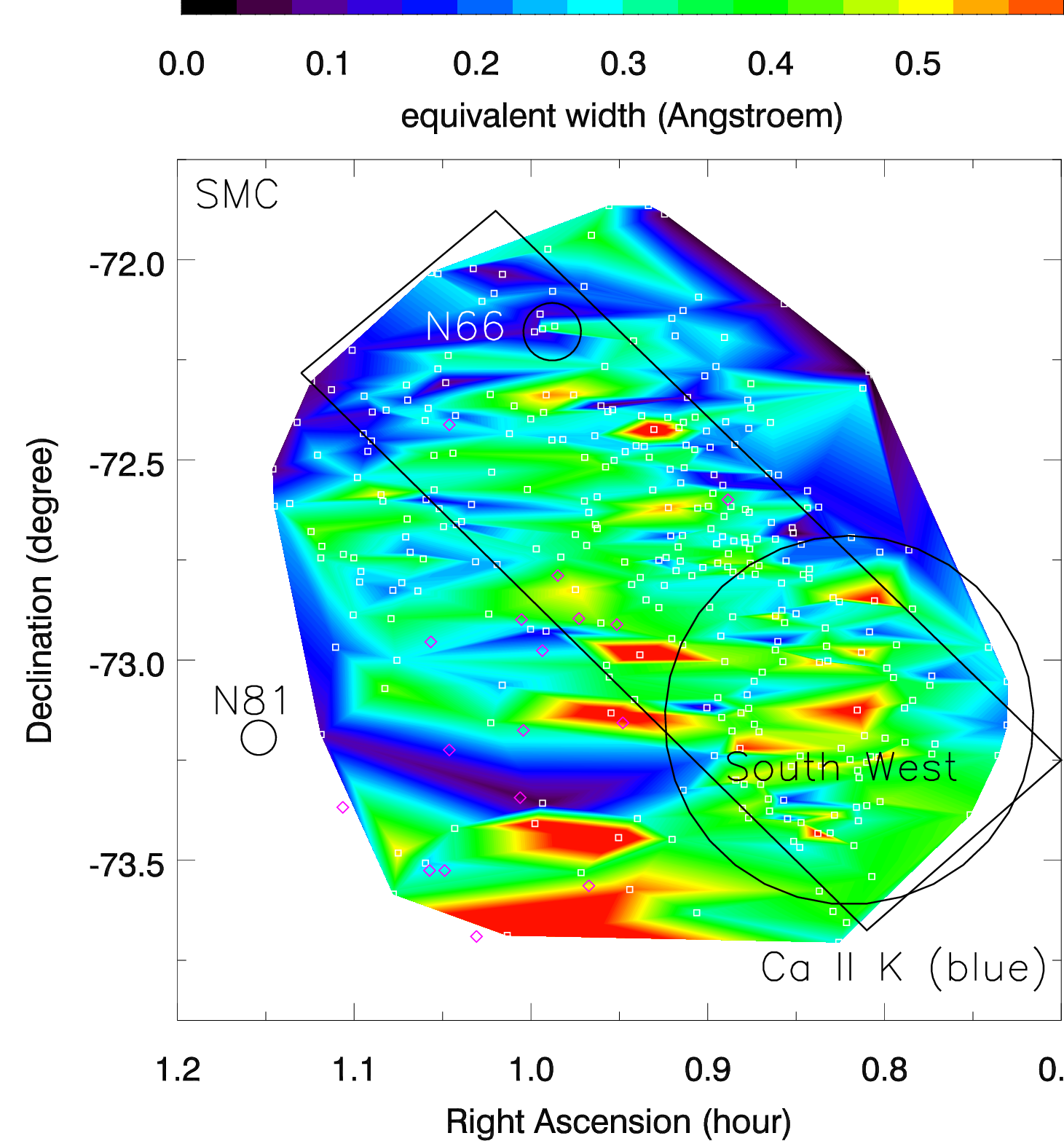,width=58mm}
\epsfig{figure=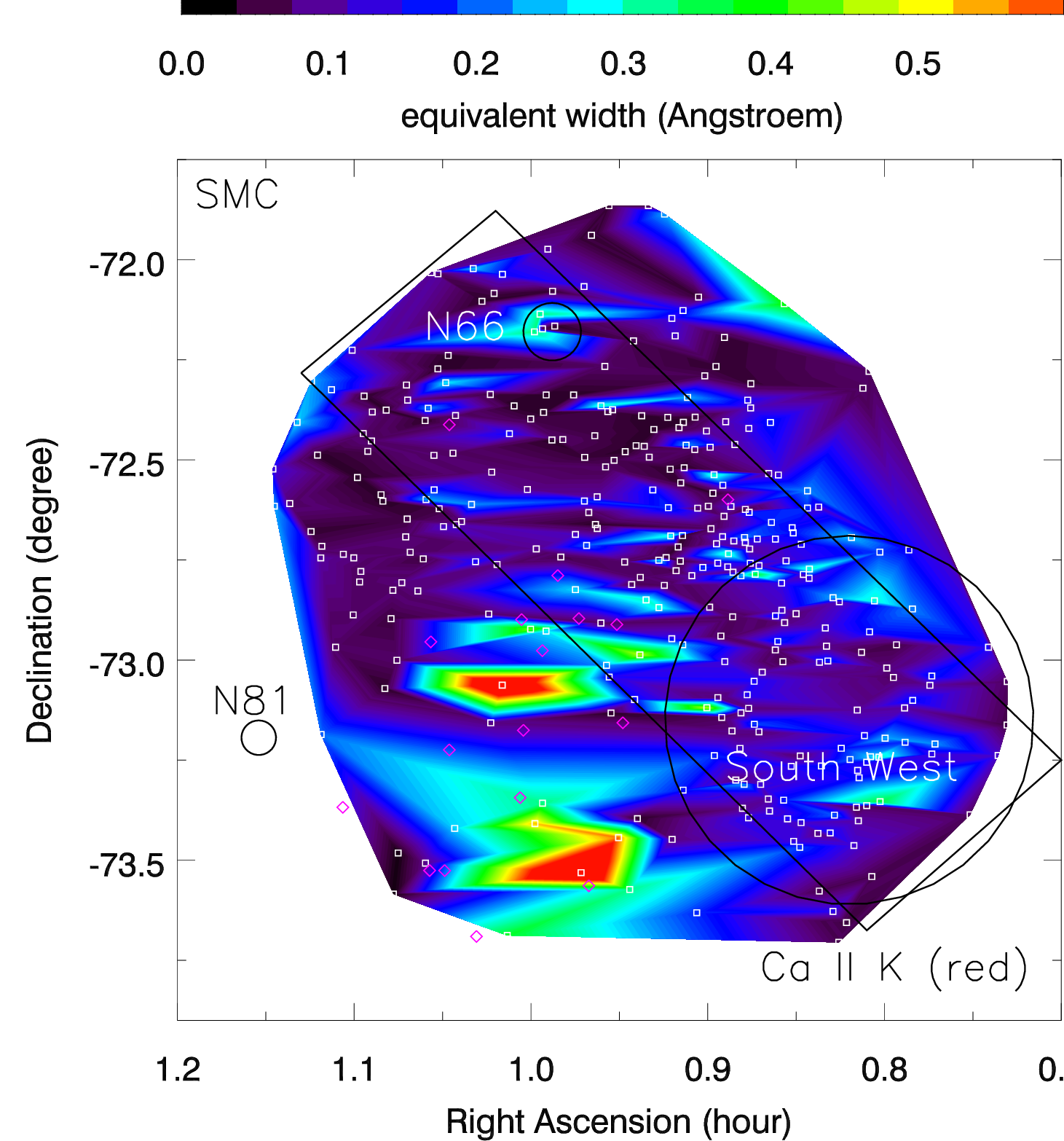,width=58mm}
\epsfig{figure=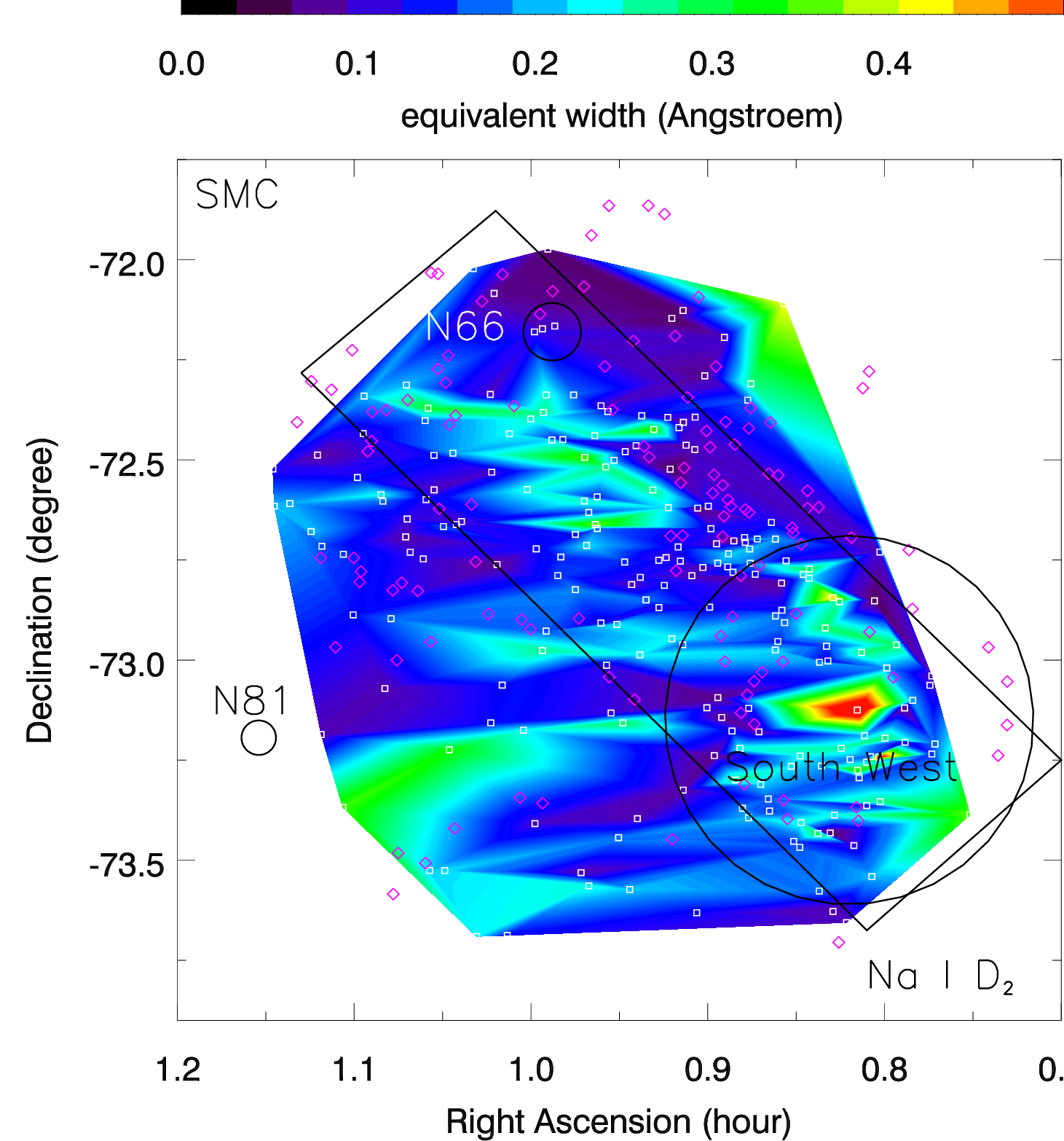,width=58mm}
}}}
\caption[]{Maps of the SMC in the equivalent width of ({\it top row:}) the
5780 and 5797 \AA\ DIBs and their ratio, and ({\it bottom row:}) the blue and
red Ca\,{\sc ii}\,K components and Na\,{\sc i}\,D$_2$. The locations of the
target stars are indicated with the small white squares (detections) and
magenta diamonds (non-detections).}
\end{figure*}

We constructed maps of the 5780 \AA\ DIB, 5797 \AA\ DIB, Ca\,{\sc ii}\,K and
Na\,{\sc i}\,D$_2$ equivalent widths within the SMC (Fig.\ 7), within the LMC
(Fig.\ 8) and within the Galactic foreground (Figs.\ 9 \& 10). We do not show
the Na\,{\sc i}\,D$_1$ maps but they look identical to the D$_2$ maps. We
overlaid the positions of the stellar probes; as absorption was not always
detected the maps are generally more sparsely sampled than the distribution of
targets -- which are already not distributed uniformly across the sky. The
maps are filled in using triangulation, which preserves the true information
(measurements) and does nothing but interpolate between measurements without
smoothing. The non-detections (magenta symbols) provide a meaningful addition
to the maps as they outline areas of very low absorption. The SMC maps of the
5780 and 5797 \AA\ DIBs are based on just 35 and 19 firm detections,
respectively; for the LMC these are 91 and 48, respectively. The
Ca\,{\sc ii}\,K line was detected in most sight-lines in both galaxies and the
Na\,{\sc i}\,D in about half of the sight-lines.

In the SMC (Fig.\ 7) the DIBs are strongly associated with the South--Western
part of the SMC ``bar'', where star formation is widespread. The
$EW_{5780}/EW_{5797}$ ratio appears to peak slightly offset from that region,
perhaps tracing harsher conditions in the periphery of the star-forming
molecular clouds as compared to those within the clouds. The South--Western
part of the bar is also  where Na\,{\sc i}\,D peaks though it is detected more
globally across the SMC. No DIBs were detected around the mini-starburst N\,66
further North in the bar. The blue velocity component of Ca\,{\sc ii}\,K is
detected across the SMC, tracing the generally hot, widely distributed gas
typical of the metal-poor ISM in the SMC, but the red velocity component is
concentrated towards the Wing, to the South--East of the bar. The red
component might be situated behind the bulk of the stars, hence the lack of
detected absorption in this gas. There is no resemblance between the
Ca\,{\sc ii}\,K maps and those of the DIBs, but it can be noted that where the
DIBs are strongest the Ca\,{\sc ii}\,K is relatively weak. This shows a
stronger affiliation of DIBs with Na\,{\sc i}-bright material than with
Ca\,{\sc ii}-bright material.

%
\begin{figure*}
\centerline{\vbox{\hbox{
\epsfig{figure=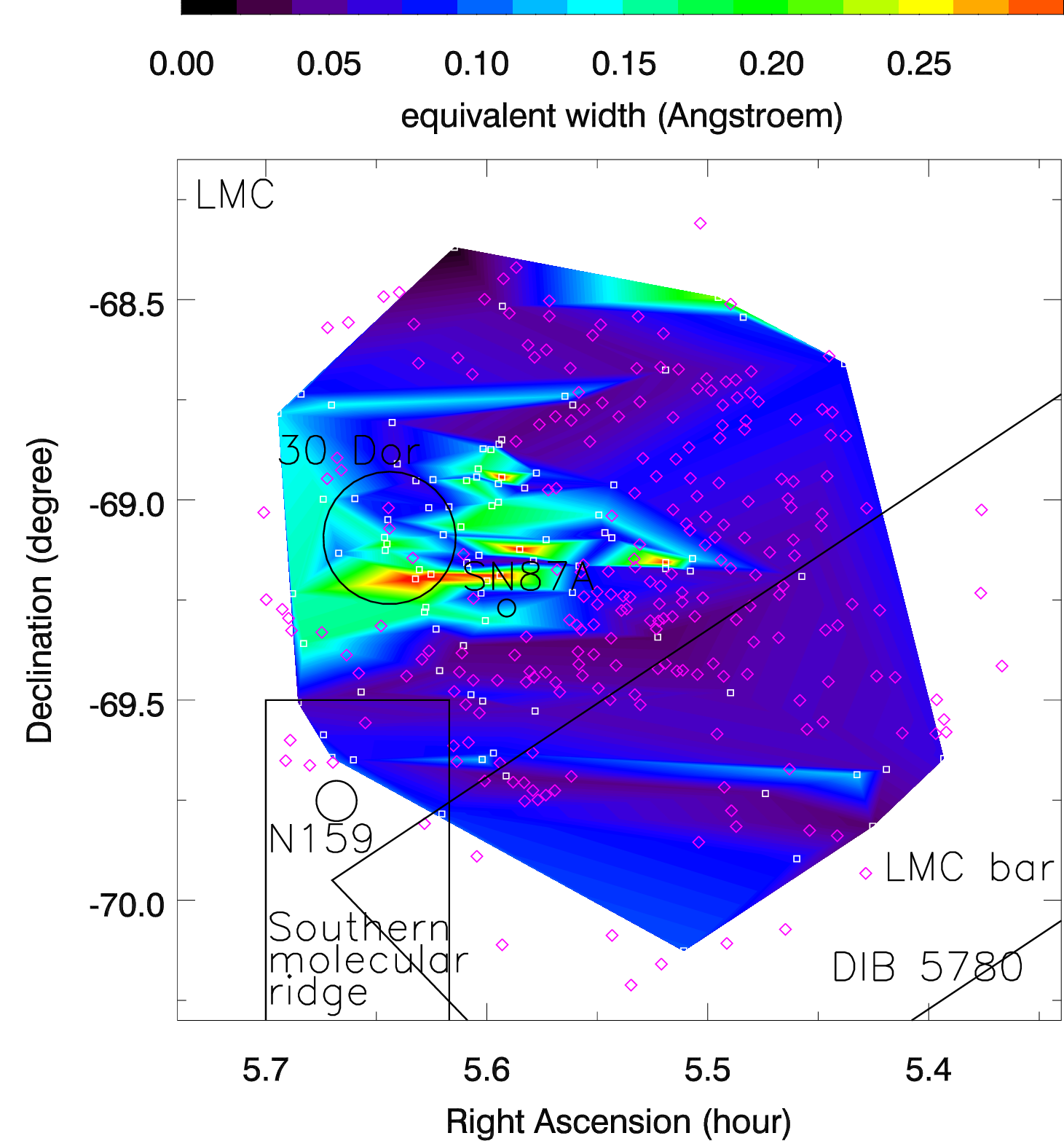,width=58mm}
\epsfig{figure=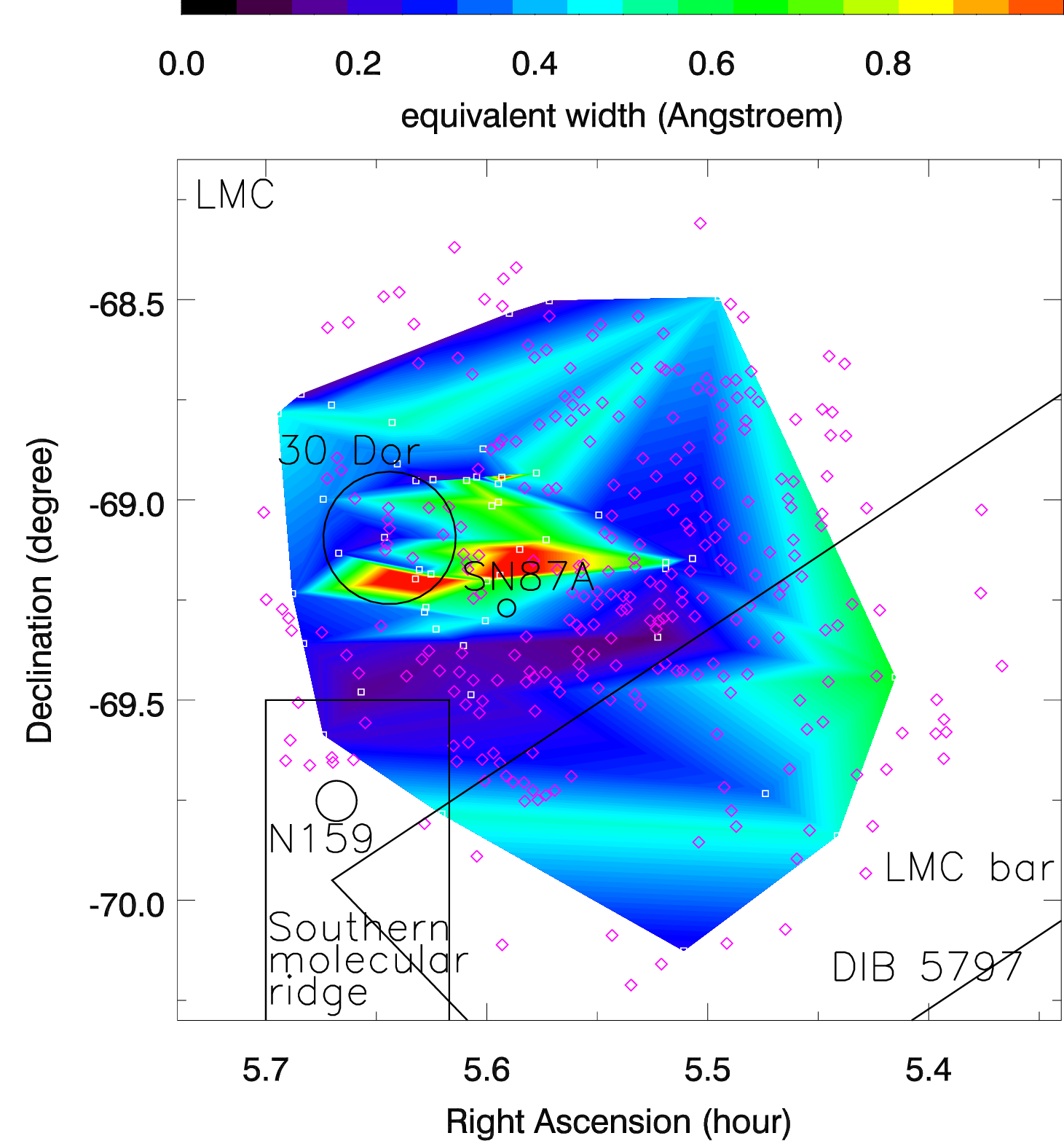,width=58mm}
\epsfig{figure=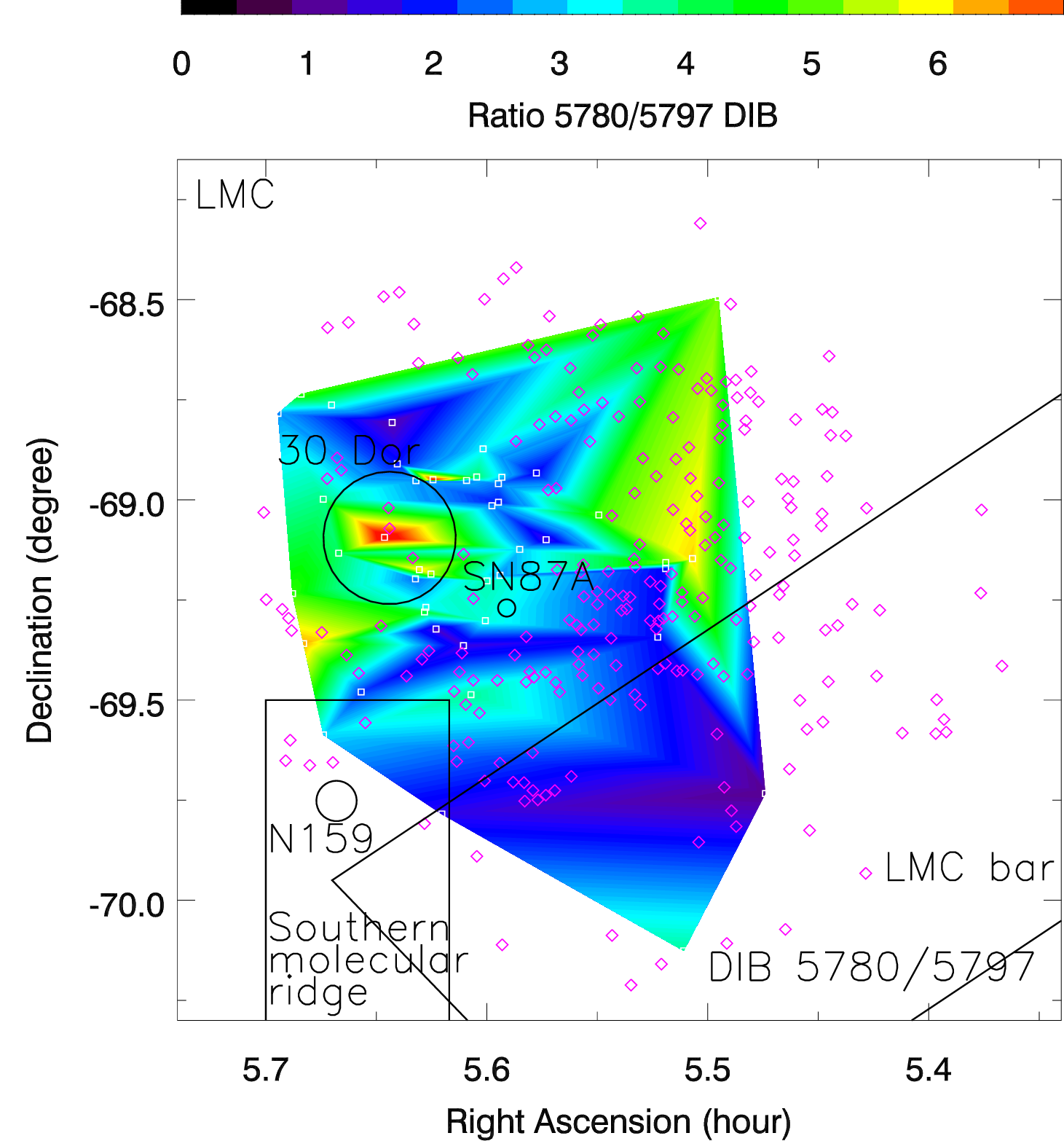,width=58mm}
}\vspace{2mm}\hbox{
\epsfig{figure=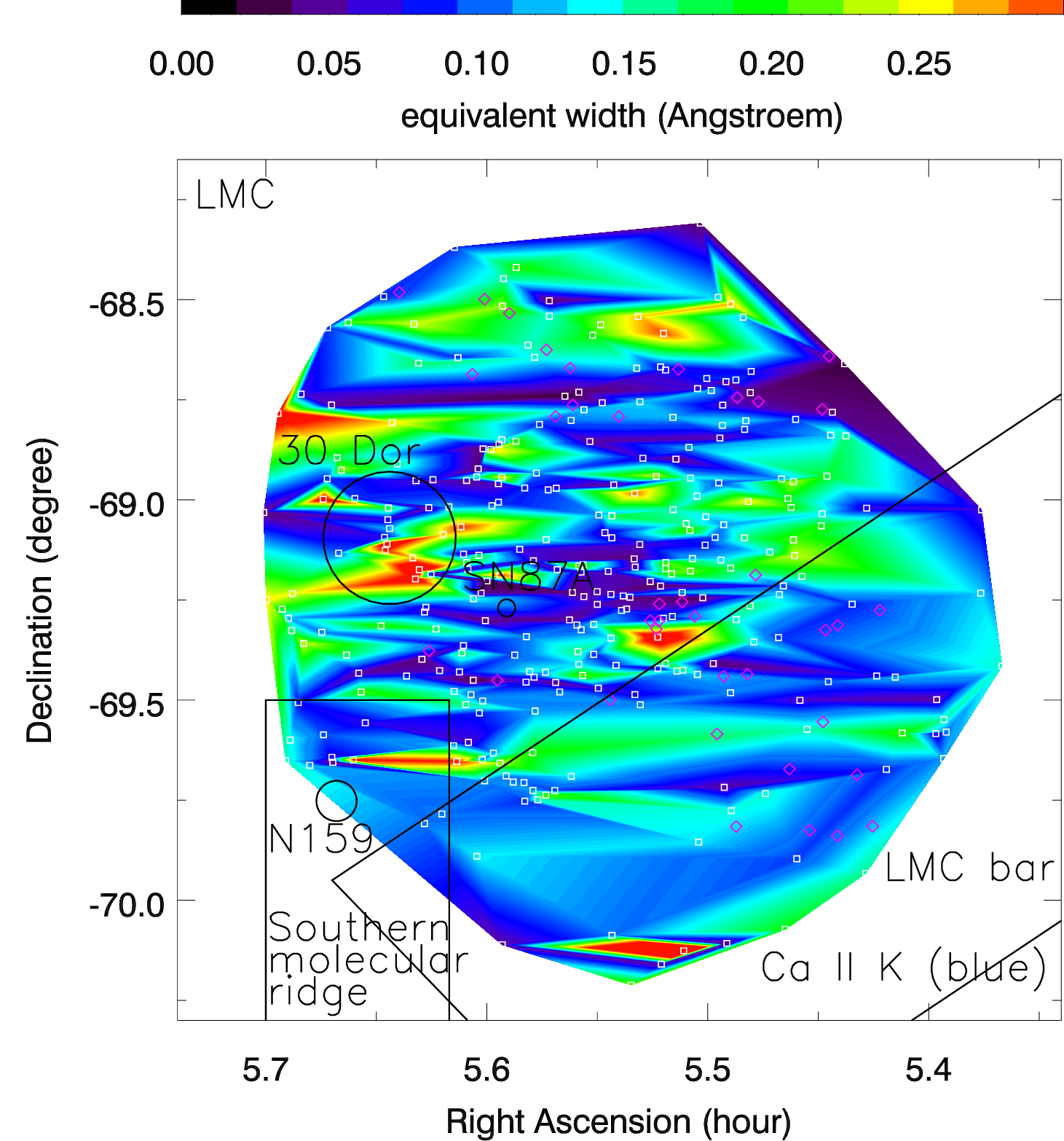,width=58mm}
\epsfig{figure=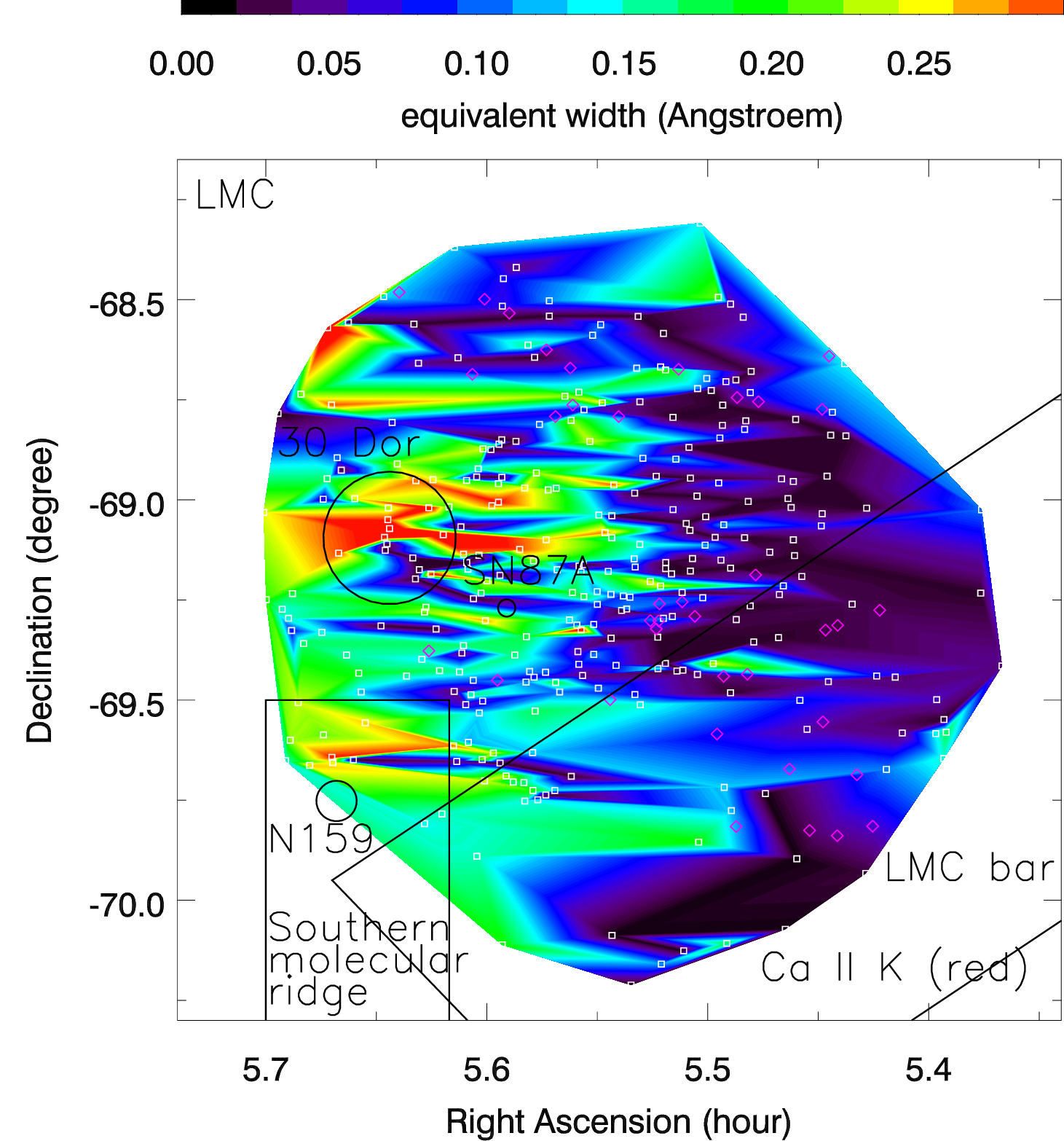,width=58mm}
\epsfig{figure=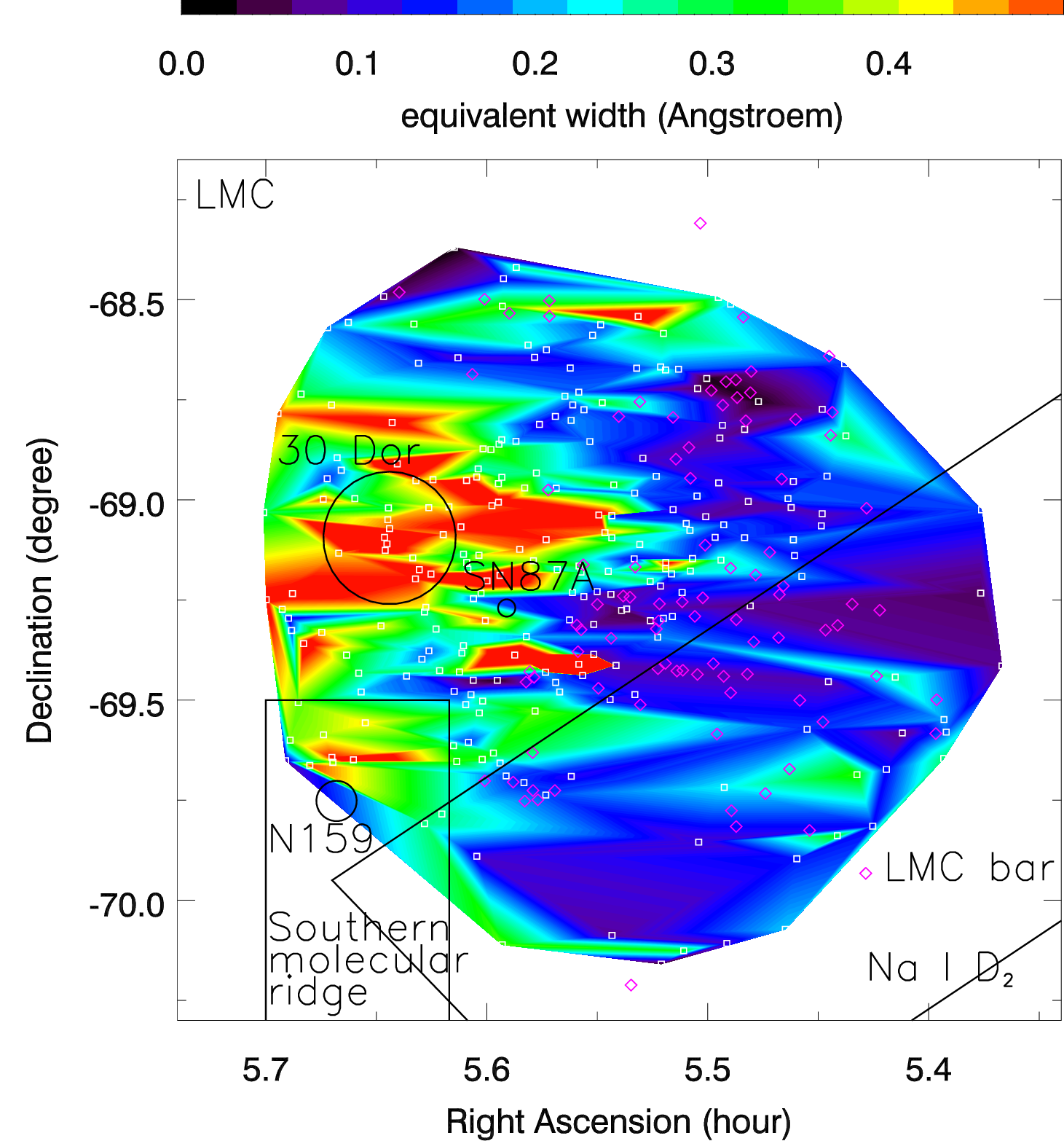,width=58mm}
}}}
\caption[]{As figure 7, but for the LMC.}
\end{figure*}

The LMC maps (Fig.\ 8) confirm this behaviour: The DIBs are strongest in
molecular clouds within star-forming regions, but not too close to recently
formed OB associations such as the central cluster R\,136 (30\,Doradus) within
the Tarantula Nebula. This was also found in the DIB maps of the Tarantula
Nebula by van Loon et al.\ (2013). The $EW_{5780}/EW_{5797}$ ratio is high
right at the location of R\,136, as well as towards the West where the ISM
becomes more diffuse (less self-shielded). The Na\,{\sc i}\,D is concentrated
in the same regions of great DIB strength, though more spread out. The
Ca\,{\sc ii}\,K data are interesting: as in the SMC, the blue component is
seen across the LMC also in areas with weak Na\,{\sc i}\,D, though the densest
parts tend to be situated in the East. The red component of Ca\,{\sc ii}\,K
resembles the Na\,{\sc i}\,D map more closely; here the East--West gradient is
particularly evident, with the strongest absorption along the Eastern, leading
edge of the LMC. Again we see a fair correspondence between DIBs and
Na\,{\sc i}\,D but not with Ca\,{\sc ii}\,K except for a few locations of
generally high column density -- e.g., the blue component of Ca\,{\sc ii}\,K
in the South of the Tarantula Nebula and the red component to the West of it.

The maps of the Galactic foreground (Figs.\ 9 \& 10) bear no resemblance to
those of the SMC and LMC, which lends credibility to both the Galactic and
Magellanic maps we created. In both the SMC and LMC sight-lines the DIBs show
gradients over scales of a degree, with possibly smaller scale structure --
similar to what was seen in the extra-planar gas in front of the globular
cluster $\omega$\,Centauri by van Loon et al.\ (2009). For instance, the 5797
\AA\ DIB (Fig.\ 9) traces a coherent presumably fairly neutral cloud lying
across the South--Eastern portion of the SMC field, whereas the 5780 \AA\ DIB
identifies material towards the South--Western corner of the LMC field. The
$EW_{5780}/EW_{5797}$ ratio is high in places where the 5797 \AA\ DIB is very
weak, not where the 5780 \AA\ DIB is strongest. This suggests that the ratio
is high in relatively diffuse gas under special conditions.

The foreground Ca\,{\sc ii}\,K and especially the Na\,{\sc i}\,D (Fig.\ 10) is
structured with high contrast on much smaller scales, $\sim0\rlap{.}^\circ1$.
This makes it hard to compare these maps with those of the DIBs (Fig.\ 9),
though where the Na\,{\sc i}\,D is strong (Southern half of the SMC field,
around $Dec\sim-69\rlap{.}^\circ7$ in the LMC field) the stronger DIBs are
seen also. It is possible that the DIBs are structured on similar small scales
as the atomic absorption, but the noise and fewer detections of the much
weaker DIBs prevents direct proof.

%
\begin{figure*}
\centerline{\vbox{\hbox{
\epsfig{figure=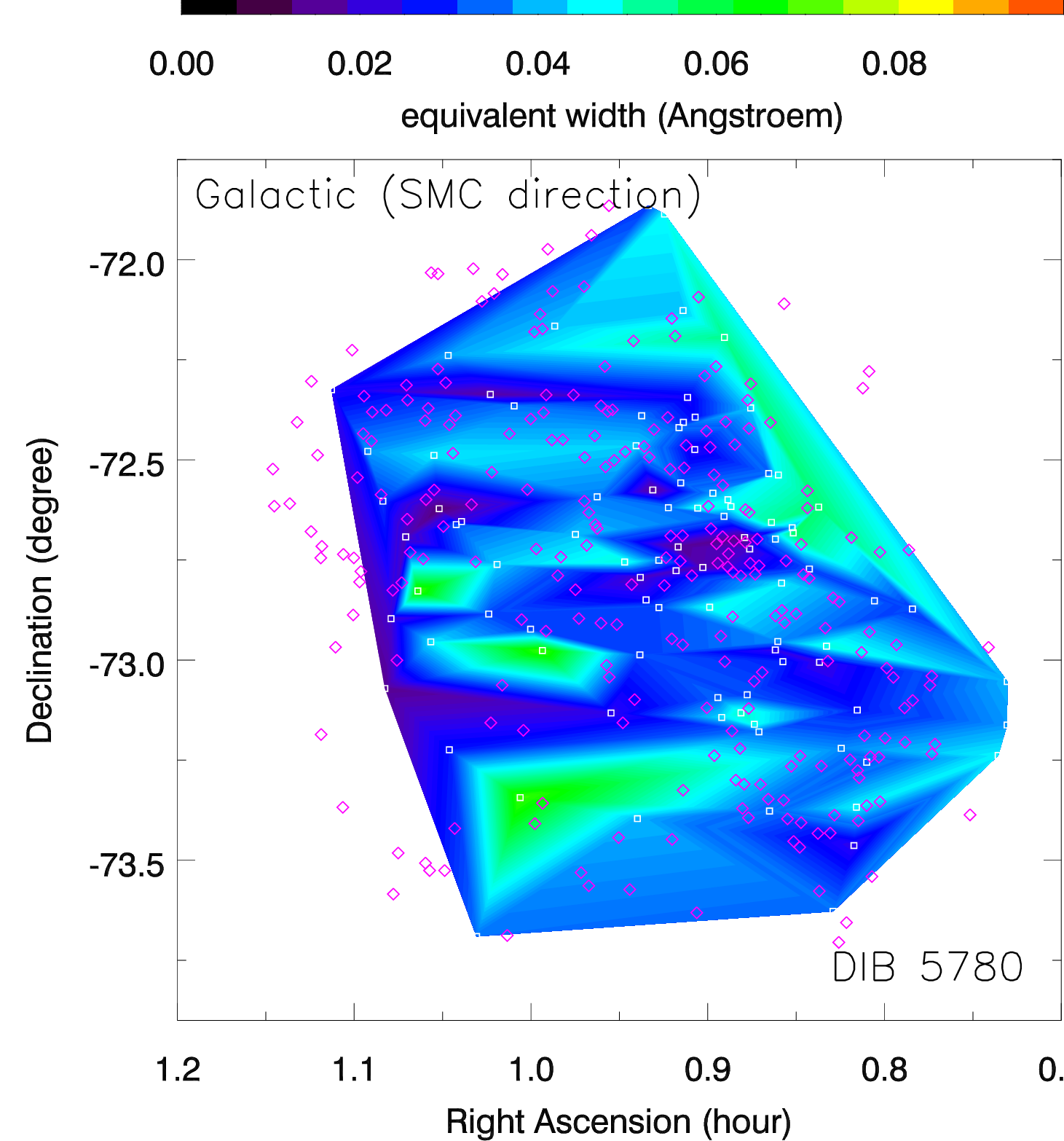,width=58mm}
\epsfig{figure=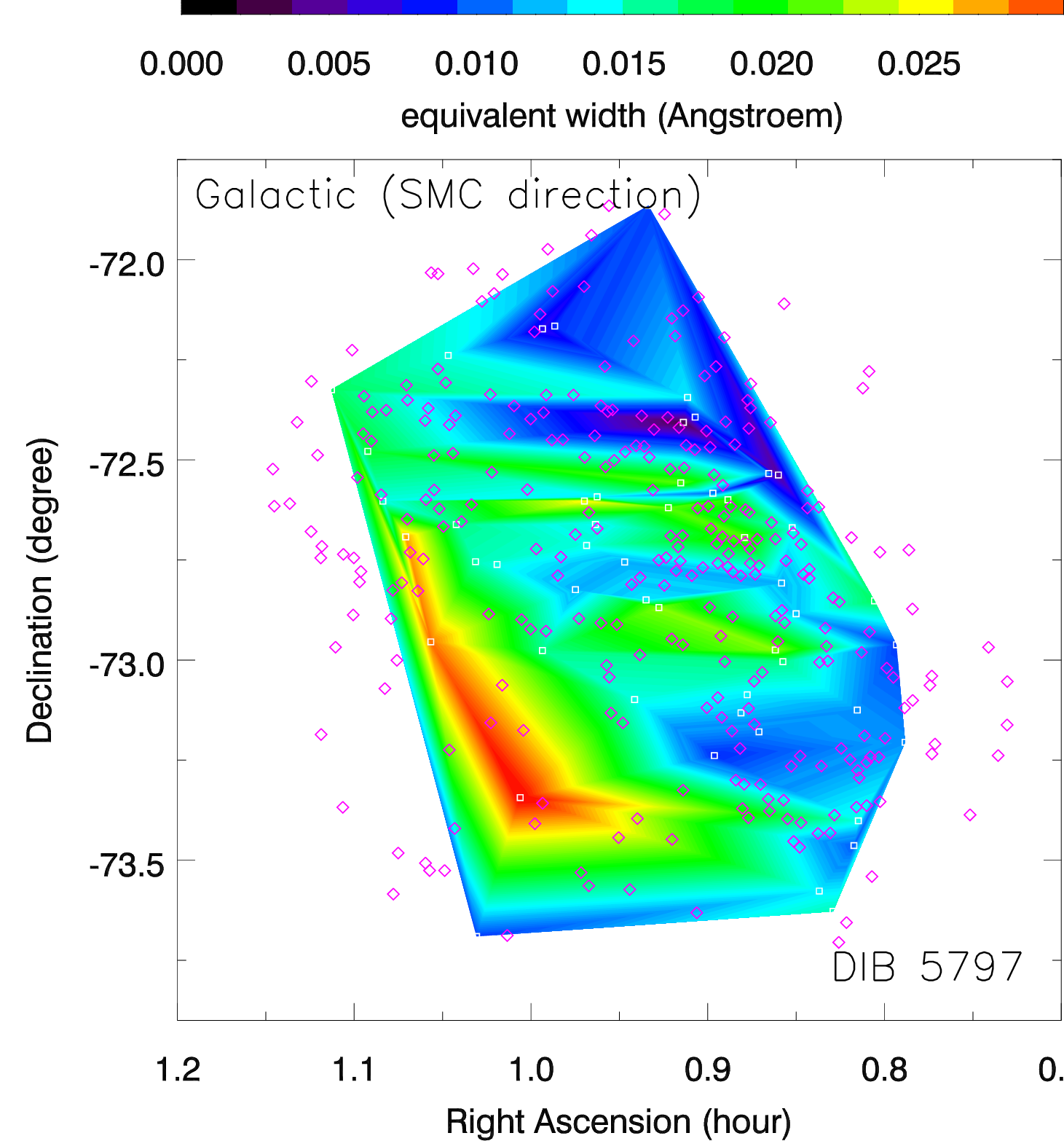,width=58mm}
\epsfig{figure=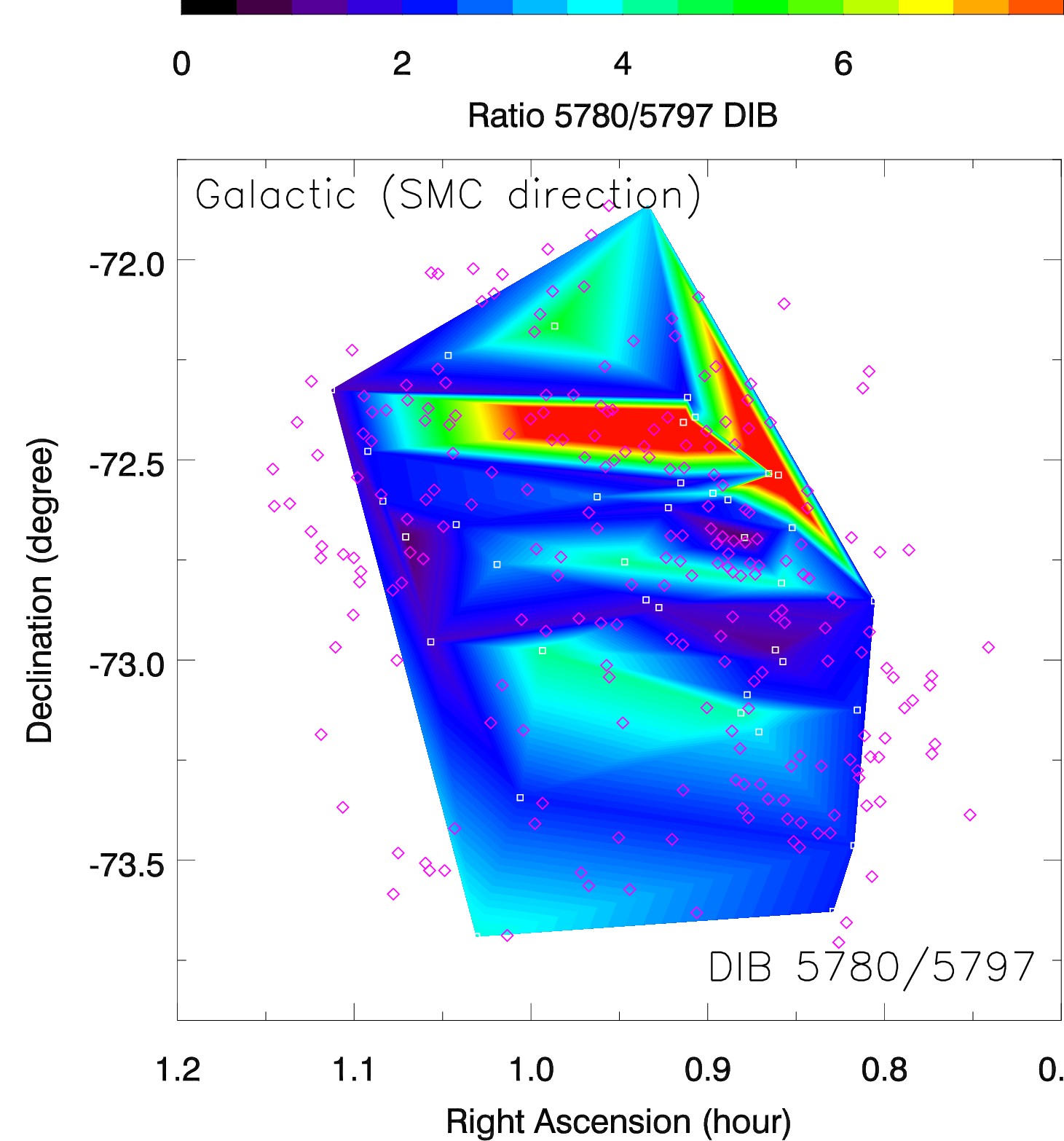,width=58mm}
}\vspace{2mm}\hbox{
\epsfig{figure=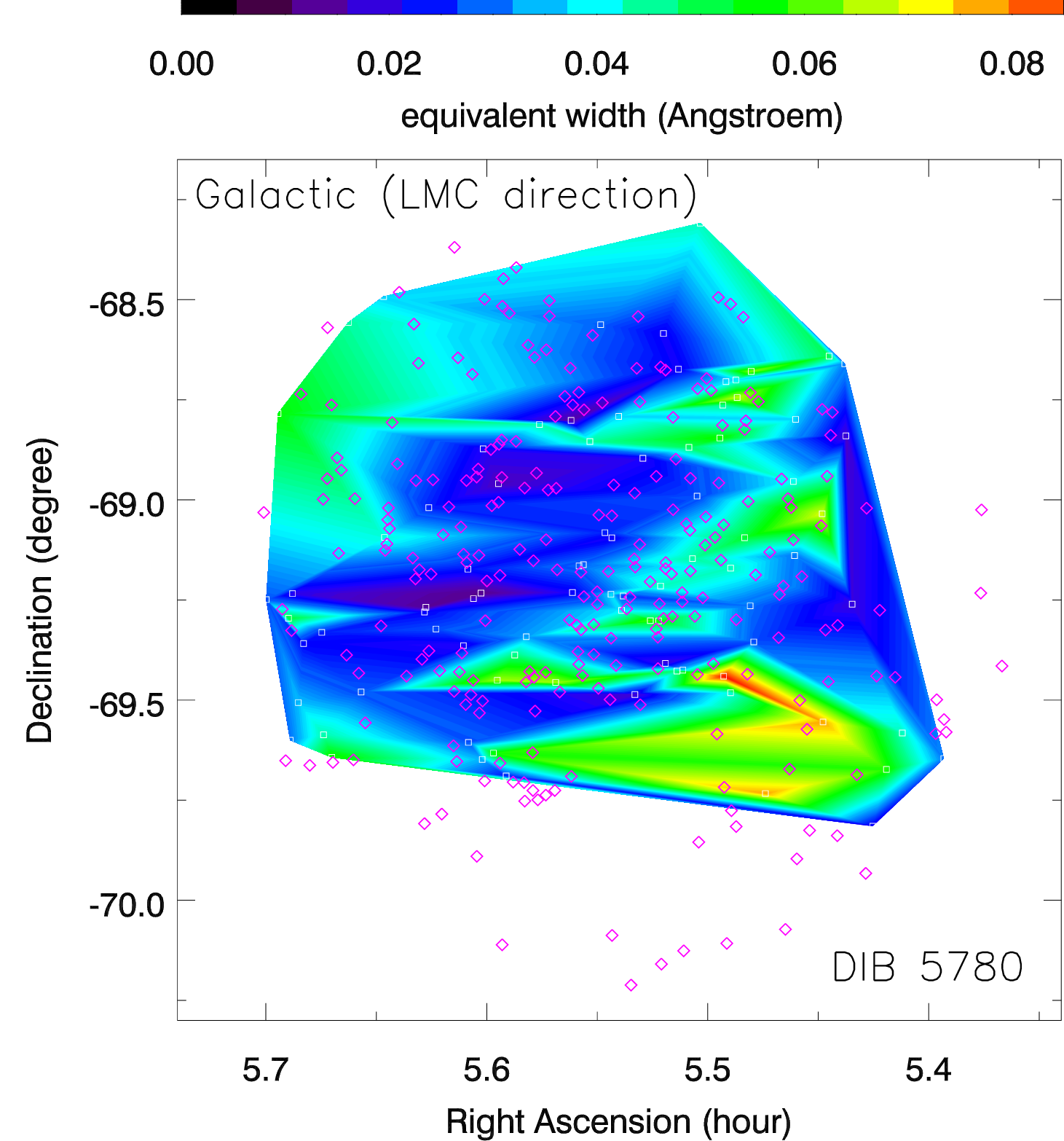,width=58mm}
\epsfig{figure=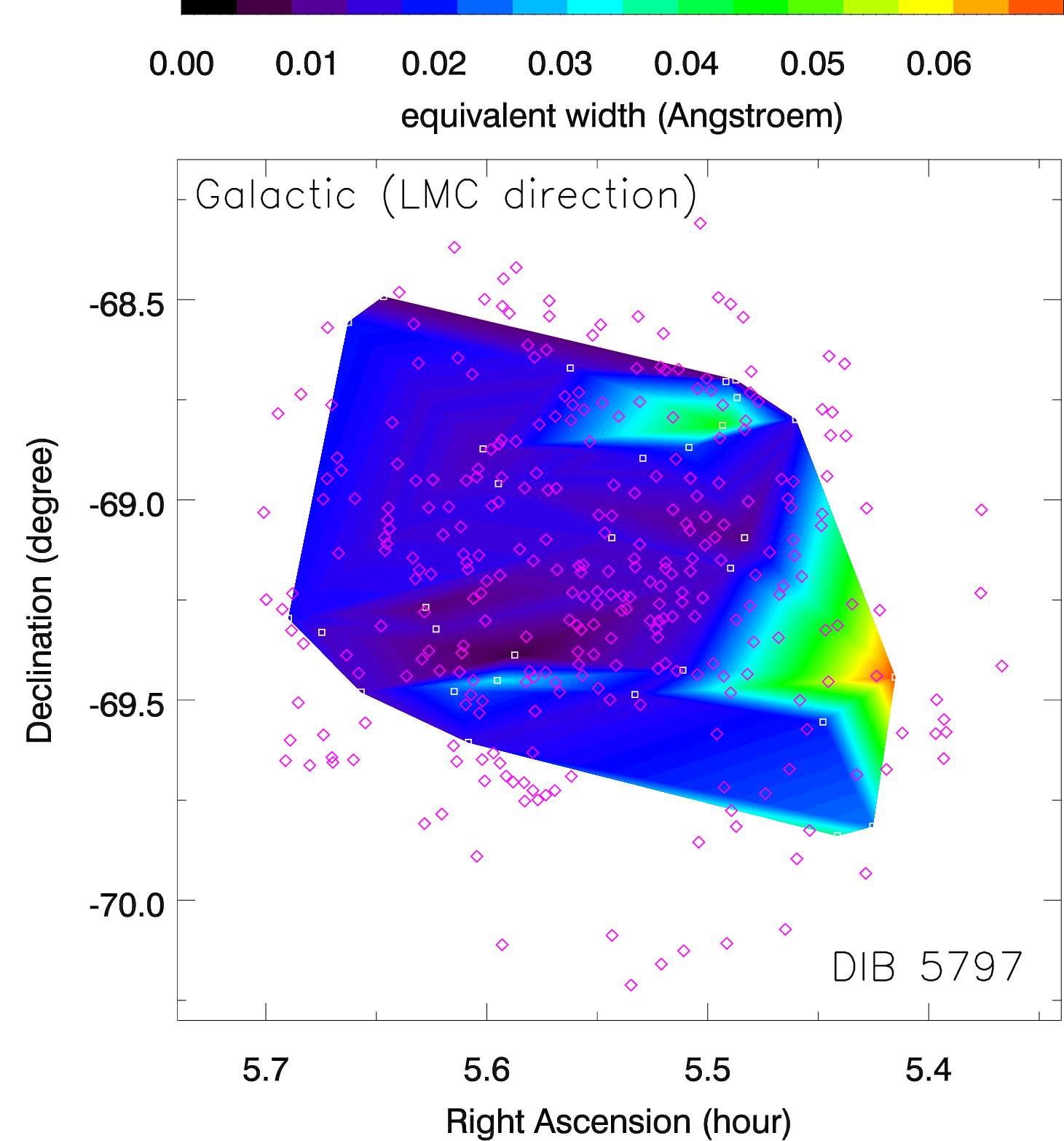,width=58mm}
\epsfig{figure=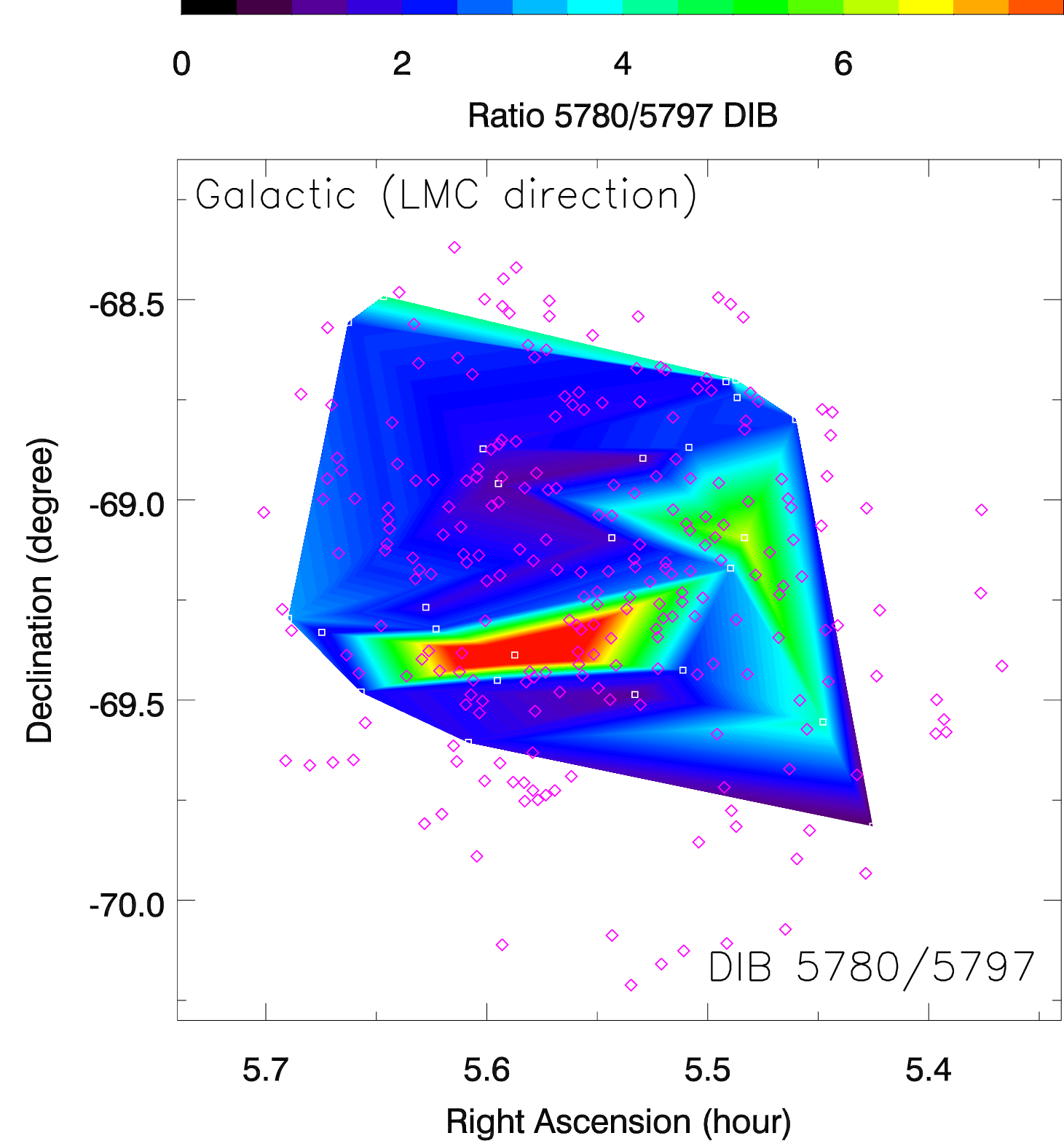,width=58mm}
}}}
\caption[]{Maps in the equivalent width of the 5780 and 5797 \AA\ DIBs, and
their ratio, for the Galactic foreground in the direction of ({\it top row:})
the SMC and ({\it bottom row:}) the LMC. The locations of the target stars are
indicated with the small white squares (detections) and magenta diamonds
(non-detections).}
\end{figure*}

%
\begin{figure*}
\centerline{\vbox{\hbox{
\epsfig{figure=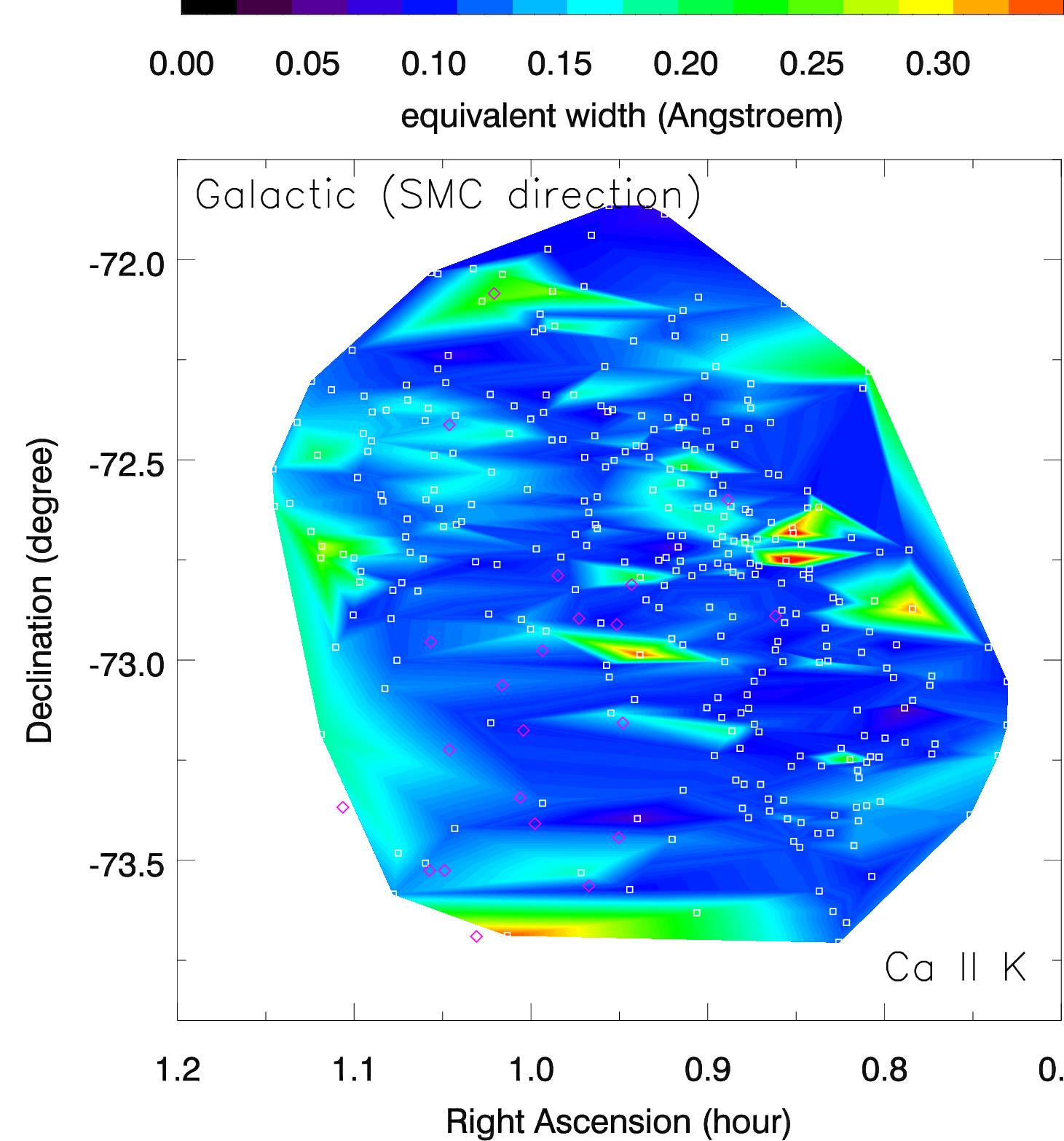,width=58mm}
\hspace{58mm}
\epsfig{figure=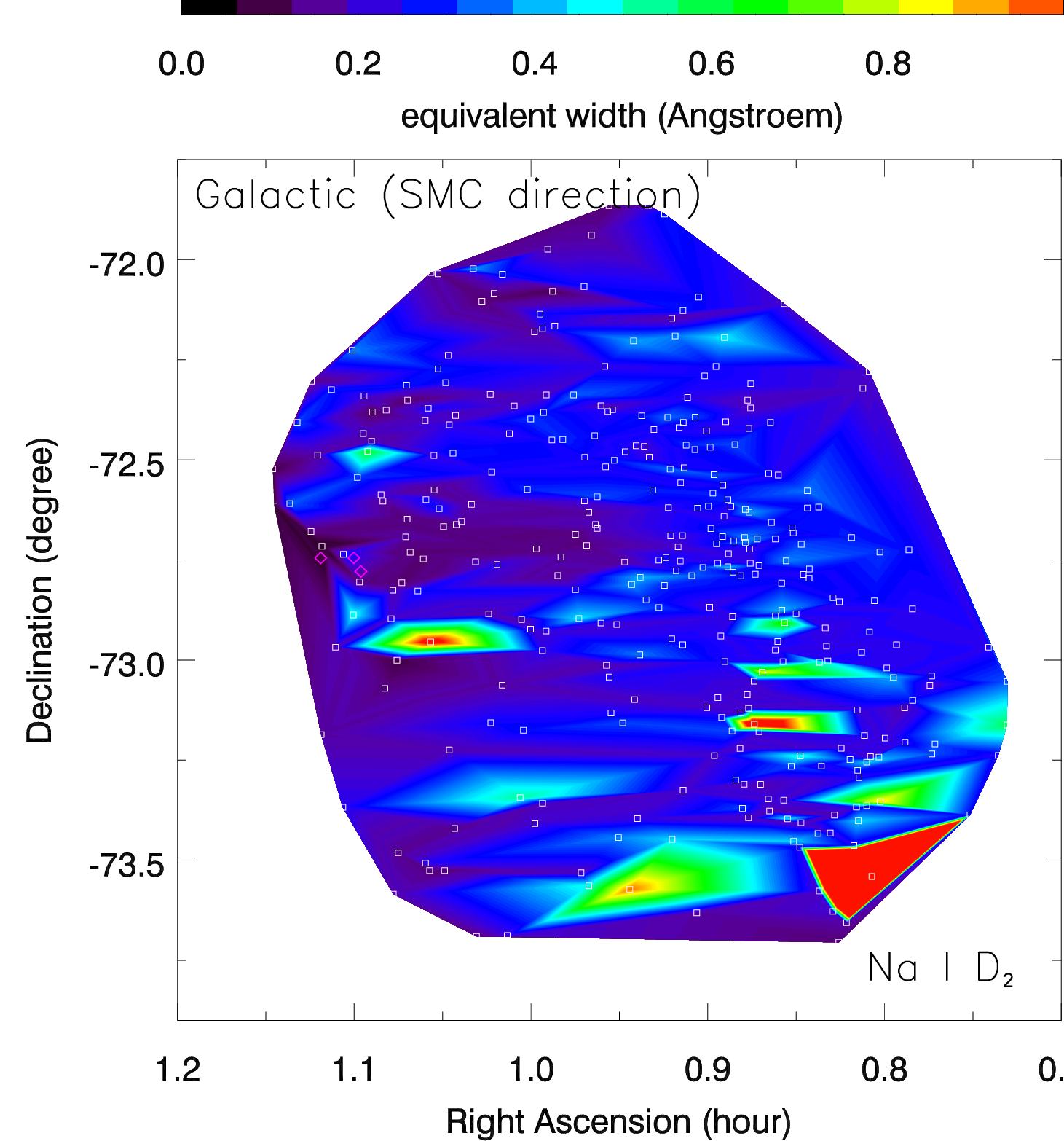,width=58mm}
}\vspace{2mm}\hbox{
\epsfig{figure=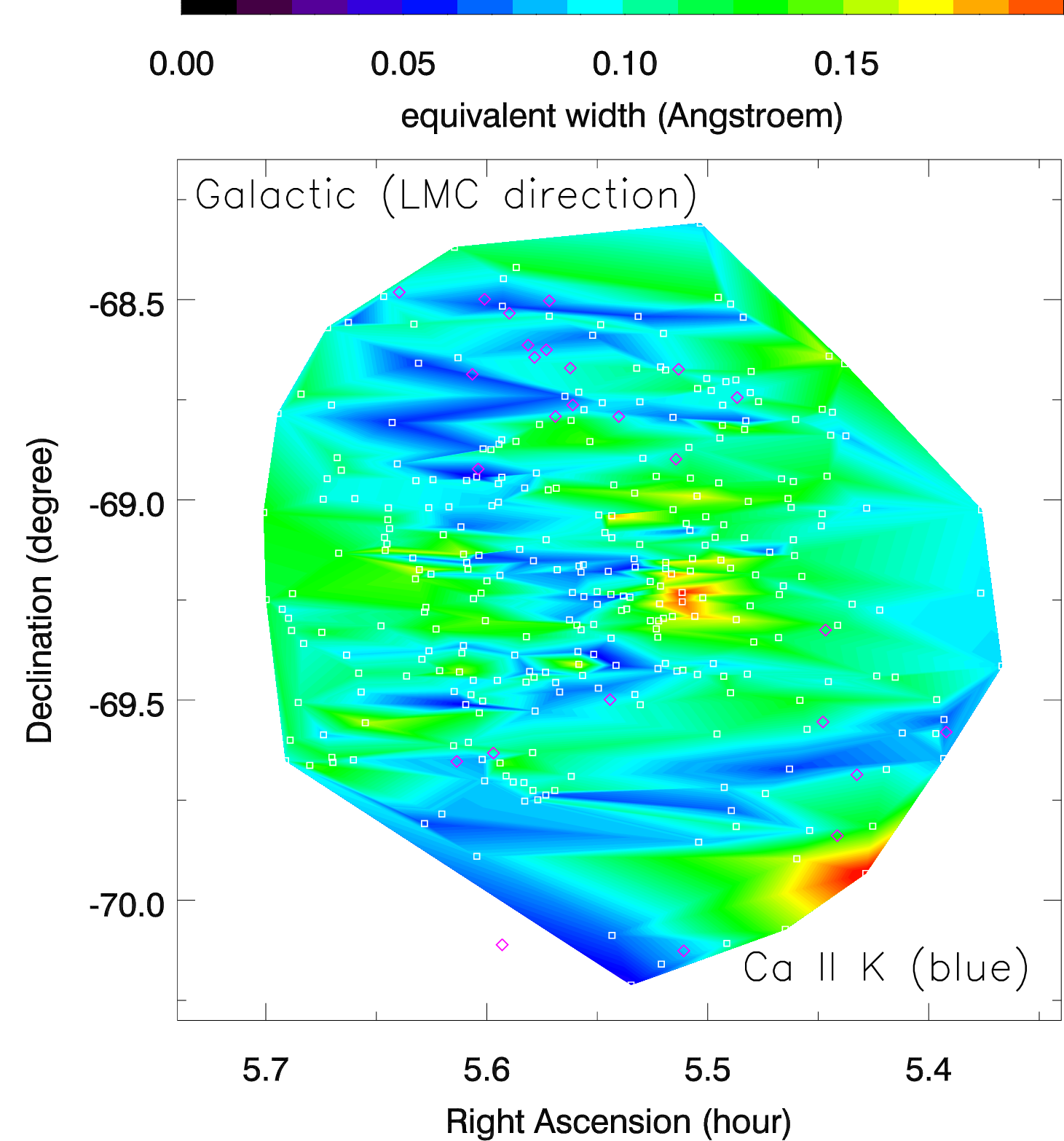,width=58mm}
\epsfig{figure=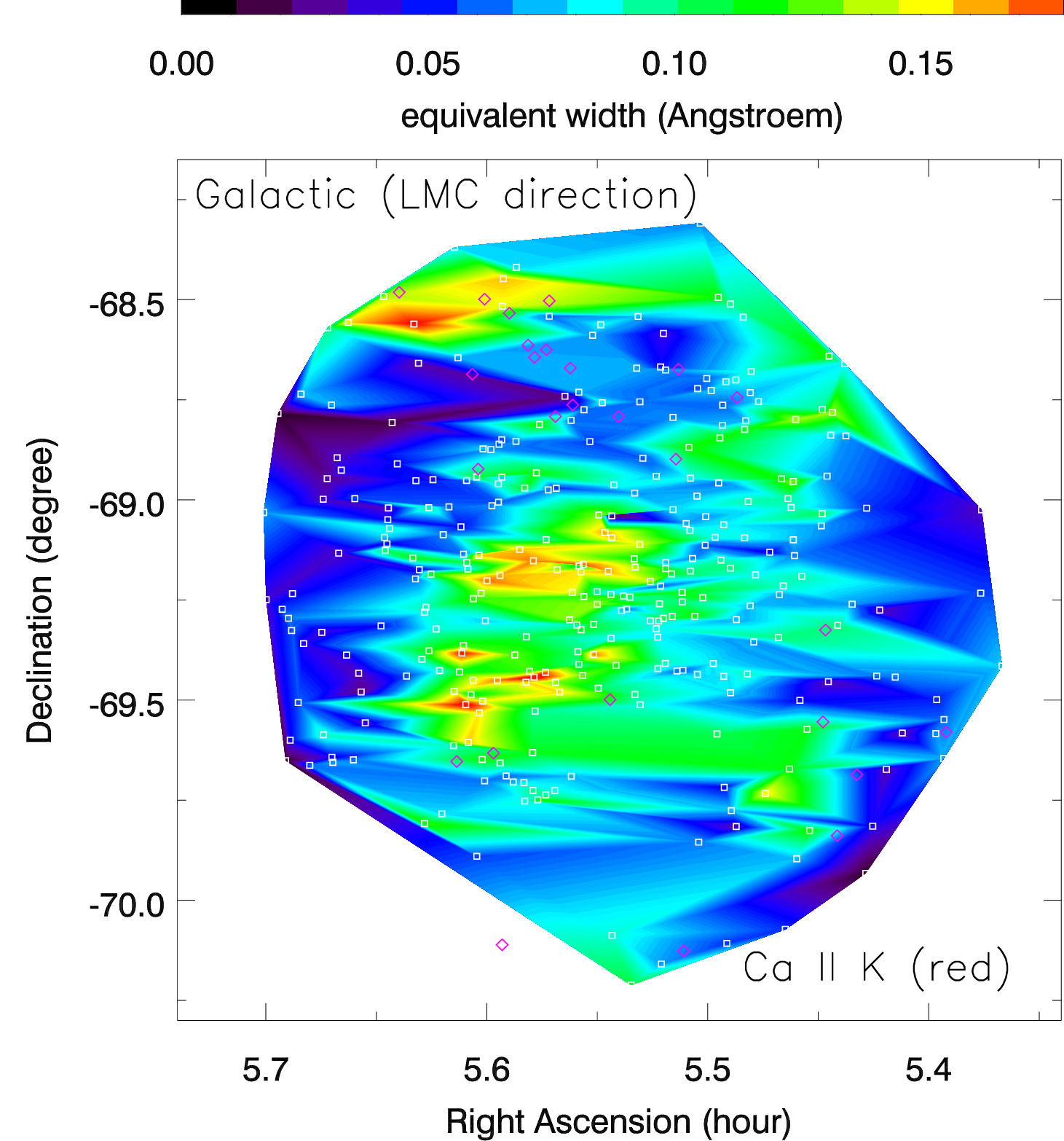,width=58mm}
\epsfig{figure=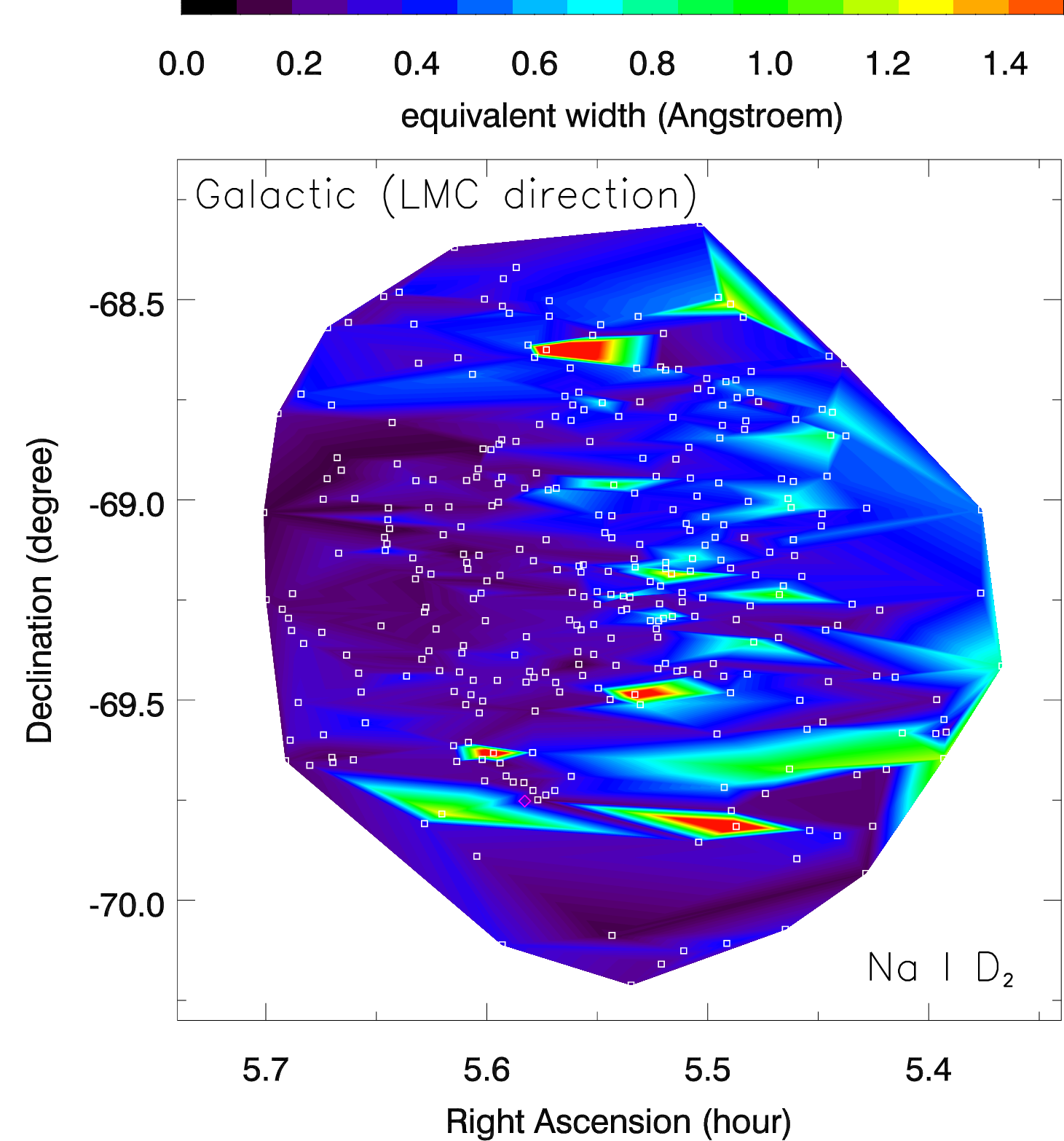,width=58mm}
}}}
\caption[]{As figure 9 but for Ca\,{\sc ii}\,K and Na\,{\sc i}\,D$_2$.}
\end{figure*}

\section{Discussion}

\subsection{The carriers of the DIBs}

The 5780 \AA\ DIB is approximately 2--3 times as strong as the 5797 \AA\ DIB
throughout most of the SMC and LMC. This contrasts with the very high ratio of
$EW_{5780}/EW_{5797}\sim6$ found towards R\,136, the ionizing cluster of the
Tarantula Nebula (van Loon et al.\ 2013). The latter is an extreme environment
and hence not typical of the ambient ISM within the Magellanic Clouds. The
$EW_{\rm NaI\,D1}/EW_{5780}\sim3$--4, on the other hand, is very similar between
our global maps and the R\,136 sight-line. The Local Bubble and its immediate
surroundings comprise a typical range of $EW_{5780}/EW_{5797}\sim2$--5 (Farhang
et al.\ 2015), concentrated around $EW_{5780}/EW_{5797}\sim3$ (Bailey et al.\
2015), which encompass the ratios found in the Magellanic Clouds. Very high
values of $EW_{5780}/EW_{5797}\sim5$--9 are seen {\it within} the Local Bubble
(Bailey et al.\ 2015); the highest reliable ratios within the Magellanic
Clouds are $EW_{5780}/EW_{5797}\sim7$. The ratio
$EW_{\rm NaI\,D1}/EW_{5780}\sim2$--3 found around the Local Bubble (Farhang et
al.\ 2015) is again commensurate with what is seen across the Magellanic
Clouds.

We note that the DIBs are conspicuously weak in the Southern molecular ridge
South from 30\,Doradus (towards N\,158) which is traced by Na\,{\sc i}\,D and
(to a lesser degree) Ca\,{\sc ii}\,K. This was also noted by van Loon et al.\
(2013) and may indicate the disappearance (or lack of excitation) of DIB
carriers in dense, UV-shielded environments.

The ionization potential of Na$^0$ is 5.1 eV, while that of Ca$^0$ and Ca$^+$
are 6.1 and 11.9 eV, respectively. Comparing this to the ionization potential
of H$^0$, 13.6 eV, it is clear that in H\,{\sc ii} regions neither Na\,{\sc i}
nor Ca\,{\sc ii} correspond to the dominant stage of ionization. However, in
cooler, generally neutral clouds Ca\,{\sc ii} will diminish with respect to
Na\,{\sc i}, which is further compounded due to depletion of calcium atoms
into grains. It is not yet clear whether DIB carriers survive when immersed in
photon baths of $>6$ eV, so mapping the DIBs alongside the Na\,{\sc i} and
Ca\,{\sc ii} may elucidate this point. Our maps show that the 5780 and 5797
\AA\ DIBs more closely trace Na\,{\sc i} than Ca\,{\sc ii}, with the 5797 \AA\
DIB tracing cooler and/or more shielded structures than the 5780 \AA\ DIB
does. This suggests that the carrier of the 5780 \AA\ DIB is removed by
particles/photons with energies in excess of 6 eV, while the carrier of the
5797 \AA\ DIB is already removed at energies in excess of 5 eV. As the 5780
\AA\ DIB stops growing in tandem with the 5797 \AA\ DIB for large column
densities, the carrier of the 5780 \AA\ DIB probably needs to be sustained by
particles/photons with energies of a few eV, making it likely to be a cation.

Remarkably, the equivalent-width diagrams of either the 5780 or 5797 \AA\ DIB
versus Na\,{\sc i}\,D (Figs.\ 5 \& 6) are indistinguishable between the SMC
and LMC but for the fewer sight-lines with strong absorption in any of these
tracers (an observation which is facilitated by our decision to plot all
diagrams between SMC, LMC and Galactic foreground on exactly the same scale).
Figure 11 shows this for the 5780 \AA\ DIB, by comparing all good
($>10\sigma$) detections between the SMC, LMC and Galactic foreground in one
graph. This implies that any metallicity dependence affects both sodium and
DIB carriers in essentially the same manner. This is very curious, especially
if the carriers of the DIBs are complex molecules. In that case one would
expect a stronger than linear relation between overall metal content and DIB
carrier. As the sodium atomic abundance scales approximately in proportion to
metallicity, apparently so does the DIB carrier. It means, in turn, that if
the DIB carriers are indeed multi-atomic compounds their formation must be
efficient and largely complete and this material must be resilient. For
instance if they are carbon-based, for instance in the form of fullerenes,
then much of the interstellar carbon must be locked within the DIB carriers.

%
\begin{figure}
\centerline{\epsfig{figure=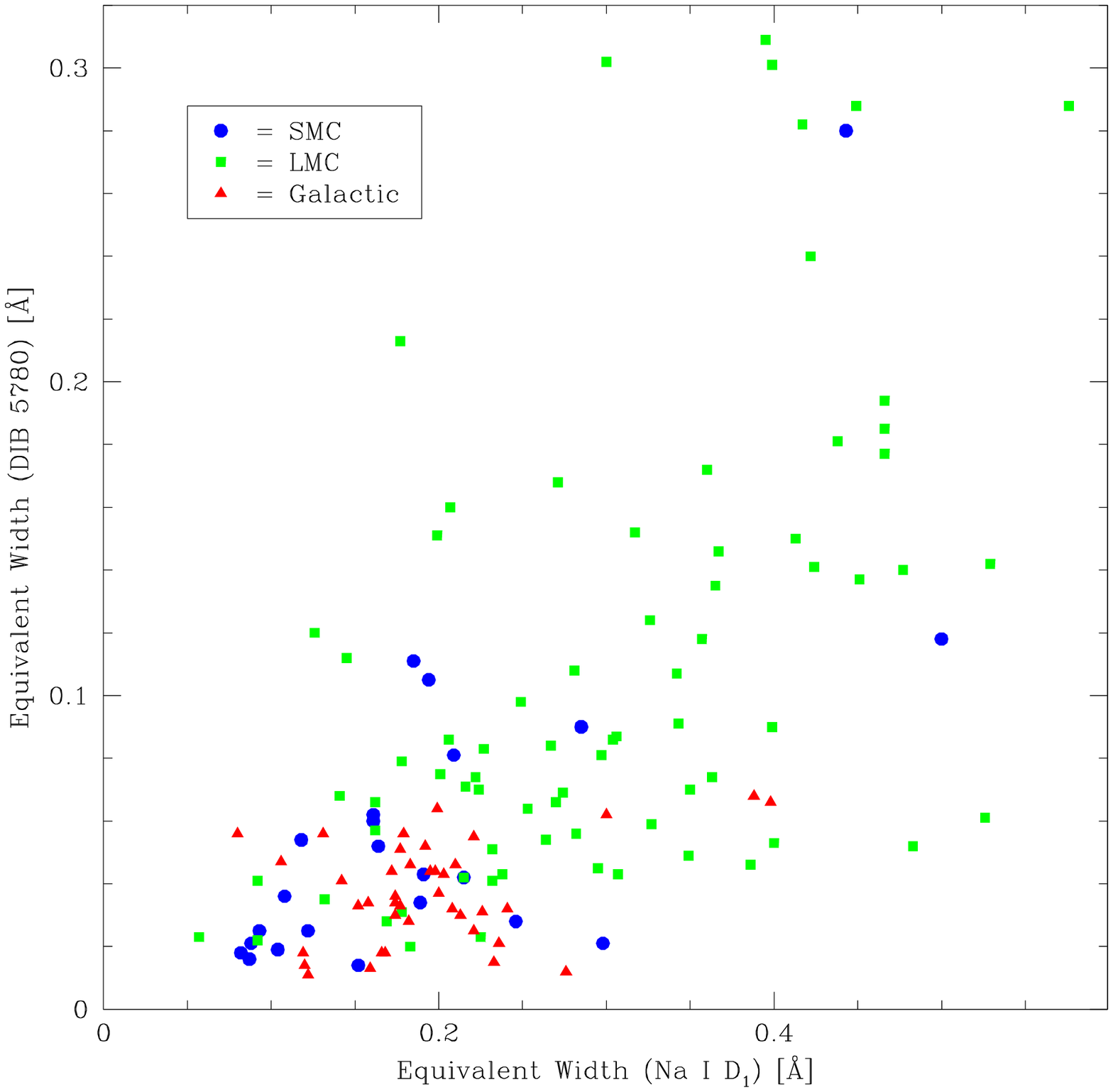,width=85mm}}
\caption[]{The equivalent width of the 5780 \AA\ DIB versus that of the
Na\,{\sc i}\,D$_1$, for the SMC (blue circles), LMC (green squares) and the
Galactic foreground (red triangles). Only $>10\sigma$ detections are plotted.}
\end{figure}

\subsection{Structure of the Galactic and Magellanic ISM}

The Galactic foreground maps of 5780 and 5797 \AA\ DIBs and Ca\,{\sc ii}\,K
and Na\,{\sc i}\,D absorption reveal structures on multiple scales -- up to a
degree down to as small as $\sim0\rlap{.}^\circ1$, which for typical distances
$d\sim200$ pc means linear scales of $\sim0.35$--3 pc. The same has been seen
in both DIBs and Na\,{\sc i}\,D through extra-planar sight-lines (van Loon et
al.\ 2009), and in Ca\,{\sc ii}\,K and Na\,{\sc i}\,D through intermediate-
and high-velocity clouds (IVCs and HVCs, respectively; Smoker, Fox \& Keenan
2015). Curiously, the correlation between the two DIBs, or Ca\,{\sc ii} and
Na\,{\sc i}, is very weak at best -- the maps show no conspicuous relationship
between each of these pairs of tracers. This suggests that where correlations
do exist it is likely due to the cumulative effect of a superposition of
structures in a multi-phase ISM in which each individual structure is
dominated by one or the other.

The leading edge of the LMC is dense in both Na$^0$ and Ca$^+$. The DIBs are
strongest in the star forming regions to the South and West of the Tarantula
Nebula; they clearly trace the warmer -- but not strongly-irradiated -- cloud
layers. This is different from both dense, UV-shielded molecular cloud tracers
such as CO and far-IR emission on the one hand (or reddening -- cf.\ Tatton et
al.\ 2013), and photo-dissociation regions on the other hand. It would be
interesting to see how well DIBs correlate with the unidentified IR emission
features commonly attributed to polycyclic aromatic hydrocarbon molecules.

The Na$^0$ velocities within the LMC cluster around $\sim280$ km s$^{-1}$, down
to $\sim240$ km s$^{-1}$. The red ``LMC'' Ca$^+$ component is mostly found at
velocities $\sim280$ km s$^{-1}$, i.e.\ associated with the bulk of the Na$^0$.
We can thus conclude that this absorption arises within the LMC. The blue
``LMC'' Ca$^+$ component is mostly found around $\sim220$ km s$^{-1}$, i.e.\
outside the Na$^0$ range. In contrast, it is the {\it blue} ``SMC'' Ca$^+$
component which coincides with the $\sim140$ km s$^{-1}$ of the Na$^0$ within
the SMC, whereas the red ``SMC'' Ca$^+$ component is found around 200--220 km
s$^{-1}$.

The similarity between the blue ``LMC'' and red ``SMC'' Ca$^+$ kinematics, and
their weaker spatial resemblance to the Magellanic Clouds, suggests that this
complex is not internal to the Magellanic Clouds, but instead arises within
the Milky Way Halo. De Boer, Koornneef \& Savage (1980) suggested that UV
absorption seen in front of the LMC at 220 km s$^{-1}$ has its origin within
coronal gas associated with the LMC. Ca\,{\sc ii} absorption was already
detected down to such low velocities towards 30\,Doradus by Blades \& Meaburn
(1980), and a $\sim200$ km s$^{-1}$ component was seen in H\,{\sc i} towards
the LMC and Magellanic Stream by McGee, Newton \& Morton (1983). Intermediate-
and high-velocity clouds such as those populating the Halo are generally
transparent in Na\,{\sc i} and show up much better in Ca\,{\sc ii} (Smoker et
al.\ 2015) -- hence perhaps why we noticed additional Ca\,{\sc ii} components
but not in Na\,{\sc i}. The fact that we detect this kinematic component even
towards the SMC suggests that the gas is not solely connected with an LMC
corona but perhaps forms part of a pan-Magellanic halo, possibly associated
with the putative common dark matter halo invoked by Bekki (2008).

\section{Summary of conclusions}

We conducted a high signal-to-noise spectroscopic survey of 666 early-type
stars within the SMC and LMC, and measured the absorption strength of the 5780
and 5797 \AA\ DIBs and of Ca\,{\sc ii}\,K and Na\,{\sc i}. We examined
correlations between these different ISM tracers and constructed for the first
time maps covering much of these galaxies. Our findings are:
\begin{itemize}
\item[$\bullet$]{The 5780 \AA\ DIB is found in more strongly irradiated and/or
hotter environments than the 5797 \AA\ DIB, but it too disappears when
conditions become too extreme. Both DIBs are depleted in dense molecular
clouds.}
\item[$\bullet$]{The 5780 and 5797 \AA\ DIBs have similar strength w.r.t.\
Na\,{\sc i} in the SMC, LMC and Milky Way. This suggests that the abundance of
their carriers is directly proportional to overall metallicity.}
\item[$\bullet$]{At low column densities, structure is seen at sub-pc scales
in all the above tracers, and their correlations exhibit large scatter. Good
correlations between different tracers seen in denser columns are probably due
at least in part to the averaging effect of multiple, distinct structures
along the sight-line even if different conditions prevail within them.}
\item[$\bullet$]{A common kinematic component around 200--220 km s$^{-1}$ is
seen in the Ca\,{\sc ii}\,K line towards both Magellanic Clouds, suggesting it
arises in a pan-Magellanic halo.}
\end{itemize}

\section*{Acknowledgments}
We thank Chris Evans for reading and commenting on an earlier version of the
manuscript, and the referee Francis Keenan for his positive (and swift)
report. MB acknowledges support from an STFC studentship at Keele University.
PJS thanks the Leverhulme Trust for the award of a Research Fellowship. The
AAT observing time was awarded under the European Union funded OPTICON
programme. This research has made use of the SIMBAD database, operated at CDS,
Strasbourg, France.

\begin{appendix}
\section{Spectral types}

%
\begin{table}
\caption{Pertinent literature sources for the spectral types, in chronological
order (by year, then alphabetical).}
\begin{tabular}{lc}
\hline\hline
{\it SMC:} & \\
Azzopardi, Vigneau \& Macquet & (1975) \\
Azzopardi \& Vigneau          & (1979) \\
Garmany, Conti \& Massey      & (1987) \\
Massey, Parker \& Garmany     & (1989) \\
Grebel, Roberts \& Brandner   & (1996) \\
Evans et al.\                 & (2004) \\
Evans et al.\                 & (2006) \\
Martayan et al.\              & (2007) \\
Evans \& Howarth              & (2008) \\
Hunter et al.\                & (2008) \\
Bonanos et al.\               & (2010) \\
Paul et al.\                  & (2012) \\
Lamb et al.\                  & (2013) \\
Ridley et al.\                & (2013) \\
\hline
{\it LMC:} & \\
Sanduleak                              & (1970) \\
Schild \& Testor                       & (1992) \\
Parker                                 & (1993) \\
Garmany, Massey \& Parker              & (1994) \\
Testor \& Niemela                      & (1998) \\
Massey, Waterhouse \& DeGioia-Eastwood & (2000) \\
Bosch et al.\                          & (2001) \\
Fari\~na et al.\                       & (2009) \\
Walborn et al.\                        & (2010) \\
Weidner \& Vink                        & (2010) \\
Massey et al.\                         & (2012) \\
Reid \& Parker                         & (2012) \\
Rivero Gonz\'alez et al.\              & (2012) \\
Sab\'{\i}n-Sanjuli\'an et al.\         & (2014) \\
McEvoy et al.\                         & (2015) \\
Evans et al.\                          & (2015) \\
\hline
\end{tabular}
\end{table}

We performed a search in the literature for know spectral types, thereby
mostly relying on the use of SIMBAD (Wenger et al.\ 2000); a summary of the
pertinent sources for these spectral types is presented in table A1. These
spectral types have been included in our catalogue, and comprise 174 SMC and
61 LMC targets. Their distribution over spectral type and luminosity class is
presented in figure A1. The more extensive coverage of the SMC includes also
more later (A) type stars, whilst the known spectral types for LMC sources
tend to include more O-type classes. Most of the targets with known spectral
type are mid-B (SMC) or early-B (LMC).

%
\begin{figure}
\centerline{\epsfig{figure=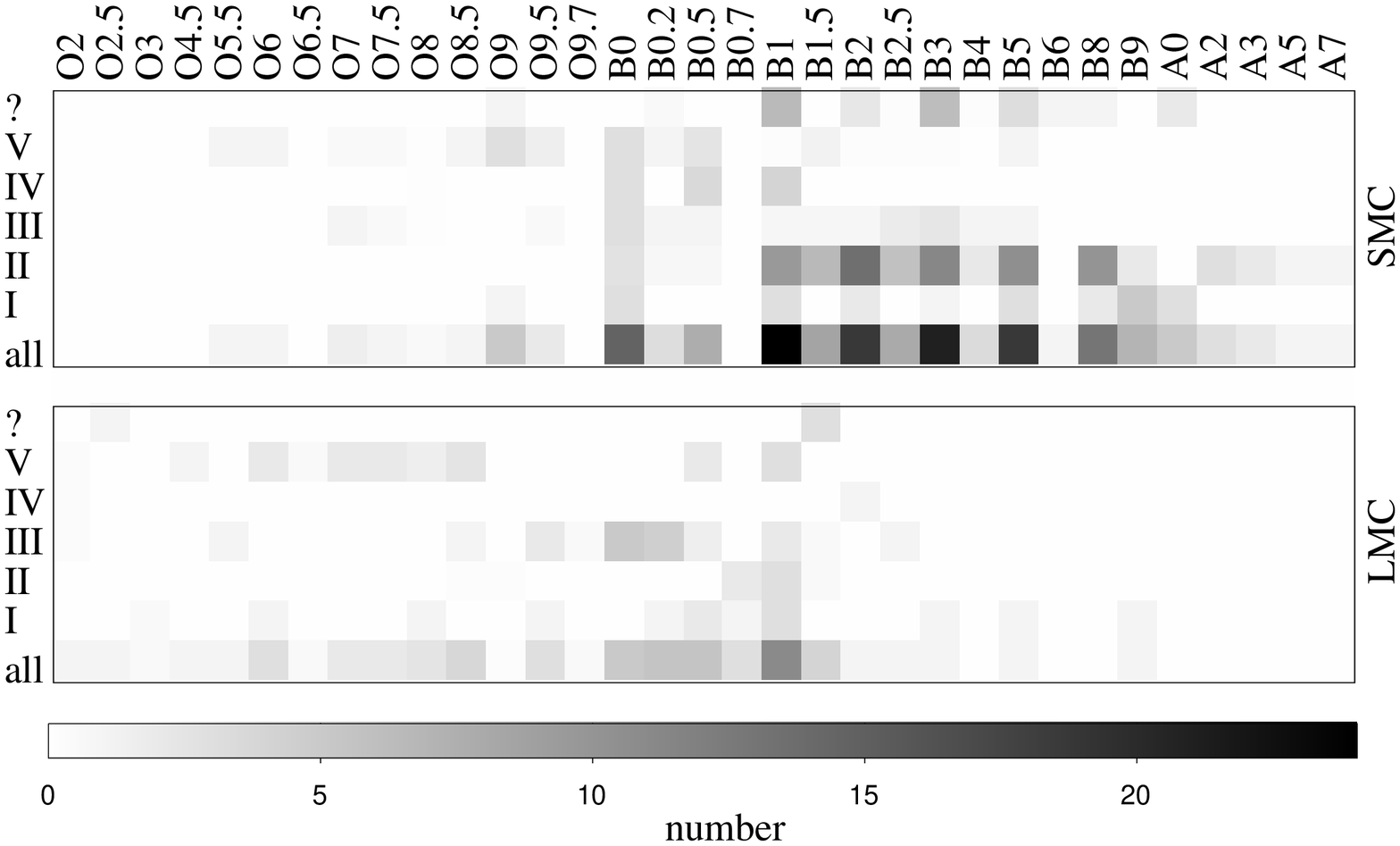,width=85mm}}
\caption[]{Distribution over spectral type and luminosity class, for the ({\it
top:}) SMC and ({\it bottom:}) LMC.}
\end{figure}

%
\begin{figure}
\centerline{\epsfig{figure=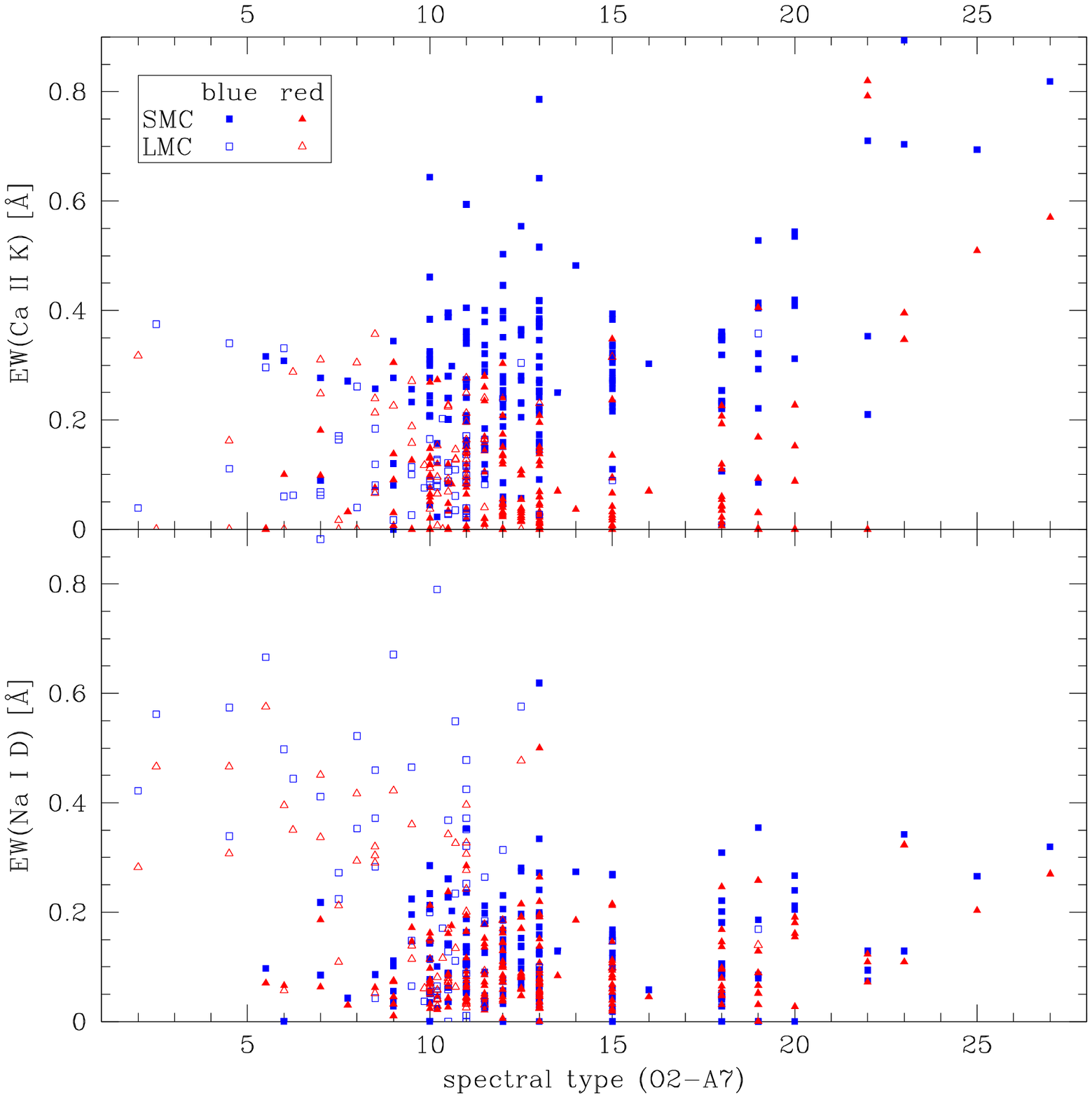,width=85mm}}
\caption[]{Distribution over spectral type and luminosity class, for the ({\it
top:}) SMC and ({\it bottom:}) LMC.}
\end{figure}

The question arises concerning the contribution from stellar photospheric
absorption to the measurements of the calcium and sodium lines. Indeed,
equivalent widths of the Ca\,{\sc ii}\,K line can reach 0.1--0.2 \AA\ at solar
metallicity (Mihalas 1973; Keenan et al.\ 1988). Figure A2, however, exhibits
no dependence of the Ca\,{\sc ii}\,K absorption on spectral type, except
perhaps for the A type stars that seem to have larger equivalent widths (but
that are few in number). The equivalent width of the stronger (blue) kinematic
component is mostly 0.2--0.4 \AA\ and thereby larger than what is expected for
the stellar photosphere of (metal-poor) O- and B-type stars; the other (red)
kinematic component is often much weaker but the same insensitivity to
spectral type holds true. The same conclusion is reached for the
Na\,{\sc i}\,D$_2$ (blue) and D$_1$ (red) lines (bottom panel in figure A2).

We can also remark that the Ca\,{\sc ii}\,K absorption is of similar strength
in the SMC and LMC, whereas we would have expected a clear difference if it
mainly originated in the stellar photospheres due to the considerable
difference in metallicity between the young populations of the SMC and LMC.
The Na\,{\sc i}\,D absorption, on the other hand, does show the expected
scaling with metallicity -- but this can also be explained if it were of
purely interstellar origin.

This all appears to be rather reassuring and lends credibility to the analysis
of the ISM absorption. Spectral typing of the remainder of the target sample
will be the subject of a forthcoming paper.

\end{appendix}

\label{lastpage}

\end{document}